\documentclass[10pt,draftcls,onecolumn]{IEEEtran}  
\IEEEoverridecommandlockouts                             

\usepackage{cite}
\usepackage{multicol}
\usepackage{amsmath}
\usepackage{amssymb}
\usepackage{graphicx,psfrag}
\usepackage{epstopdf}
\usepackage{caption}
\usepackage{subcaption}
\usepackage{tikz,verbatim}

\newtheorem{result}{Result}

\title{Right buffer sizing matters:\\ some dynamical and statistical studies on Compound TCP}

\author{Debayani Ghosh, Krishna Jagannathan and Gaurav Raina
\thanks{D.~Ghosh, K.~Jagannathan and G.~Raina are with the Department of Electrical Engineering, IIT
Madras, Chennai 600036, India. Email: \{ee12s052,   krishnaj, gaurav\}@ee.iitm.ac.in
        }
\thanks{A part of this work appeared in~\cite{Jagannathan}.}
}

\begin{document}
\maketitle
\begin{abstract}
Motivated by recent concerns that queuing delays in the Internet are on the rise, we conduct a performance evaluation of Compound TCP (C-TCP) in two topologies: a single bottleneck and a multi-bottleneck topology, under different traffic scenarios. The first topology consists of a single bottleneck router, and the second  consists of two distinct sets of TCP flows, regulated by two edge routers, feeding into a common core router. We focus on some dynamical and statistical properties of the underlying system. From a dynamical perspective, we develop fluid models in a regime wherein the number of flows is large, bandwidth-delay product is high, buffers are dimensioned small (independent of the bandwidth-delay product) and routers deploy a Drop-Tail queue policy. A detailed local stability analysis for these models yields the following key insight: smaller buffers favour stability. Additionally, we highlight that larger buffers, in addition to increasing latency, are prone to inducing limit cycles in the system dynamics, via a Hopf bifurcation. These limit cycles in turn cause synchronisation among the TCP flows, and also result in a loss of link utilisation. For the topologies considered, we also empirically analyse some statistical properties of the bottleneck queues. These statistical analyses serve to validate an important modelling assumption: that in the regime considered, each bottleneck queue may be approximated as either an $M/M/1/B$ or an $M/D/1/B$ queue. This immediately makes the modelling perspective attractive and the analysis tractable. Finally, we show that smaller buffers, in addition to ensuring stability and low latency, would also yield fairly good system performance, in terms of throughput and flow completion times.


\end{abstract}
\IEEEpeerreviewmaketitle
\section{Introduction}
There is an increasing concern regarding large queuing delays in today's Internet. This rise in queuing delays has primarily been attributed to a phenomenon called \emph{bufferbloat}; \emph{i.e.}, the presence of large and persistently full buffers in Internet routers \cite{Cerf}, \cite{Gettys}. Excessive queuing delays, caused by these large buffers, would be a hindrance to the efficient functioning of various real-time delay-sensitive applications such as Voice over IP (VoIP), live streaming video and online gaming. Several factors impact the end-to-end latency, and hence the quality of service in the Internet: namely, size of buffers in routers, the choice of TCP, and the queue management scheme implemented at the routers. Currently, three buffer sizing regimes have been proposed in the literature \cite{Wischik}: a large, an intermediate, and a small buffer regime. In practice, today's router buffers follow the large buffer rule. In particular, this rule mandates the buffer size $B=C\times \overline{RTT},$ where $C$ is the link capacity of the router, and $\overline{RTT}$ is the harmonic mean of the round trip times of the flows accessing the router~\cite{Dovrolis, Villamizar}. In practice, $\overline{RTT}$ is typically chosen to be $250$ ms. This rule leads to larger buffers as the capacity of the router increases. The intermediate buffer rule mandates the buffer size $B=C\times \overline{RTT}/\sqrt{N},$ where $N$ is the number of long-lived flows multiplexed at the router~\cite{McKeown}. In the small buffer regime, the buffer size at the router is chosen independent of the bandwidth-delay product~\cite{Wischik}.


There have also been numerous proposals for TCP flavours in the literature. However, Compound TCP \cite{Tan} (C-TCP) is the default protocol in the Windows operating system and Cubic TCP \cite{Ha} is used in Linux. Recent studies \cite{Peng} have shown that $15\%\sim25\%$ of $30,000$ web servers implement Compound TCP. Given the large fraction of web servers that currently use Compound TCP, we primarily focus on Compound TCP for our study. 

As far as queue management is concerned, solutions have been proposed in an attempt to eradicate the pervasive problem of excessively large queuing delays. The primary aim of an active queue management strategy is to maintain the bottleneck buffers small by dropping or marking packets before the buffers become full. Some common examples of active queue management strategies are RED~\cite{Jacobson}, CODEL~\cite{Nichols}, and PIE~\cite{Rong}. However, in practice, router buffers widely deploy a simple Drop-Tail policy which drops incoming packets if the buffer is full.

In this paper, we conduct a performance evaluation of Compound TCP, in conjunction with small Drop-Tail buffers, in two topologies. We start with a single bottleneck topology and then proceed towards a more complex topology with three bottleneck routers. At a broad level, we distinguish between dynamical and statistical properties of the underlying system.  We wish to emphasise that a single bottleneck topology has been widely used in the literature to understand the properties of TCP~\cite{Hollot,Manjunath,Wischik,Raja}. However, to the best of our knowledge, this work is the first to analyse the properties of Compound TCP in a multiple bottleneck topology. 
\subsection{Dynamical properties}

The first topology we consider consists of a single bottleneck router. A large number of TCP flows feed into this router, either with equal round trip times or with heterogeneous round trip times. For the traffic, we consider three scenarios. In the first scenario, the traffic consists of a large number of only long-lived flows. In the second scenario, apart from the presence of many long-lived flows, short flows arrive and depart the network. In the third scenario, we assume file sizes to be drawn from a heavy-tailed distribution~\cite{Ayesta,Joo}. This is motivated by the fact that measurements on real Internet traffic show the presence of high variability at the connection level~\cite{Willinger}.

For the single bottleneck topology, we first outline the fluid models for various scenarios. We then conduct a local stability analysis, in the small buffer regime, and derive conditions that ensure local stability and non-oscillatory convergence. A key insight obtained from our stability analysis is that \emph{smaller buffers are favourable for local stability}. In fact, even minor variations in sizing router buffers would drive the underlying dynamical systems into a locally unstable regime.

The second topology is comprised of two edge routers fed by two sets of long-lived TCP flows, each with a different round trip time. The outputs from the edge routers feed into a core router. For this topology, deriving the necessary and sufficient condition for local stability in full generality is rather hard. To make progress, we conduct the stability analysis with two simplifying assumptions, and outline necessary and sufficient conditions for local stability. 
We also outline a rather simple sufficient condition for local stability, which could provide design guidelines to dimension router buffers in a decentralised manner. For the multiple bottleneck topology, our analysis highlights the importance of \emph{smaller buffers in ensuring local stability}. Further, larger buffers would drive the system from a locally stable to an unstable regime.


Another important contribution of this work lies in determining the behaviour of the system as it transits from a locally stable to an unstable regime. In this work, the fluid models outlined for Compound TCP are parameterised, non-linear, time-delayed dynamical systems. Such time-delayed systems can readily lose local stability if either the feedback delay or other system parameters vary~\cite{Hale}. In our models, we show that the transition from stability to instability occurs via a Hopf bifurcation, which alerts us to the emergence of limit cycles~\cite{Guckenheimer,Hassard}. Such limit cycles were indeed observed in our packet-level simulations, conducted via the Network Simulator version 2.35 (NS2)~\cite{ns2}. In the context of TCP, the emergence of limit cycles manifest as: $(i)$ synchronisation effects among TCP windows, $(ii)$ periodic oscillations in the queue-size occupancy and $(iii)$ loss of link utilisation. 

\subsection{Statistical properties}

In this work, we also empirically investigate some statistical properties of the bottleneck queues, in the presence of a large number of TCP flows. Numerous studies have empirically shown that real Internet traffic exhibits long range dependence~\cite{Paxson,Willinger}, and a Poisson modelling for the packet arrivals might not be appropriate~\cite{Paxson}. However, these studies are applicable to core links which typically use bandwidth-delay worth of buffering. In \cite{Ayesta,Vishwanath}, the authors have shown that with small Drop-Tail queues and a large number of TCP Reno flows, the packet arrival process can be approximated as Poisson. A statistical analysis with Compound TCP is in order. In particular, we pay attention to the arrival process to the queue as well as the queue size distribution.

For the single bottleneck topology, our empirical study reveals that in the absence of synchronisation, \emph{the bottleneck queue may be well approximated by either an $M/M/1/B$ or an $M/D/1/B$ queue}. We would like to emphasise that this approximation holds reasonably well even in the presence of high variability at the TCP connection level. Thus, for the analysis, this allows us to approximate the drop probability of the bottleneck queue using the blocking probability of an $M/M/1/B$ queue. Notably, this insight carries over even to the multiple bottleneck topology. In particular, our empirical study highlights that, when a large number of flows feed into each of the edge routers, \emph{each bottleneck queue can be well approximated by either an $M/M/1/B$ or an $M/D/1/B$ queue.} This validates our theoretical approximation of the drop probability at each queue using the loss probability expression of an $M/M/1/B$ queue, for our local stability analysis. 

Our analysis highlights the importance of smaller buffers in ensuring stability. Hence, it becomes imperative to study the impact of buffer sizing on both network and user performance. To that end, we consider two performance metrics: throughput and Average Flow Completion Time (AFCT), and show that it is indeed possible to significantly reduce buffers at routers while guaranteeing acceptable network and user performance.

In summary, the primary contributions of our work are the following:

(1) From a dynamical perspective, we highlight the interplay between buffer sizes and stability of the systems in the presence of feedback delays. In particular, we show that smaller buffers aid stability. Additionally, larger buffers would increase queuing delay, in addition to inducing limit cycles in the system dynamics. We then show that this insight holds true in each of the topologies, and traffic scenarios considered in our work. This lends credence to the fact that the loss of stability and consequently the emergence of limit cycles is primarily influenced by the buffer sizes at the routers, in the presence of large feedback delays.

(2) From a statistical perspective, we show that the bottleneck queues can be well approximated by either an $M/M/1/B$ or an $M/D/1/B$ queue, with smaller buffers. Notably, this insight remains consistent across the topologies considered. This validates our modelling assumption, and makes our system models amenable to analysis.

(3) We show that smaller buffers can indeed be realised without degrading the system performance, namely, throughput and flow completion times.


The rest of this paper is organised as follows. In Section \ref{compound}, we briefly outline the congestion avoidance algorithm of Compound TCP. In Section \ref{model_a}, we analyse a single bottleneck topology with long-lived, and a combination of long- and short-lived flows, with a single feedback delay. In Section \ref{model_ah}, we study the single bottleneck topology with heterogeneous  delays. In Section \ref{model_aheavy}, we analyse the single bottleneck topology under a traffic scenario wherein users generate heavy-tailed files. Some of our analytical insights are corroborated by packet-level simulations, conducted using NS2. In Section \ref{model_b}, we study a multiple bottleneck topology using a combination of analysis and packet-level simulations. In Section \ref{performance_multi}, we investigate the impact of our buffer sizing recommendations on the system performance. Finally, in Section \ref{conclusions} we summarise our contributions.

\section{Compound TCP}
\label{compound}
Compound TCP (C-TCP) \cite{Tan} is a widely implemented Transmission Control Protocol (TCP) in the Windows operating system. Transport protocols like Reno and HighSpeed  (HSTCP) use packet loss as the only indication of congestion, and protocols like Vegas \cite{Brakmo} uses only queuing delay as the measure of network congestion. C-TCP is a synergy of both loss and delay-based feedback. The motivation behind incorporating both forms of feedback in C-TCP is to achieve high link utilisation and also to provide fairness to other competing TCP flows.\\
\indent Compound TCP incorporates a scalable delay-based component into the congestion avoidance algorithm  of TCP Reno. C-TCP controls its packet sending rate by maintaining two windows, a loss window $cwnd$ and a delay window $dwnd$. In a time period of one round trip time, C-TCP updates its sending window $w$ as follows:
\begin{equation}
\label{eq:window}
w = \min\left(cwnd+dwnd,awnd\right).
\end{equation}
Here, $awnd$ is the advertised window at the receiver side. The decision function \eqref{eq:window} governing the evolution of the sending window guarantees flow control between the source and the destination if the end systems operate at different speeds. In our paper, we assume that the sending window is constrained only by the congestion in the network path and not by the congestion at the receiver. Hence, for C-TCP, the source's sending window will always be $cwnd+dwnd$. The loss window $cwnd$, behaves similar to the loss window of TCP Reno and follows the Additive Increase and Multiplicative Decrease (AIMD) rule whereas the delay window $dwnd$, controls the delay-based component. If there is no loss detected, $cwnd$ increases by one packet in one round trip time and reduces by half whenever a loss is signalled. The algorithm for the delay-based component of Compound TCP is motivated from TCP Vegas. A state variable, $baseRTT$ gives the transmission delay of a packet in the network path. If the current round trip time of the TCP flow is RTT, then
\begin{equation*}
diff =\left( \frac{{w}}{{ baseRTT}} - \frac{{w}}{{RTT}}\right) baseRTT,
\end{equation*}
gives the amount of backlogged data in the bottleneck queue. If $diff$ is less than the congestion threshold $\gamma$, the network is considered underutilised and the TCP flow increases its packet sending rate. If $diff$ exceeds $\gamma$, congestion is detected in the network path which prompts the TCP flow to decrease its delay-based component. In C-TCP implementation, the default value of $\gamma$ is fixed to be $30$ packets. The overall behaviour of the window size of a C-TCP sender can then be summarised by combining the loss window and the delay window. When there is no congestion in the network path, neither in terms of increased queuing delay nor packet losses, a C-TCP sender increases its window size in its congestion avoidance phase as follows:
\begin{equation}
w(t+1)=w(t) + {\alpha w(t)^k }.
\end{equation}
Here, $\alpha$, $k$ are the increase parameters and their default values are $\alpha=0.125$ and $k=0.75$ \cite{Tan} respectively. If a packet loss is detected, the window size is multiplicatively reduced as follows:
\begin{equation}
w(t+1)=w(t) {(1-\beta)}.
\end{equation}
Here, $\beta$ is the decrease parameter and its default value is $\beta=0.5$ \cite{Tan}.

\section{ Single bottleneck with homogeneous delay}
\label{model_a}

This topology consists of a single bottleneck link with \emph{many} TCP flows feeding into a bottleneck router (see Fig.~1). We consider the case where the buffer at the core router is sized \emph{small}~\cite{Wischik} and deploys a Drop-Tail queue policy. The flows are subject to a common round trip time $\tau$, and the bandwidth-delay product is assumed to be \emph{large}. Let the \emph{average} window size of the flows be $w(t)$. Then the average rate at which packets are sent is approximately $x(t)=w(t)/\tau$. Let the average congestion window increase by $i(w(t))$ for each received acknowledgement and decrease by $d(w(t))$ for each packet loss detected. The following non-linear, time-delayed differential equation describes the evolution of the \emph{average} window size in the congestion avoidance phase \cite{Raja}
\begin{align}
\label{eq:modela}
\frac{dw(t)}{dt} =& \frac{w(t-\tau)}{\tau}\bigg(i\left(w(t)\right)\Big(1-p(t-\tau)\Big)- d\left((w(t)\right)p(t-\tau)\bigg), 
\end{align}
where $p(t)$ denotes the loss probability experienced by packets sent at time $t$. We assume that the packet losses are independent across all TCP flows. If the server capacity is high, then it is easy to see that the packet loss probability would depend on the instantaneous arrival rate and consequently on the window size $w(t).$  We can then rewite \eqref{eq:modela} as 
\begin{align}
\label{eq:modela2}
\dot{w}(t) = & \frac{w(t-\tau)}{\tau}\bigg(i\left(w(t)\right)\Big(1-p(w(t-\tau))\Big)- d\left((w(t)\right)p(w(t-\tau))\bigg).
\end{align}
We analyse and study system \eqref{eq:modela2} in three scenarios. In the first scenario, we assume that all TCP flows are long-lived, \emph{i.e.}, each TCP source sends an infinite sized file. In the second scenario, the traffic is a mix of long and short-lived flows. In the third scenario, each TCP source sends a Poisson stream of finite sized connections, and the size of each connection is sampled from a heavy-tailed dstribution.

\begin{figure}
\begin{center} 
\vspace{2mm}
{
\begin{tikzpicture}[scale=0.80]
\draw (0,0) -- (-2,0.7);
\draw (-2,-0.7) -- (0,0);
\draw (0,-0.5) -- (4,-0.5);
\draw (4,-0.5) -- (4,0.5);
\draw (4,0.5) -- (0,0.5);
\draw (0,0.5) -- (0,-0.5);
\draw (3.75,-0.5) -- (3.75,0.5);
\draw (3.5,-0.5) -- (3.5,0.5);
\draw (3.25,-0.5) -- (3.25,0.5);
\draw (4.5,0) circle(0.5 cm);
\draw [loosely dotted,very thick] (3.1,0) -- (2.6,0);
\draw [loosely dotted,very thick,] (-1.7,0.4) -- (-1.7,-0.4);
\draw[->] (5,0) -- (6,0);
\draw (5.5,0) -- (5.5,1.2);
\draw (5.5,1.2) -- (-3,1.2);
\draw (-3,1.2) -- (-3,0);
\draw[->] (-3,0) -- (-2,0);

\node at (1,0) {$B$};
\node at (4.5,0) {$C$};
\node at (1.5, 1) {$\tau$};
\node at (-2.2, 0.7) {$1$};
\node at (-2.2, -0.7) {$N$};

\end{tikzpicture}
}
\caption{Single bottleneck topology with many TCP flows feeding into a router. The flows have a common round trip time $\tau$. The bottleneck queue has buffer size $B$ and link capacity $C$.}
\label{Figure.1}
\end{center}
\end{figure}
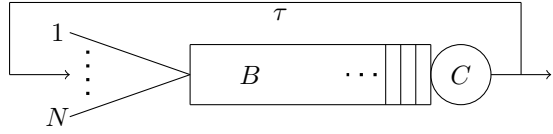
Our focus will be on C-TCP, however other variants of TCP, like TCP Reno and HighSpeed TCP, can also be analysed via \eqref{eq:modela2}. The following are the  functional forms of $i\left(w(t)\right)$ and $d\left(w(t)\right)$ for Compound, Reno and HighSpeed TCP.
\begin{itemize}
\item Compound TCP
\begin{align}
\label{eq:Compound}
i\left(w\left(t\right)\right)=\frac{\alpha\left(w(t)\right)^{k}}{w(t)} \hspace{1ex} \mbox{and} \hspace{1ex} d\left(w\left(t\right)\right)=\beta w(t).
\end{align}
\item TCP Reno
\begin{align}
\label{eq:Reno}
i\left(w\left(t\right)\right)=\frac{1}{w(t)} \hspace{1ex} \mbox{and} \hspace{1ex} d\left(w\left(t\right)\right)=\frac{w(t)}{2}.
\end{align}
\item HighSpeed TCP
\begin{align}
\label{eq:HSTCP}
i\left(w\left(t\right)\right)=\frac{f_{1}\left(w(t)\right)}{w(t)} \hspace{1ex} \mbox{and} \hspace{1ex} d\left(w\left(t\right)\right)=f_{2}\left(w(t)\right)w(t),
\end{align}
\end{itemize}
where $f_1(.)$ and $f_2(.)$ are continuous functions of the window size, and are given as
\begin{align*}
f_1\left(w(t)\right) &=\frac{0.156w^2(t)f_2\left(w(t)\right)}{w^{1.2}(t)\left(2-f_2\left(w(t)\right)\right)},\hspace{1ex}\text{and}\\
f_2\left(w(t)\right) &=\frac{-0.4\left(\log\left(w(t)\right)-\log\left(38\right)\right)}{\left(\log(83000-\log(38)\right)}+0.5.
\end{align*}

The equilibrium of system \eqref{eq:modela2} satisfies 
\begin{equation}
\label{eq:equilibrium}
i(w^{*})(1-p(w^{*}))=d(w^{*})p(w^{*}).
\end{equation}
We let $u(t)=w(t)-w^{*}$, and linearise \eqref{eq:modela2} about its non-trivial equilibrium point $w^{*}$ to get
\begin{align}
\dot{u}(t) = -au(t) - bu(t-\tau),\label{eq:lineara}
\end{align}
where
\begin{eqnarray}
\label{eq:a_b}
&a& = -\frac{w^{*}}{\tau}\left(i'(w^{*})(1-p(w^{*}))-d'(w^{*})p(w^{*})\right), \hspace{1ex} \mbox{and}\notag\\
&b& = \frac{w^{*}}{\tau}p'(w^{*})\left(i(w^{*})+d(w^{*})\right).
\end{eqnarray}
Looking for exponential solutions of \eqref{eq:lineara}, we get 
\begin{align}
s+a+be^{-s\tau}=0.\label{eq:chareq}
\end{align}
We now outline the necessary and sufficient and sufficient conditions for \eqref{eq:lineara} to be asymptotically stable. This would then yield the corresponding local stability conditions for the original non-linear system \eqref{eq:modela2}. 
\subsection{Local stability and Hopf bifurcation analysis with long-lived flows}

As the number of traffic sources feeding into a small buffer router increases, in the limiting regime, the buffer overflow probability tends to the corresponding probability with Poisson arrivals~\cite{Poisson_limit}. This finding motivates us to approximate the packet drop probability of the core router by the corresponding probability of an $M/M/1/B$ queue, where $B$ is the buffer size. For analytical purposes, we use the buffer exceedance probability of an $M/M/1$ queue as a surrogate for the overflow probability of an $M/M/1/B$ queue. This approach has been commonly used in the literature. This yields the packet loss probability at the core router as  
\begin{align}
\label{eq:loss}
p(w(t))=\left( \frac{w(t)}{C'\tau}\right)^{B},
\end{align}
where $C'$ is the service rate per flow of the bottleneck link, and $B$ is the buffer size. A comment is in order. Note that owing to TCP's congestion control algorithm, our scenario constitutes a \emph{closed loop} feedback system. However, we use the loss probability of an \emph{open loop} queueing system as a substitute for the packet loss probability at the bottleneck queue. A justification for this assumption is provided in~\cite{Wischik}, which says that for a queue with a small buffer, traffic characteristics over very short timescales matter, and if the average arrival rate to the queue does not exhibit much variation in such a short timescale, then this approximation is benign. We will empirically validate this approximation later. 

We now outline some stability conditions for \eqref{eq:lineara} to be asymptotically stable. From \cite{Gaurav}, if $a\geq0$, $b>0$, $b>a$ and $\tau>0$, a sufficient condition for stability of \eqref{eq:lineara} is 
\begin{align}
\label{eq:sufficient}
b\tau<\frac{\pi}{2},
\end{align}
the necessary and sufficient condition for stability of \eqref{eq:lineara} is
\begin{align}
\label{eq:necessary}
\tau\sqrt{b^{2}-a^{2}}<\cos^{-1}(-a/b),
\end{align}
and the system undergoes the first Hopf bifurcation at
\begin{align}
\label{eq:hopf}
\tau\sqrt{b^{2}-a^{2}}=\cos^{-1}(-a/b).
\end{align}
We can now easily particularise the above conditions for Compound, Reno and HighSpeed TCP. However, we outline conditions for local stability only for Compound TCP, with long-lived flows. The necessary and sufficient condition for local stability with Compound TCP flows is \cite{Raja}
\begin{align}
\label{eq:ns_compound}
\alpha \left(w^{*}\right)^{k-1}\sqrt{B^{2}-\left((k-2)(1-p(w^{*}))\right)^{2}}<\cos^{-1}\left(\frac{(k-2)(1-p(w^*))}{B}\right),
\end{align}
and a Hopf bifurcation would occur at
\begin{align}
\alpha \left(w^{*}\right)^{k-1}\sqrt{B^{2}-\left((k-2)(1-p(w^{*}))\right)^{2}}=\cos^{-1}\left(\frac{(k-2)(1-p(w^*))}{B}\right).
\label{eq:hopf_compound}
\end{align}
A sufficient condition for local stability with Compound TCP flows is 
\begin{align*}
\alpha B \left(w^{\ast}\right)^{k-1}<\frac{\pi}{2}.
\label{eq:suff_compound}
\end{align*}
Clearly, buffer thresholds and protocol parameters all greatly influence stability. In particular, if condition \eqref{eq:ns_compound} gets violated, the system would lose local stability via a Hopf bifurcation. This would lead to the emergence of limit cycles in the system dynamics. In the context of TCP, these limit cycles manifest as oscillations in the packet arrival process and queue size dynamics~\cite{Joo,Wischik}. This would lead to synchronisation effects among TCP windows, and loss of link utilisation, and hence would be detrimental to the overall network performance.  

\begin{figure}[t!]

        \begin{subfigure}{\columnwidth}
        \centering
        \psfrag{aaaa}{\hspace{6mm}\begin{scriptsize}Time (seconds)\end{scriptsize}}
        \psfrag{cccc}{\hspace{1mm}\begin{scriptsize}Queue size (pkts)\end{scriptsize}}
        \psfrag{dddd}{\hspace{4mm}\begin{scriptsize}Round trip time = 200 ms\end{scriptsize}}
        \psfrag{eeee}{\hspace{4mm}\begin{scriptsize}Round trip time = 10 ms\end{scriptsize}}
         \psfrag{bbbb}{\hspace{3mm}\begin{scriptsize}Utilisation (\%)\end{scriptsize}}
         \psfrag{275}{\begin{scriptsize}$275$\end{scriptsize}}
         \psfrag{300}{\begin{scriptsize}$300$\end{scriptsize}}
         \psfrag{0}{\begin{scriptsize}$0$\end{scriptsize}}
         \psfrag{60}{\begin{scriptsize}$60$\end{scriptsize}}
         \psfrag{100}{\begin{scriptsize}$100$\end{scriptsize}}
          \psfrag{15}{\begin{scriptsize}$15$\end{scriptsize}}
                \includegraphics[height=3in,width=2in,angle=-90]{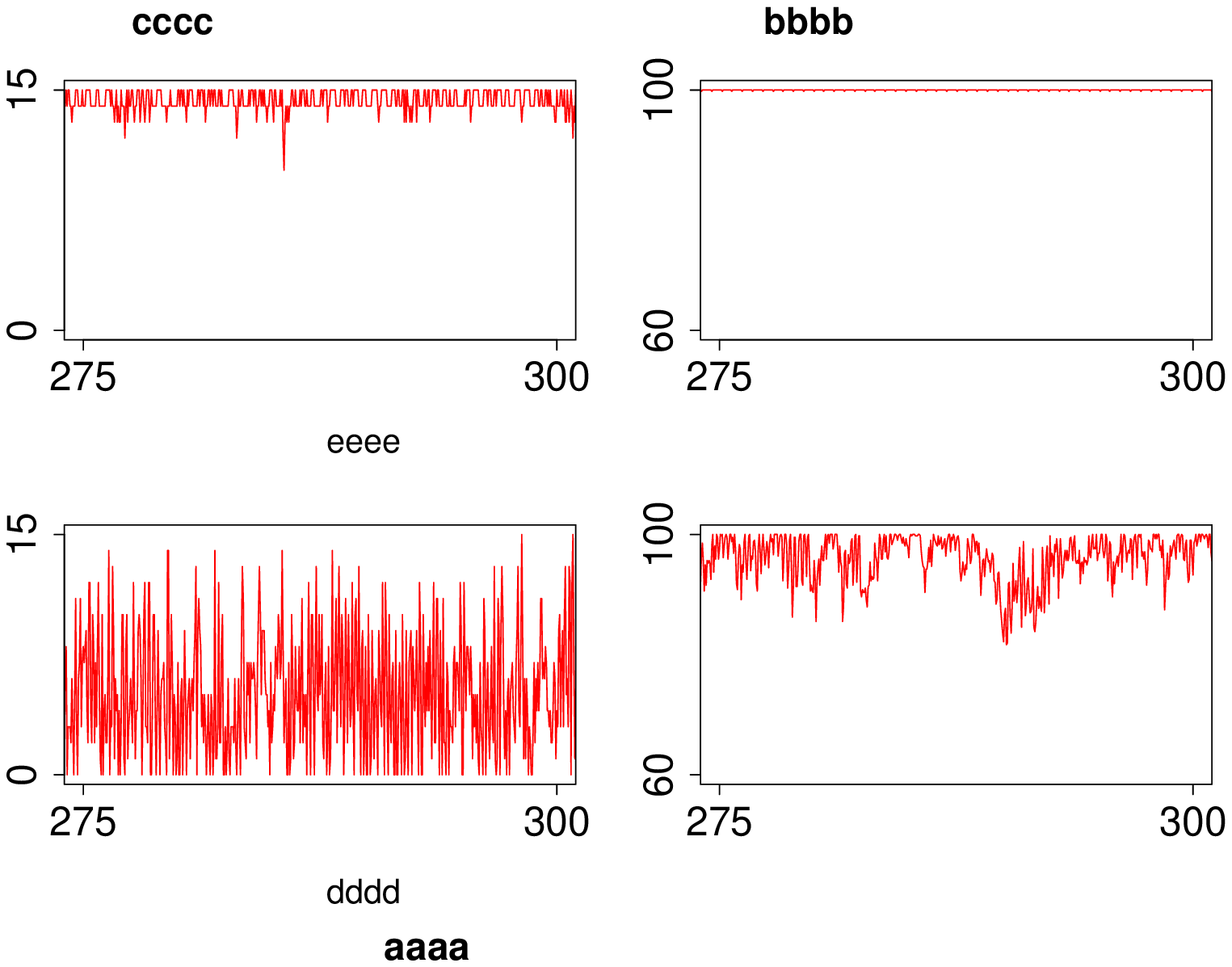}
                \caption{Buffer size = 15}
                \label{fig:long15}
        \end{subfigure}%
        
         \vspace{1mm} 
        \begin{subfigure}{\columnwidth}
        \centering
        \psfrag{aaaa}{\hspace{6mm}\begin{scriptsize}Time (seconds)\end{scriptsize}}
        \psfrag{cccc}{\hspace{1mm}\begin{scriptsize}Queue size (pkts)\end{scriptsize}}
        \psfrag{dddd}{\hspace{4mm}\begin{scriptsize}Round trip time = 200 ms\end{scriptsize}}
        \psfrag{eeee}{\hspace{4mm}\begin{scriptsize}Round trip time = 10 ms\end{scriptsize}}
         \psfrag{bbbb}{\hspace{3mm}\begin{scriptsize}Utilisation (\%)\end{scriptsize}}
         \psfrag{275}{\begin{scriptsize}$275$\end{scriptsize}}
         \psfrag{300}{\begin{scriptsize}$300$\end{scriptsize}}
         \psfrag{0}{\begin{scriptsize}$0$\end{scriptsize}}
         \psfrag{60}{\begin{scriptsize}$60$\end{scriptsize}}
         \psfrag{100}{\begin{scriptsize}$100$\end{scriptsize}}
                \includegraphics[height=3in,width=2in,angle=-90]{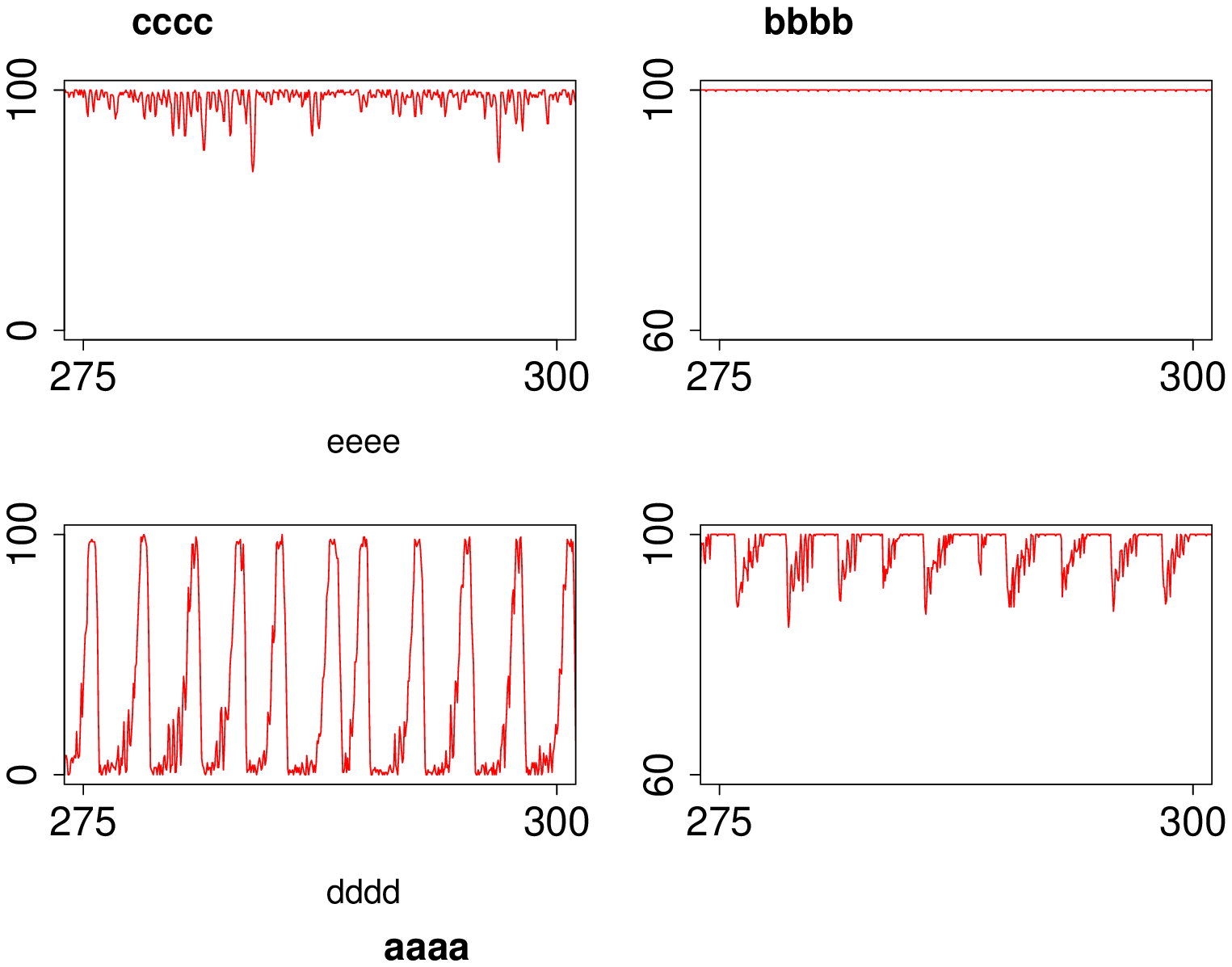}
                \caption{Buffer size = 100}
                \label{fig:long270}
        \end{subfigure}

        \caption{ \emph{Long-lived flows}. 60 long-lived Compound TCP flows over a 2 Mbps link feeding into a single bottleneck queue with link capacity 100 Mbps. Note that as the buffer threshold of the bottleneck router increases, we see the emergence of limit cycles in the queue size.}
\label{fig:long1}
\end{figure}

\begin{figure}[t!]
        
        \begin{subfigure}{\columnwidth}
        \centering
         \psfrag{aaaa}{\hspace{6mm}\begin{scriptsize}Time (seconds)\end{scriptsize}}
        \psfrag{cccc}{\hspace{1mm}\begin{scriptsize}Queue size (pkts)\end{scriptsize}}
        \psfrag{dddd}{\hspace{4mm}\begin{scriptsize}Round trip time = 200 ms\end{scriptsize}}
        \psfrag{eeee}{\hspace{4mm}\begin{scriptsize}Round trip time = 10 ms\end{scriptsize}}
         \psfrag{bbbb}{\hspace{3mm}\begin{scriptsize}Utilisation (\%)\end{scriptsize}}
         \psfrag{275}{\begin{scriptsize}$275$\end{scriptsize}}
         \psfrag{300}{\begin{scriptsize}$300$\end{scriptsize}}
         \psfrag{0}{\begin{scriptsize}$0$\end{scriptsize}}
         \psfrag{60}{\begin{scriptsize}$60$\end{scriptsize}}
         \psfrag{100}{\begin{scriptsize}$100$\end{scriptsize}}
          \psfrag{15}{\begin{scriptsize}$15$\end{scriptsize}}
                \includegraphics[height=3in,width=2in,angle=-90]{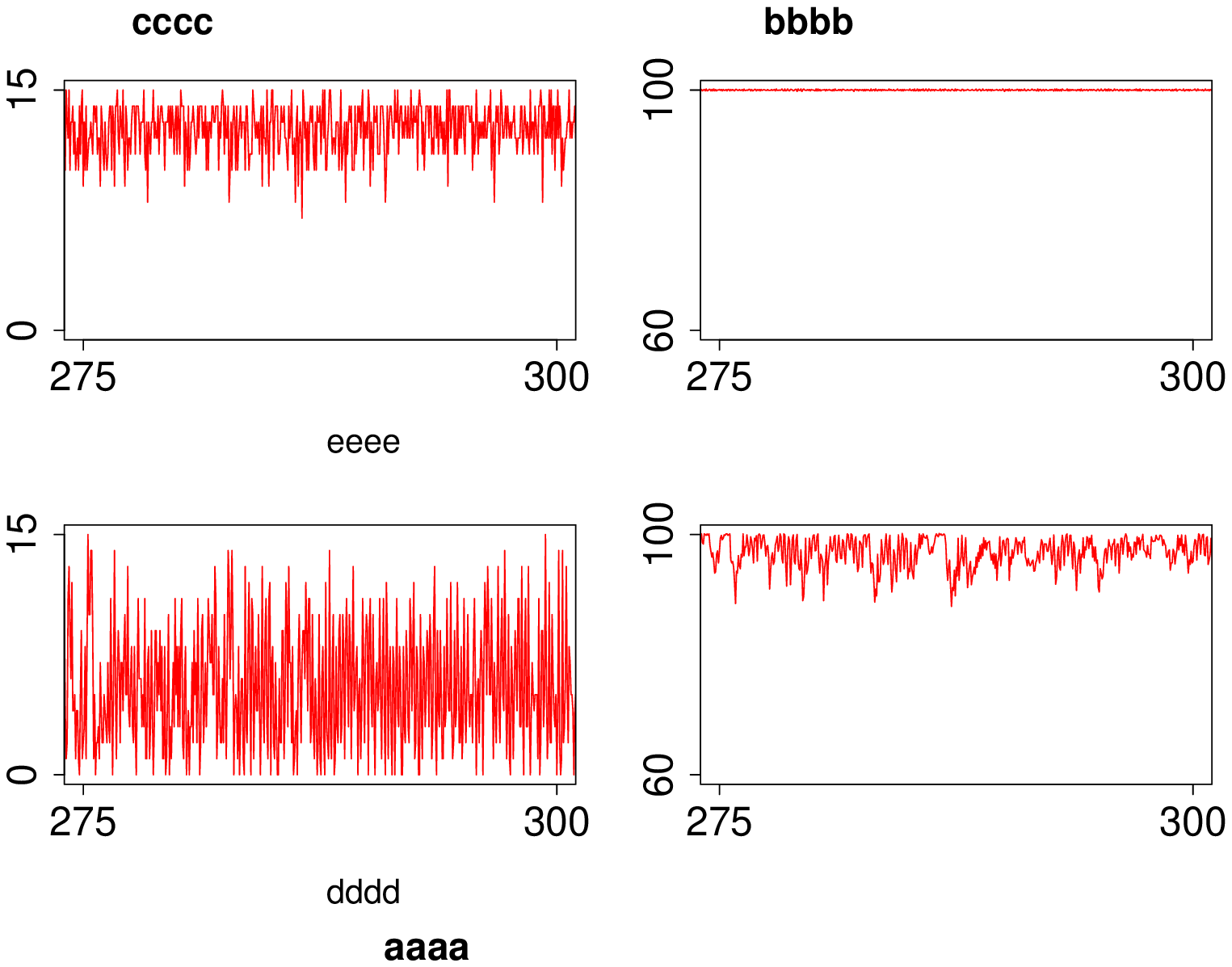}
                
                \caption{Buffer size = 15}
                \label{fig:short15}
                \end{subfigure}

          \vspace{1mm}
                \begin{subfigure}{\columnwidth}
                \centering
                 \psfrag{aaaa}{\hspace{6mm}\begin{scriptsize}Time (seconds)\end{scriptsize}}
        \psfrag{cccc}{\hspace{1mm}\begin{scriptsize}Queue size (pkts)\end{scriptsize}}
        \psfrag{dddd}{\hspace{4mm}\begin{scriptsize}Round trip time = 200 ms\end{scriptsize}}
        \psfrag{eeee}{\hspace{4mm}\begin{scriptsize}Round trip time = 10 ms\end{scriptsize}}
         \psfrag{bbbb}{\hspace{3mm}\begin{scriptsize}Utilisation (\%)\end{scriptsize}}
         \psfrag{275}{\begin{scriptsize}$275$\end{scriptsize}}
         \psfrag{300}{\begin{scriptsize}$300$\end{scriptsize}}
         \psfrag{0}{\begin{scriptsize}$0$\end{scriptsize}}
         \psfrag{60}{\begin{scriptsize}$60$\end{scriptsize}}
         \psfrag{100}{\begin{scriptsize}$100$\end{scriptsize}}
          \psfrag{270}{\begin{scriptsize}$100$\end{scriptsize}}
                \includegraphics[height=3in,width=2in,angle=-90]{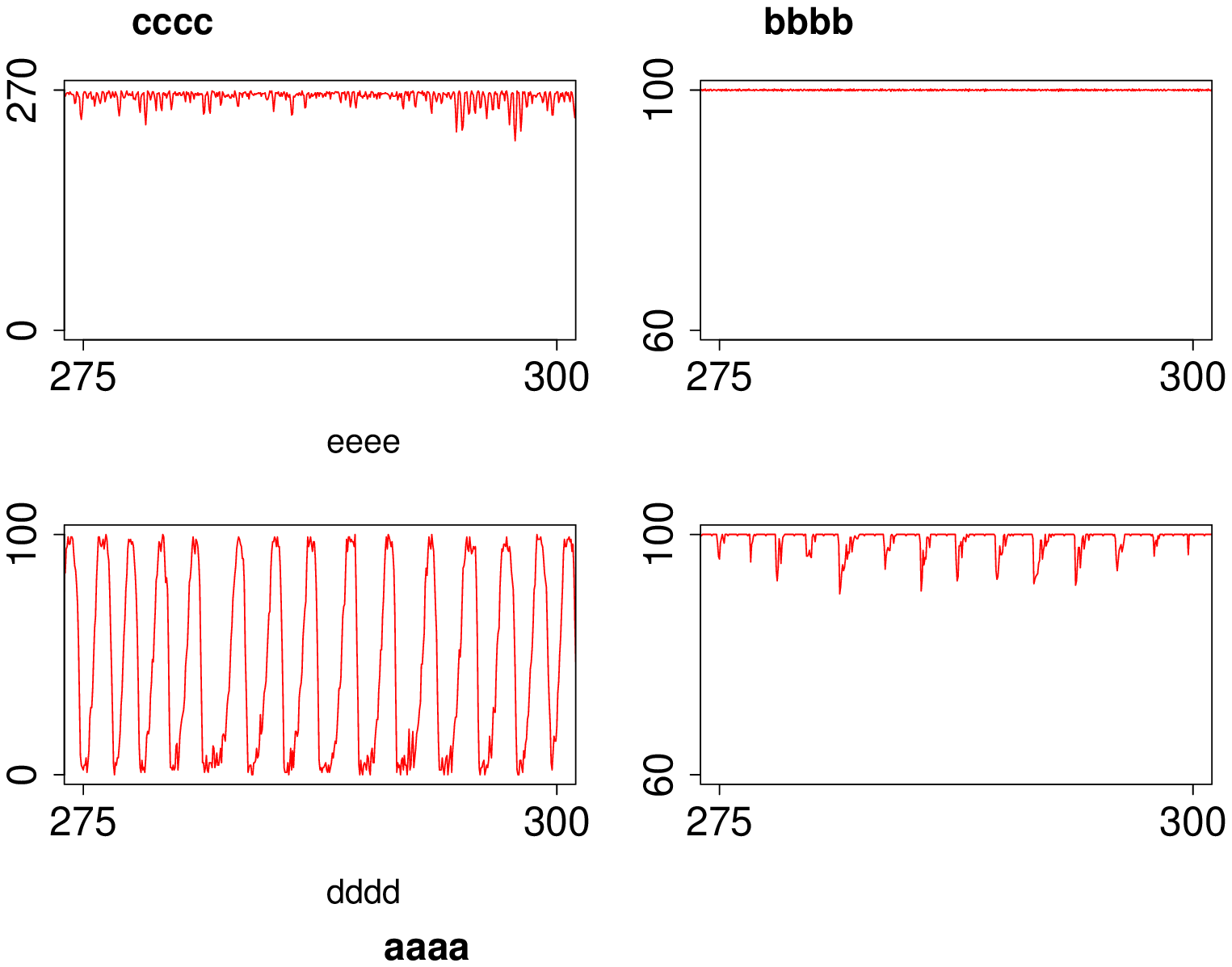}
                \caption{Buffer size = 100}
                \label{fig:short270}
        \end{subfigure}
        \caption{\emph{Long-lived and short-lived flows}. 55 long-lived Compound TCP flows over a 2 Mbps link, and \emph{exponentially distributed short files}, feeding into a single bottleneck queue with link capacity 100 Mbps. Observe the emergence of limit cycles in the queue size, as the buffer threshold at the bottleneck router increases.}
\label{fig:short1}
\end{figure}
\begin{figure}[t!]
        
        \begin{subfigure}{\columnwidth}
        \centering
         \psfrag{aaaa}{\hspace{6mm}\begin{scriptsize}Time (seconds)\end{scriptsize}}
        \psfrag{cccc}{\hspace{1mm}\begin{scriptsize}Queue size (pkts)\end{scriptsize}}
        \psfrag{dddd}{\hspace{4mm}\begin{scriptsize}Round trip time = 200 ms\end{scriptsize}}
        \psfrag{eeee}{\hspace{4mm}\begin{scriptsize}Round trip time = 10 ms\end{scriptsize}}
         \psfrag{bbbb}{\hspace{3mm}\begin{scriptsize}Utilisation (\%)\end{scriptsize}}
         \psfrag{275}{\begin{scriptsize}$275$\end{scriptsize}}
         \psfrag{300}{\begin{scriptsize}$300$\end{scriptsize}}
         \psfrag{0}{\begin{scriptsize}$0$\end{scriptsize}}
         \psfrag{60}{\begin{scriptsize}$60$\end{scriptsize}}
         \psfrag{100}{\begin{scriptsize}$100$\end{scriptsize}}
          \psfrag{15}{\begin{scriptsize}$15$\end{scriptsize}}
                \includegraphics[height=3in,width=2in,angle=-90]{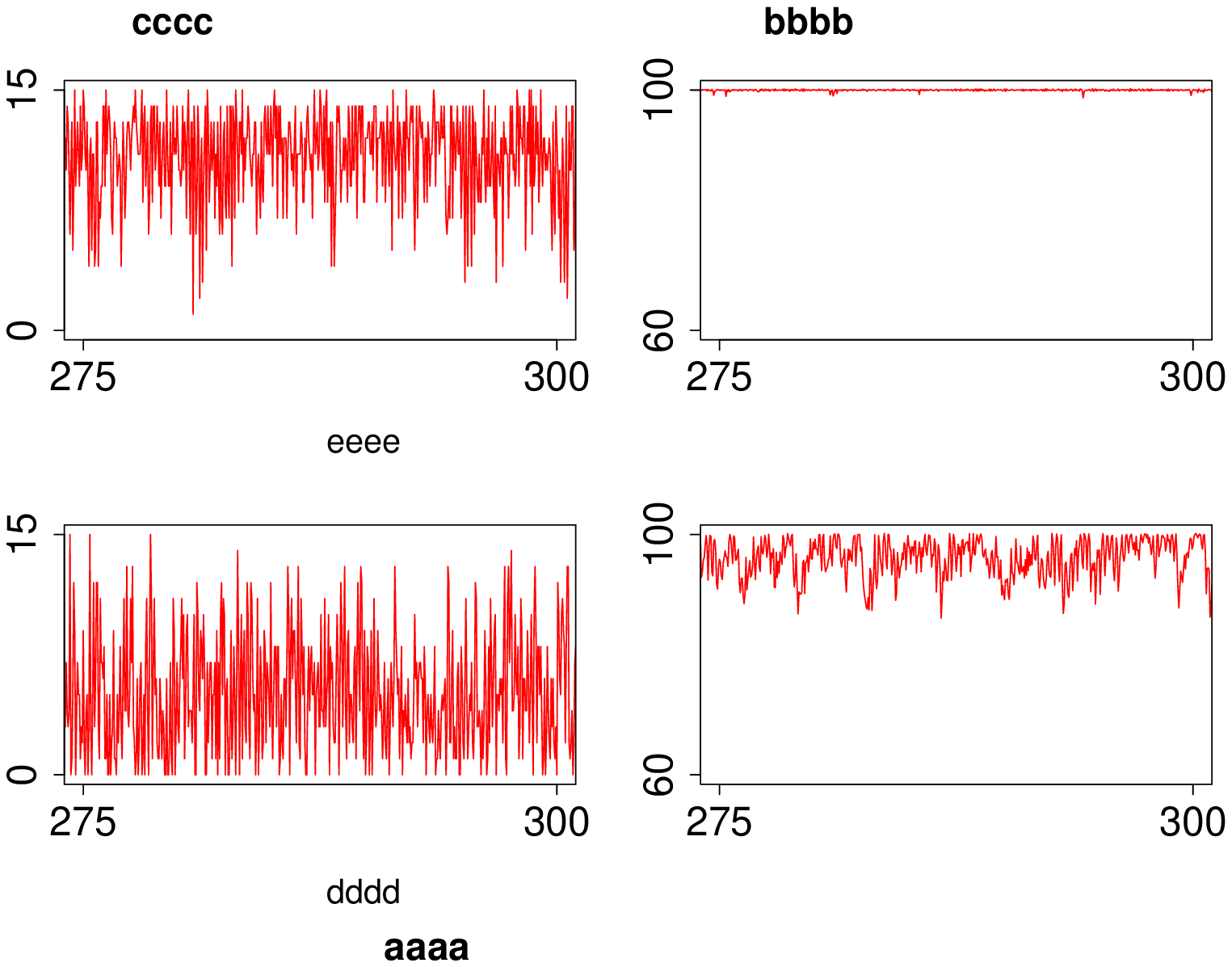}
                
                \caption{Buffer size = 15}
                \label{fig:short15}
                \end{subfigure}

          \vspace{1mm}
                \begin{subfigure}{\columnwidth}
                \centering
                 \psfrag{aaaa}{\hspace{6mm}\begin{scriptsize}Time (seconds)\end{scriptsize}}
        \psfrag{cccc}{\hspace{1mm}\begin{scriptsize}Queue size (pkts)\end{scriptsize}}
        \psfrag{dddd}{\hspace{4mm}\begin{scriptsize}Round trip time = 200 ms\end{scriptsize}}
        \psfrag{eeee}{\hspace{4mm}\begin{scriptsize}Round trip time = 10 ms\end{scriptsize}}
         \psfrag{bbbb}{\hspace{3mm}\begin{scriptsize}Utilisation (\%)\end{scriptsize}}
         \psfrag{275}{\begin{scriptsize}$275$\end{scriptsize}}
         \psfrag{300}{\begin{scriptsize}$300$\end{scriptsize}}
         \psfrag{0}{\begin{scriptsize}$0$\end{scriptsize}}
         \psfrag{60}{\begin{scriptsize}$60$\end{scriptsize}}
         \psfrag{100}{\begin{scriptsize}$100$\end{scriptsize}}
          \psfrag{270}{\begin{scriptsize}$100$\end{scriptsize}}
                \includegraphics[height=3in,width=2in,angle=-90]{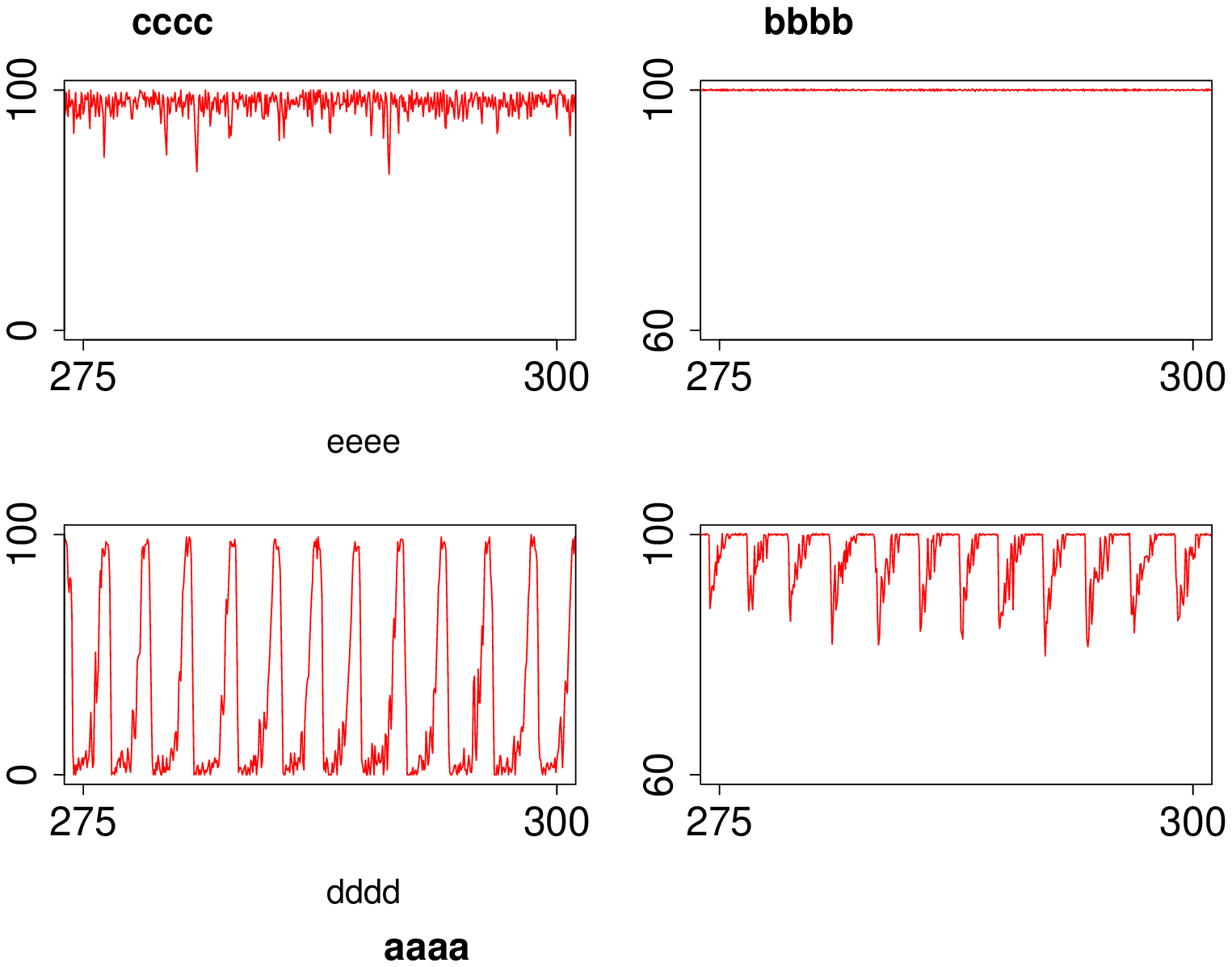}
                \caption{Buffer size = 100}
        \end{subfigure}
         \caption{ \emph{TCP and UDP flows}. 55 long-lived Compound TCP flows over a 2 Mbps link feeding into a single bottleneck queue with link capacity 100 Mbps. We consider 10 UDP flows each over a 1 Mbps link. As the  buffer threshold at the bottleneck router increases, observe that the queue size dynamics exhibits limit cycles.}
\label{fig:longudp}
\end{figure} 

\begin{figure}[t!]
\begin{center}
\vspace{-5mm}
  \psfrag{b}{\hspace{-10.5mm}Window size}
  \psfrag{a}{\hspace{-7mm}Queue size}
   \psfrag{c}{\hspace{-5mm}Average window}
  \psfrag{0}{\begin{scriptsize}$0$\end{scriptsize}}
  \psfrag{15}{\begin{scriptsize}$15$\end{scriptsize}}
  \psfrag{100}{\begin{scriptsize}$100$\end{scriptsize}}
  \psfrag{80}{\begin{scriptsize}$80$\end{scriptsize}}
  \psfrag{200}{\begin{scriptsize}$200$\end{scriptsize}}
  \psfrag{150}{\begin{scriptsize}$275$\end{scriptsize}}
  \psfrag{175}{\begin{scriptsize}$300$\end{scriptsize}}
  \psfrag{fffff}{\hspace{-0.2cm}Buffer size = $15$ packets}
  \psfrag{eeeee}{\hspace{-0.5cm}Buffer size = $100$ packets}
  \psfrag{aa}{\hspace{-1cm}Time (seconds)}
  
  \includegraphics[width=2.75in,height=3.75in,angle=270]{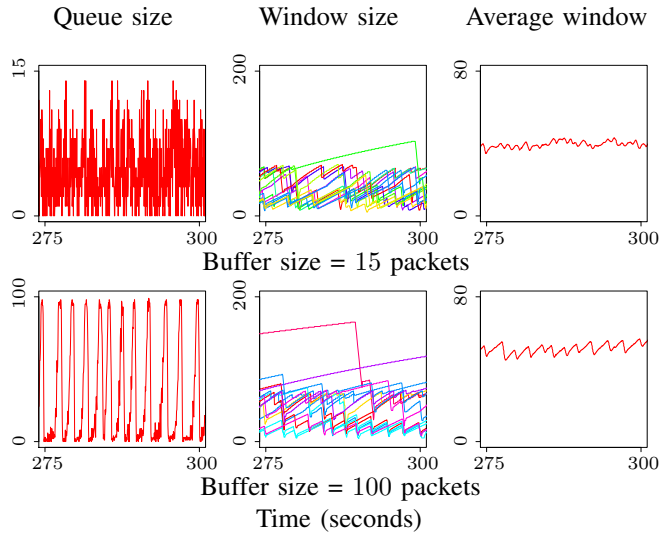}

  \caption {\emph{Compound TCP in single bottleneck topology.} 60 long-lived Compound TCP flows each with an access speed of $2$ Mbps, over a single bottleneck link with a capacity $100$ Mbps. Observe the emergence of limit cycles in the queue size, and synchronisation among Compound TCP windows, as the buffer threshold at the core router increases. }
  \label{fig:window_sync}
 \end{center}
 \end{figure}
\begin{figure}
\begin{center}
\vspace{-5mm}
  \psfrag{b}{\hspace{4mm}$\log_2$(Time in $\mu s$)}
  \psfrag{c}{\hspace{-10mm}Coefficient of variation}
  \psfrag{0.8}{\begin{scriptsize}$0.8$\end{scriptsize}}
  \psfrag{20}{\begin{scriptsize}$20$\end{scriptsize}}
  \psfrag{12}{\begin{scriptsize}$12$\end{scriptsize}}
  \psfrag{13}{\begin{scriptsize}$13$\end{scriptsize}}
  \psfrag{a}{\vspace{-1mm}\begin{scriptsize}$\times 10^{-2}$\end{scriptsize}}
  \psfrag{50}{\begin{scriptsize}$50$\end{scriptsize}}
  \psfrag{h}{\begin{tiny}Buffer size = 2084 pkts\end{tiny}}
  \psfrag{x}{\begin{tiny}Buffer size = 100 pkts\end{tiny}}
  \psfrag{y}{\begin{tiny}Buffer size = 50 pkts\end{tiny}}
  \psfrag{z}{\begin{tiny}Buffer size = 15 pkts\end{tiny}}
  \psfrag{ffff}{\hspace{3.25cm}CCDF}
  \psfrag{cccc}{\hspace{-5mm}\begin{scriptsize}Round trip time = $100$ ms \end{scriptsize}}
  \psfrag{bbbb}{\hspace{-5mm}\begin{scriptsize}Round trip time = $200$ ms \end{scriptsize}}
   \psfrag{eeee}{\hspace{12mm}\begin{scriptsize}No. of flows = $60$  \end{scriptsize}}
   \psfrag{dddd}{\hspace{12mm}\begin{scriptsize}No. of flows = $120$  \end{scriptsize}}
   \psfrag{aaaa}{\hspace{6mm}Queue threshold}

  \includegraphics[width=2.85in,height=3.85in,angle=270]{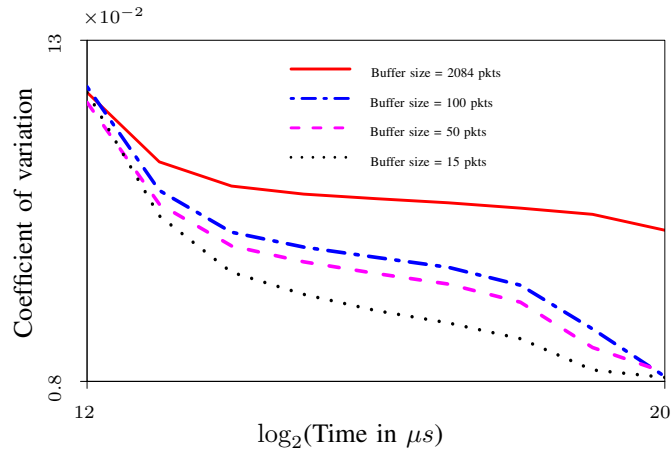}

  \caption{\emph{Statistics of the arrival process.} $60$ long-lived Compound TCP flows each over a $2$ Mbps link. The round trip time is fixed at $200$ ms. The capacity of the bottleneck router is $100$ Mbps. We consider three regimes: stable ($B = 15$ packets), presence of synchronisation ($B = 50$, and $100$ packets), and bandwidth-delay product rule ($B = 2084$ packets). Observe that for smaller buffers ($15$ packets), the aggregate arrival process exhibits less burstiness or variability.}
  \label{fig:statistics_arrival}
 \end{center}
\end{figure}
\begin{figure}
\begin{center}
\vspace{-5mm}
  \psfrag{b}{Time scale(seconds)}
  \psfrag{T}{\hspace{2mm}Protocol parameter, $\alpha$}
  \psfrag{0}{\begin{scriptsize}$0$\end{scriptsize}}
  \psfrag{1}{\begin{scriptsize}$1$\end{scriptsize}}
  \psfrag{15}{\begin{scriptsize}$15$\end{scriptsize}}
  \psfrag{50}{\begin{scriptsize}$50$\end{scriptsize}}
  \psfrag{h}{\begin{tiny}Empirical\end{tiny}}
  \psfrag{x}{\begin{tiny}$M/M/1/B$\end{tiny}}
  \psfrag{y}{\begin{tiny}$M/D/1/B$\end{tiny}}
  \psfrag{ffff}{\hspace{3.25cm}CCDF}
  \psfrag{cccc}{\hspace{-5mm}\begin{scriptsize}Round trip time = $100$ ms \end{scriptsize}}
  \psfrag{bbbb}{\hspace{-5mm}\begin{scriptsize}Round trip time = $200$ ms \end{scriptsize}}
   \psfrag{eeee}{\hspace{12mm}\begin{scriptsize}Number of flows = $60$  \end{scriptsize}}
   \psfrag{dddd}{\hspace{10mm}\begin{scriptsize}Number of flows = $120$  \end{scriptsize}}
   \psfrag{aaaa}{\hspace{6mm}Queue threshold}

  \includegraphics[width=2.85in,height=3.85in,angle=270]{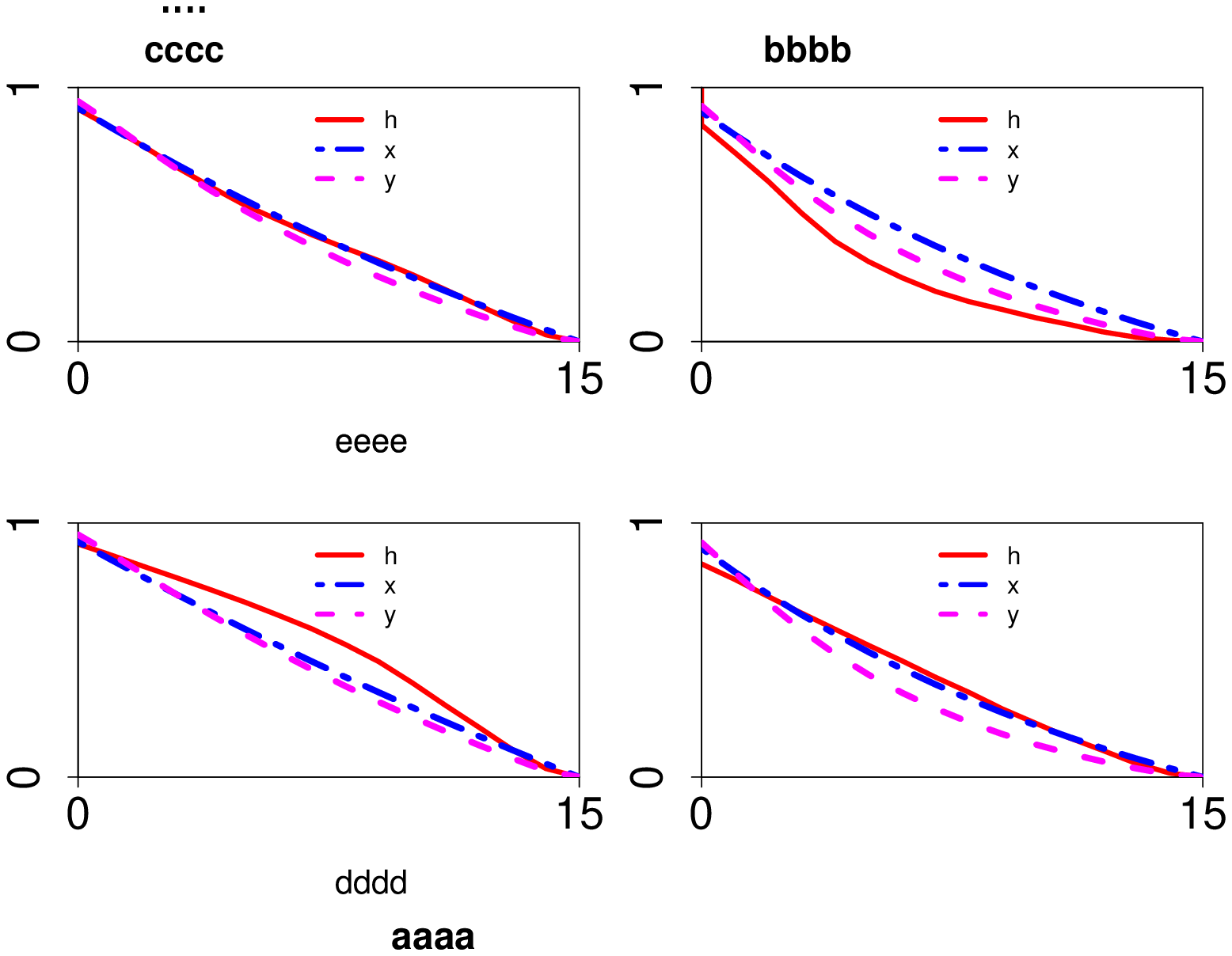}

  \caption{ \emph{Statistics of the queue size.} Empirical queue length distribution for single bottleneck topology  with $60$ and $120$ long-lived flows each having an access speed of $2$ Mbps and $1$ Mbps, with round trip times $100$ ms and $200$ ms respectively.  We compare the empirical queue statistics with the queue length distributions of $M/M/1/B$, and $M/D/1/B$ for two different round trip times. Observe that as the number of long-lived flows increases, the approximation becomes better at a larger bandwidth-delay product. }
  \label{fig:statistics_long_single}
 \end{center}
\end{figure}

 \subsection{Robust stability analysis with long-lived flows}
 We now investigate the dynamical properties of system \eqref{eq:modela2} under parametric uncertainties. Note that for networks with small buffer routers, the queueing delay can be assumed to be negligible as compared to the propagation delay, which is constant for fixed source destination pair. Hence, in this scenario, the feedback delay or the round trip time is also fixed. Further, the capacity of the network is known apriori, and protocol parameters are fixed. However, a protocol designer would not be aware of the exact buffer size at the core router. Hence, we allow uncertainties in the buffer size at the core router and assume that it lies in an interval, say $B\in[\underline{B},\overline{B}].$ Here, both $\underline{B}$ and $\overline{B}$ lie in the small buffer regime. We then derive bounds on the protocol parameters and network parameters which would ensure that the system is locally asymptotically stable for all values of $B\in[\underline{B},\overline{B}].$
 
 Recall that the characteristic equation of the linearised system \eqref{eq:lineara} is 
 \begin{align*}
 s+a+be^{-s\tau} = 0
 \end{align*}
 
 Note that the above can be written in the form $P(s)+Q(s)e^{-s\tau},$ where $P(s)=s+a$ and $Q(s)=b.$ Since $a>0,$ it is clear that $P(s)$ is a stable polynomial, and $deg(P)>deg(Q).$ As stated in~\cite{Kharitonov}, system \eqref{eq:lineara} would be robust stable independent of the delay if and only if $P(j\omega)>Q(j\omega),\,\forall\,\omega\geq0.$ This condition translates into $b<a,\,\forall\,\omega\geq0.$ However, we can easily show that this is not true for default values of protocol parameters of Compound, Reno and HighSpeed TCP, and physically relevant values of network parameters. Hence, system \eqref{eq:modela2} cannot be locally robust stable independent of the feedback delay. This motivates us to look for conditions for delay dependent robust stability.
 
As stated in~\cite{Kharitonov}, a sufficient condition for stability of \eqref{eq:lineara} is $b\tau<1.$ It is easy to show that the equilibrium coefficient $b$ is a monotonically increasing function of the buffer size of the core rourter $B.$ This implies that if $B$ lies in the interval $[\underline{B},\overline{B}],$ then $b$ would also lie in some interval $[\underline{b},\overline{b}].$ Here, $\underline{b}$ and $\overline{b}$ can be evaluated by substituting $\underline{B}$ and $\overline{B}$ respectively in \eqref{eq:a_b}. Then, a sufficient condition for robust stability of system \eqref{eq:modela2} is 
\begin{align}
\overline{b}\tau<1.
\end{align}
Using the functional forms \eqref{eq:Compound}, a sufficient condition for robust stability with Compound TCP flows can be outlined as 
\begin{align*}
\label{eq:robustcompound}
\alpha \overline{B} \left(w^{*}\right)^{k-1}<1.
\end{align*}
Observe that, condition \eqref{eq:robustcompound} could provide guidlines to protocol designers to design transport protocols at end-systems, which could ensure stability of the system for a wide range of values of buffer thresholds at core routers.    
\subsection{Local stability and Hopf bifurcation analysis with long-lived and short-lived flows}
 We now deviate from the assumption that the system has only long-lived flows and consider the scenario where in addition to a large number of long-lived flows, short flows arrive and depart the network. On a short time scale, short TCP connections may act as an uncontrolled and random background load on the network. Suppose the workload per flow arriving at the bottleneck queue over a time period $T$ is modelled as Gaussian with mean $x^{*}T$ and variance $x^{*}\sigma_{1}^{2}T$ and the background load per flow due to the short transfers over the time period $T$ is also modelled as Gaussian with mean $vT$ and variance $v\sigma_{2}^{2}T$. Then the loss probability at the  bottleneck queue can be expressed as \cite{Kelly}
\begin{align}
p(w^{*})=\exp \left(\frac{-2B\left(C'\tau-w^{*}-v\tau\right)}{w^{*}\sigma_{1}^{2}+v\sigma_{2}^{2}\tau}\right).
\end{align}

 Recall that \eqref{eq:sufficient} gives a sufficient condition for local stability and \eqref{eq:hopf} gives the condition for which the system undergoes a Hopf-type bifurcation. We now particularise the sufficient condition only for Compound TCP, however conditions for TCP Reno and HighSpeed TCP also can easily be outlined. We are also in a position to state the Hopf bifurcation conditions, which are left out due to space constraints. 

%
%
%

For Compound TCP, a sufficient condition for local stability of the system is  
\begin{align}
\label{eq:shortcompound}
2B \alpha \left(w^{*}\right)^{k}\tau\frac{v\sigma_{2}^{2}+\left(C'-v\right)\sigma_{1}^{2}}{\left(w^{*}\sigma_{1}^{2}+v\sigma_{2}^{2}\tau\right)^2}<\frac{\pi}{2}.
\end{align}

Condition \eqref{eq:shortcompound} captures the relationship between the various protocol and network parameters. It is interesting to note that in general, larger the value of parameter $B$, greater the possibility of driving the system to an unstable state. In Compound TCP, there appears to be an intrinsic trade off in the choice of the parameter $\alpha$ and the queue threshold parameter $B$. Indeed, it can be easily shown that increasing buffer sizes would prompt the system to lose local stability. The presence of short-lived flows, which are modelled here as random uncontrolled traffic, does not change the requirement of choosing smaller values of $B$ to ensure stability. We now present some packet-level simulations, which will enable us to comment on the dynamical and statistical properties of the system.
\subsection{Simulations}
\subsubsection*{Dynamical Properties}

We now conduct packet-level simulations, using NS2 \cite{ns2}, for the single bottleneck topology in a small buffer sizing regime. With small buffers, we employ $15$ and $100$ packets. We consider two scenarios (i) only long-lived flows, and (ii) a combination of long and short flows. The bottleneck link has a capacity of $100$ Mbps. The packet size is fixed at $1500$ bytes. In the scenario wherein only long-lived flows are present, we consider $60$ and $120$ long-lived flows where each flow has an access link speed of 2 Mbps and 1 Mbps respectively. With a combination of long- and short-lived flows, we consider $55$ long flows each with an access speed of $2$ Mbps. The file size of each of the short flows is exponentially distributed with a mean file size of $12.5$ KB. The mean rate at which short flows arrive is $200$ flows per second, and the total data rate contributed by these short flows is restricted to $20$ Mbps.

Fig.~\ref{fig:long1} depicts the simulations where the system only has long-lived flows. With a buffer size of $15$ packets, as expected, the queuing delay is negligible and the system is stable in the sense that there are no limit cycles in the queue size. With buffer size of $100$ packets, with smaller round trip times, the queues are full which yields full link utilisation but at the cost of extra latency. With larger delays, limit cycles will emerge in the queue size which also start to hurt link utilisation. Fig.~\ref{fig:short1} depicts the simulation results where the system has a combination of long and short-lived flows. Qualitatively, the results are very similar to those shown in Fig.~\ref{fig:long1}. This is expected as the models did indeed predict that despite the presence of short flows the system could readily lose stability if key system parameters were not properly dimensioned.

Today's Internet is heterogeneous in nature. While applications like File Transfer Protocol (FTP) uses the services of TCP (closed loop), real-time applications like VoIP and online gaming use UDP (open loop). Further, the use of real-time applications is rapidly increasing. Hence, it is imperative to understand the impact of buffer sizes on the system stability, when both TCP and UDP traffic co-exist \cite{Vishwanath}. To that end, we consider a scenario with $55$ long-lived Compound TCP flows each over an access link with a speed of $2$ Mbps, and $10$ UDP flows each over a $1$ Mbps link. Fig. \ref{fig:longudp} shows the packet-level simulations for this scenario. We can immediately observe that the results obtained are qualitatively similar to the scenario with only long-lived flows. This suggests that the requirement of choosing buffer sizes carefully does not change even when open-loop UDP traffic is present. 

The loss of local stability of the underlying dynamical system and hence the emergence of limit cycles indicates synchronisation among the TCP flows. For smaller buffer thresholds, all TCP flows would be totally de-synchronised and the mean window size would have small oscillations, see Fig.~\ref{fig:window_sync}. As the buffer
thresholds are increased, the mean window size would exhibit bigger oscillations owing to the
synchronisation of the flows. We now empirically analyse some statistical properties of the bottleneck queue.

\subsubsection*{Statistical Properties}
\emph{Statistical properties of the arrival process:}  We first conduct an empirical study on the statistical properties of the aggregate arrival process to the bottleneck queue. In particular, we capture the impact of the buffer size at the bottleneck queue on the burstiness or variability of the arrival process to the queue at different time scales. One way to characterise this is to measure the coefficient of variation  (ratio of standard deviation to mean) of the arrival traffic at different time scales. In particular, we closely follow the method presented in \cite{Vishwanath} for our study.

For our empirical study, we consider three representative regimes: $(i)$ $B=15$ packets, wherein the underlying dynamical system is stable, $(ii)$ $B=50,\,100$ packets, wherein the system dynamics exhibits limit cycles and synchronisation among TCP windows, and $(iii)$ $B=2084$ packets, which corresponds to the case of bandwidth-delay product worth of buffering. Note that for this buffer sizing rule, we have used a delay value of $250$ ms, which is typically used in practice. This study would enable us to understand the buffer sizing regime in which the Poisson approximation for the aggregate arrival process seems justified. For our simulations, we consider the number of long-lived flows in the system to be $60,$ and the average round trip time to be $200$ ms. Note that we are interested in measuring the burstiness of the arrival traffic at short time scales. Hence, we vary the time scale of aggregation from $2^{12}\,\mu s=4$ ms to $2^{20}\,\mu s=1$ second. 

Fig.~\ref{fig:statistics_arrival} depicts the coefficient of variation curves for the aggregate traffic arriving at the bottleneck queue, at different time scales and buffer sizes. For a buffer size of $15$ packets, we can observe that the coefficient of variation curve falls quite rapidly for very short time scales, and does not change its slope significantly for larger time scales. On the contrary, as we gradually increase the buffer size, the coefficient of variation curve exhibits a relatively slower decay over very short time scales, and flattens over larger time scales. Further, with a buffer size of $15$ packets, the coefficient of variation values are lower than that of $50,$ $100$ and $2084$ packets. A larger value of coefficient of variation signifies the presence of higher variability or burstiness in the traffic arrival process to the bottleneck queue. This implies that when buffers at the bottleneck queue are large enough to cause synchonisation, the arrival traffic to the queue would be bursty. However, for a buffer size of $15$ packets, there is no synchronisation among the TCP windows, and we can observe reduced burstiness in the traffic arrival process over short time scales. This suggests that only when the buffer size is sized small enough to avoid synchronisation, the aggregate arrival process behaves qualitatively similar to short-range dependent processes, a typical example of which is a Poisson process. This lends credence to the fact that packet arrivals to the bottleneck queue can be reasonably approximated as Poisson, with smaller buffers when there is no synchronisation. 

Next, we empirically validate that the queue size distribution can be reasonably approximated by that of either an $M/M/1/B$ or an $M/D/1/B$ queue, with smaller buffers.

\emph{Statistical properties of the queue size:} 
We have already established that when the buffer size at the bottleneck router is small enough to mitigate synchronisation effects, \emph{i.e.},  $15$ packets, the packet arrival process can be approximated by a Poisson process. Hence, to study the statistical properties of the queue size, we fix the buffer size at $15$ packets.  


In Fig.~\ref{fig:statistics_long_single}, we demonstrate the empirical Complementary Cumulative Distribution Function (CCDF) of the queue size. We consider $60$ long-lived flows, each with an access speed of $2$ Mbps. We also consider a scenario with $120$ long-lived flows each with an access speed of $1$ Mbps. We consider the time scale to be $50$ ms, consistent with our time scale of interest. The round trip times which we choose are $100$ ms and $200$ ms. 
 
For each of these cases, we perform a comparative study of the empirical queue length distribution with the theoretical queue distributions of  $M/M/1/B$ and $M/D/1/B$ queues. From the simulations we can infer that, with $60$ long-lived flows, the empirical queue distribution can be reasonably approximated by the corresponding queue distribution of either an $M/M/1/B$ or an $M/D/1/B$ queue when the bandwidth-delay product is large. Notably, as the number of flows is increased to $120$, this approximation holds true for larger round trip times perhaps due to increased statistical multiplexing. Indeed, we verify that, as the number of long-lived flows is increased further, this approximation holds true for even larger bandwidth-delay product values. Hence, with increased statistical multiplexing, the approximation holds at a larger bandwidth-delay product.

Thus, our empirical study serves to validate a very important modelling assumption: even with TCP (closed loop) controlled traffic, the packet drop probability at the bottleneck router can be reasonably approximated using that of an $M/M/1/B$ queue, in the absence of synchronisation.

\section{ Single bottleneck with heterogeneous delays}
\label{model_ah}
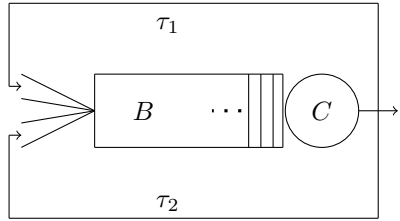
\begin{figure*}
\centering
{
\begin{tikzpicture}[scale=0.65]
\draw (0,0) -- (-1.5,0.75);
\draw (0,0) -- (-1.5,0.25);
\draw (0,0) -- (-1.5,-0.25);
\draw (-1.5,-0.75) -- (0,0);
\draw (0,-0.75) -- (3.85,-0.75);
\draw (3.85,-0.75) -- (3.85,0.75);
\draw (3.85,0.75) -- (0,0.75);
\draw (0,0.75) -- (0,-0.75);
\draw (3.65,-0.75) -- (3.65,0.75);
\draw (3.4,-0.75) -- (3.4,0.75);
\draw (3.15,-0.75) -- (3.15,0.75);
\draw (4.65,0) circle(0.75 cm);
\draw [loosely dotted,very thick] (3.0,0) -- (2.4,0);
\draw[->] (5.4,0) -- (6.2,0);
\draw (5.8,0) -- (5.8,2.2);
\draw (5.8,2.2) -- (-1.75,2.2);
\draw (-1.75,2.2) -- (-1.75,0.5);
\draw [->](-1.75,0.5) -- (-1.5,0.5);
\draw (5.8,0) -- (5.8,-2.2);
\draw (5.8,-2.2) -- (-1.75,-2.2);
\draw (-1.75,-2.2) -- (-1.75,-0.5);
\draw [->](-1.75,-0.5) -- (-1.5,-0.5);
\node at (1,0) {$B$};
\node at (4.65,0) {$C$};
\node at (1.5, 1.75) {$\tau_1$};
\node at (1.5,-1.9) {$\tau_2$};
\end{tikzpicture}
}
\caption{\emph{Single bottleneck topology with heterogeneous round trip times.} Two sets of TCP flows with round trip times $\tau_1$ and $\tau_2,$ feeding into a common bottleneck router. The buffer size at the router is $B$ and the link capacity is $C.$}
\label{fig:secscenario}
\end{figure*}
Till now, we have highlighted the interplay between buffer thresholds and stability of the underlying dynamical systems in a scenario wherein all flows are subject to a common round trip time. At this juncture, a natural question which might arise is: are the synchronisation effects of TCP windows for larger buffer thresholds an artefact of the assumption that all flows have a common round trip time? To answer this, we now investigate the impact of buffer thresholds on the dynamical properties of the system in the single bottleneck topology for the following scenario.

This model consists of a single bottleneck link with two distinct sets of \emph{many} long-lived TCP flows feeding into a common core router, as shown in Fig.~\ref{fig:secscenario}. The core router has a buffer size of $B$, with service rate per flow as $C'$. From a modelling perspective, both sets of TCP flows can be of different flavours and hence, can have different increase and decrease rules to govern the evolution of the corresponding window sizes. However, in this paper, we assume that the traffic in both sets is controlled by Compound TCP. Let the average window sizes of the two sets of flows be $w_{1}(t)$ and $w_{2}(t)$ respectively. For each acknowledgement received, the average window sizes increase by $i(w_1(t))$ and $i(w_2(t))$, and for each packet loss detected, the average window sizes decrease by  $d(w_1(t))$ and $d(w_2(t))$ respectively. Thus, for generalised TCP flows, the non-linear, time-delayed, fluid model of the system can be outlined
\begin{align}
\dot{w}_{j}(t) = \frac{w_{j}(t-\tau_{j})}{\tau_{j}}\bigg(i\left(w_{j}(t)\right)\Big(1-q(t,\tau_1,\tau_2)\Big) - d\left((w_{j}(t)\right)q(t,\tau_1,\tau_2)\bigg), \notag\\
\hspace{10ex}j=1,2,
\label{eq:modela_2}
\end{align}
where $q(t,\tau_1,\tau_2)$ represents the packet loss probability at the core router, and depends on the sending rates of both sets of TCP flows. Recall that the buffer size at the core router is dimensioned small, and the router deploys a Drop-Tail queue policy. When the bandwidth-delay product is large, the fluid model for the loss probability at the core router can be approximated as
\begin{align}
q(t)=\Bigg(\frac{w_1(t) / \tau_1 + w_2(t) / \tau_2}{\widetilde{C}}\Bigg)^B.
\label{eq:case1_loss}
\end{align} 
Here, $\widetilde{C}=2C'.$ Using this functional form for the loss probability at the core router, we now perform a local stability analysis for the system given by \eqref{eq:modela_2}. This would enable us to understand the impact of buffer thresholds on stability of the underlying dynamical system in presence of heterogeneous feedback delays. 
\subsection{Local stability and Hopf bifurcation analysis with long-lived flows}
Suppose $(w_1^{*}, w_{2}^{*})$ is a non-trivial equilibrium of \eqref{eq:modela_2} and let $u_{1}(t)=w_{1}(t)-w_{1}^{*}$ and $u_{2}(t)=w_{2}(t)-w_{2}^{*}$ be small perturbations about $w_{1}^{*}$ and $w_{2}^{*}$ respectively. Linearising \eqref{eq:modela_2} about this equilibrium, we obtain
\begin{align}
\label{eq:linearmodela_2}
&\dot{u}_1(t) = -\mathcal{M}_{1}u_{1}(t)-\mathcal{N}_{1}u_{1}(t-\tau_{1})-\mathcal{P}_{1}u_{2}(t-\tau_2),\notag\\
&\dot{u}_2(t) = -\mathcal{M}_{2}u_2(t)-\mathcal{N}_{2}u_{2}(t-\tau_2)-\mathcal{P}_{2}u_{1}(t-\tau_{1}).
\end{align}
Here, the increase and decrease functions for Compound TCP given by \eqref{eq:Compound}, and the functional form of the loss probability at the core router given by~\eqref{eq:case1_loss} yield the following equilibrium coefficients: 
\begin{align}
\label{eq:coefficients}
\mathcal{M}_{j} &=-\frac{\alpha}{\tau_{j}}\left(k-2\right)\ \left(w_{j}^{*}\right)^{k-1}\left(1-\frac{1}{(2C')^B}\left(\frac{w_{1}^{*}}{\tau_1}+\frac{w_{2}^{*}}{\tau_2}\right)^{B}\right),\notag\\
\mathcal{N}_{j} &=\frac{ B \left(w_{j}^{*}\right)^{2}}{\tau_{j}^{2}\left(2C'\right)^{B}}\left(\alpha \left(w_{j}^{*}\right)^{k-2}+\beta\right)\left(\frac{w_{1}^{*}}{\tau_{1}}+\frac{w_{2}^{*}}{\tau_{2}}\right)^{B-1},\notag\\
\mathcal{P}_{j}&=\frac{ B \left(w_{j}^{*}\right)^{2}}{\tau_{1}\tau_{2}\left(2C'\right)^{B}}\left(\alpha \left(w_{j}^{*}\right)^{k-2}+\beta\right)\left(\frac{w_{1}^{*}}{\tau_{1}}+\frac{w_{2}^{*}}{\tau_{2}}\right)^{B-1},
\end{align}
for $j=1,2$. Looking for exponential solutions, we obtain the characteristic equation for the linearised system \eqref{eq:linearmodela_2} as
\begin{align}
\lambda^2 &+\lambda\left(\mathcal{N}_1 e^{-\lambda\tau_1}+\mathcal{N}_2 e^{-\lambda\tau_2}\right)+\left(\mathcal{M}_1\mathcal{N}_2 e^{-\lambda\tau_1}+\mathcal{M}_2\mathcal{N}_1 e^{-\lambda\tau_2}\right)\notag\\
&+\lambda\left(\mathcal{M}_1+\mathcal{M}_2\right)+\mathcal{M}_1\mathcal{M}_2=0.
\label{eq:charac_hrtt}
\end{align}
Now, for the linearised system \eqref{eq:linearmodela_2} to be asymptotically stable, all roots of the characteristic equaion \eqref{eq:charac_hrtt} should have negative real parts. We then aim to find conditions on different system parameters which would ensure asymptotic stability of the linearised system \eqref{eq:linearmodela_2}. However, obtaining the necessary and sufficient condition for the asymptotic stability of a system having a characteristic equation of the form \eqref{eq:charac_hrtt} analytically seems rather hard. Hence, to investigate how different system parameters impact the asymptotic stability of \eqref{eq:linearmodela_2}, we outline a sufficient condition for stability for this system. To that end, we make use of a result which was derived in \cite{Kharitonov}. This would then yield a sufficient condition for stability of \eqref{eq:modela_2} about its equilibrium.

Note that the equilibrium $(w_1^\ast,w_2^{\ast})$ satisfy the following conditions:
\begin{align}
\alpha \left(w_1^{\ast}\right)^{k-1}=\left(\alpha \left(w_1^{\ast}\right)^{k-1}+\beta w_1^{\ast}\right)\frac{1}{\left(2C'\right)^B}\left(\frac{w_1^{\ast}}{\tau_1}+\frac{w_2^{\ast}}{\tau_2}\right)^B,\notag\\
\alpha \left(w_2^{\ast}\right)^{k-1}=\left(\alpha \left(w_2^{\ast}\right)^{k-1}+\beta w_2^{\ast}\right)\frac{1}{\left(2C'\right)^B}\left(\frac{w_1^{\ast}}{\tau_1}+\frac{w_2^{\ast}}{\tau_2}\right)^B.
\label{eq:eqmodela_2}
\end{align}
From \eqref{eq:eqmodela_2}, it can be shown that 
\begin{align}
\frac{\alpha (w_1^{\ast})^{k-1}}{\alpha (w_1^{\ast})^{k-1}+\beta w_1^{\ast}}=\frac{\alpha (w_2^{\ast})^{k-1}}{\alpha (w_2^{\ast})^{k-1}+\beta w_2^{\ast}}
\end{align}
This implies that $\left(w_1^{\ast}\right)^{2-k}=\left(w_2^{\ast}\right)^{2-k}.$ It can also be shown that this equilibrium is unique. At this equilibrium, the coefficients \eqref{eq:coefficients} reduce to 
\begin{align}
\mathcal{M}_j=\frac{\mathcal{A}}{\tau_j},\,\,\,\mathcal{N}_j=\frac{\mathcal{B}}{\tau_j^2}, \,\,\, \text{and}\,\,\, \mathcal{P}_j=\frac{\mathcal{B}}{\tau_1\tau_2},\,\,\,j=1,2.
\end{align}
Here,
\begin{align}
\mathcal{A} &=-\alpha\left(k-2\right)\ \left(w^{*}\right)^{k-1}\left(1-\frac{1}{(2C')^B}\left(\frac{w^{*}}{\tau_1}+\frac{w^{*}}{\tau_2}\right)^{B}\right),\,\,\,\text{and}\notag\\
\mathcal{B} &=\frac{ B \left(w^{*}\right)^{2}}{\left(2C'\right)^{B}}\left(\alpha \left(w^{*}\right)^{k-2}+\beta\right)\left(\frac{w^{*}}{\tau_{1}}+\frac{w^{*}}{\tau_{2}}\right)^{B-1}.
\end{align}
Observe that, the linearised system \eqref{eq:linearmodela_2} can be re-written in the following matrix form 
\begin{align*}
\underbrace{\begin{bmatrix}
\dot{u}_1(t) \\
\dot{u}_2(t) 
\end{bmatrix}}_{\dot{\textbf{U}}(t)}
& =
\underbrace{\begin{bmatrix}
-\mathcal{M}_1 & 0 \\
 0 & -\mathcal{M}_2
\end{bmatrix}}_{A}
\underbrace{\begin{bmatrix}
u_1(t) \\
u_2(t) 
\end{bmatrix}}_{\textbf{U}(t)}
+
\underbrace{\begin{bmatrix}
-\mathcal{N}_1 & 0 \\
-\mathcal{P}_2 & 0
\end{bmatrix}}_{A_1}
\underbrace{\begin{bmatrix}
u_1(t - \tau_1) \\
u_2(t - \tau_1)
\end{bmatrix}}_{\textbf{U}(t - \tau_1)} \\
& +
\underbrace{\begin{bmatrix}
0 & -\mathcal{P}_1  \\
0 & -\mathcal{N}_2
\end{bmatrix}}_{A_2}
\underbrace{\begin{bmatrix}
u_1(t - \tau_2) \\
u_2(t - \tau_2) 
\end{bmatrix}}_{\textbf{U}(t- \tau_2)}.
\end{align*}
Succinctly, the above can be written as 
\begin{align}
\dot{\textbf{U}}(t)= A\textbf{U}(t)+\sum_{i=1}^{2}A_i\textbf{U}(t-\tau_i)
\label{eq:matlinearmodela_2}
\end{align}
As stated in \cite{Kharitonov}, a sufficient condition for the asymptotic stability of the linear system \eqref{eq:linearmodela_2} is 
\begin{align}
\sum_{i=1}^{2} \Vert A_i\Vert \tau_i<1.
\label{eq:suff_hrtt}
\end{align}
To obtain a sufficient condition for the asymptotic stability of  \eqref{eq:linearmodela_2}, we now use the Frobenius norm of a matrix in \eqref{eq:suff_hrtt} which yields 
\begin{align}
\mathcal{B}\sqrt{\frac{1}{\tau_1^2}+\frac{1}{\tau_2^2}}<\frac{1}{2}.
\label{eq:suff_hrtt2}
\end{align}
Substituting $\mathcal{B}$ in \eqref{eq:suff_hrtt2} and simplifying yields the following condition:
\begin{align}
\alpha B \left(w^{\ast}\right)^{k-1}\sqrt{1-\frac{2}{\frac{\tau_1}{\tau_2}+\frac{\tau_2}{\tau_1}+2}}<\frac{1}{2}.
\label{eq:suff_hrtt2_final}
\end{align}
The above would then yield a sufficient condition for the local stability of system \eqref{eq:modela_2} about its equilibrium with Compound TCP flows. The above condition evidently highlights that buffer thresholds need to be dimensioned rather carefully to ensure stability. In particular, larger buffer thresholds might prompt the system to lose local stability and transit into a locally unstable regime.

Observe that in addition to buffer threshold of the core router and protocol parameters, the condition \eqref{eq:suff_hrtt2_final} depends on the ratios of the round trip times. To facilitate a better understanding of the impact of heterogeneous feedback delays on local stability, we consider two cases: $(i)$ both $\tau_1$ and $\tau_2$ are comparable to each other, and $(ii)$ either $\tau_1$ or $\tau_2$ is significantly smaller than the other. In both cases, increasing buffer thresholds might destabilise the system. This suggests that, even if one round trip time is large, local stability might be lost with increasing buffer sizes. However, condition \eqref{eq:suff_hrtt2_final} highlights that in the latter case, the stability region might be smaller as compared to the former.  
\subsection{Numerical computations}
Since \eqref{eq:suff_hrtt2_final} is a sufficient condition for local stability of \eqref{eq:modela_2}, violating the same by varying any model parameter would not guarantee the loss of local stability. We now numerically illustrate through DDE-BIFTOOL (version 2.03) \cite{Engelborghs}, that system \eqref{eq:modela_2} indeed loses local stability if the buffer threshold at the core router is increased beyond a critical value. Specifically, local stability is lost via a Hopf bifurcation, when exactly one pair of complex conjugate roots crosses over the imaginary axis from left half to the right half of the complex plane. At the point of criticality, the system has exactly one pair of complex conjugate roots on the imaginary axis.

For the numerical computation, we consider the following values of the protocol parameters: $\alpha=0.125,\,k=0.75,\,\text{and}\,\beta = 0.5.$ Further, we fix $\widetilde{C} = 140$ packets/second. Additionally, we consider the following values for the round trip times: $(i)$ $\tau_1=0.1$ seconds and $\tau_2=0.2$ seconds $(ii)$ $\tau_1=0.01$ seconds and $\tau_2=0.2$ seconds. We then vary the buffer threshold $B$ in the interval $[10,100].$ For case $(i),$ system \eqref{eq:modela_2} undergoes a Hopf bifurcation at $B=25$ packets. For case $(ii),$ it occurs at $B=15$ packets.
\subsubsection*{Stability charts}
\begin{figure}[t!]
\begin{center}
  \psfrag{b}{\hspace{1mm}Buffer size, $B$}
  \psfrag{T}{\hspace{2mm}Protocol parameter, $\alpha$}
  \psfrag{0}{\begin{scriptsize}$0$\end{scriptsize}}
  \psfrag{0.2}{\begin{scriptsize}$0.2$\end{scriptsize}}
  \psfrag{15}{\begin{scriptsize}$15$\end{scriptsize}}
  \psfrag{50}{\begin{scriptsize}$50$\end{scriptsize}}
  \psfrag{h}{\scalebox{0.85}{Hopf condition}}
  \psfrag{x}{\scalebox{0.85}{Sufficient condition}}
  \psfrag{y}{Non-oscillatory convergence}
  
  \includegraphics[width=2.75in,height=3.75in,angle=270]{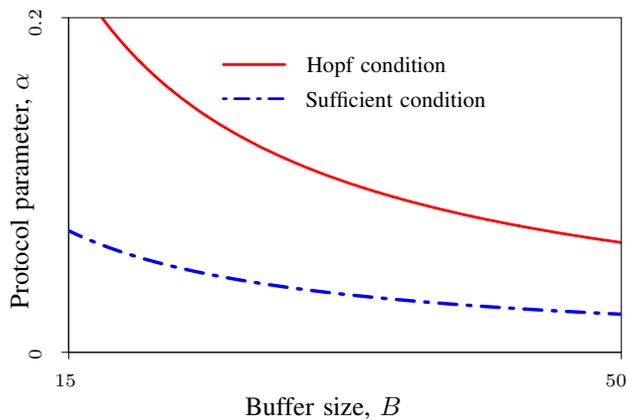}

   \caption {\emph{Stability Chart.} Shows the Hopf condition and sufficient condition for local stability with Compound TCP, with $\tau_1=0.1$ and $\tau_2=0.2$ seconds. The stability chart highlights the impact of the buffer size $B$ and the protocol parameter $\alpha$ to ensure local stability.}
  \label{fig:stability_hrtt1}
 \end{center}
 \end{figure}

\begin{figure}[h!]
\begin{center}
  \psfrag{b}{\hspace{1mm}Buffer size, $B$}
  \psfrag{T}{\hspace{1.5mm}Protocol parameter, $\alpha$}
  \psfrag{0}{\begin{scriptsize}$0$\end{scriptsize}}
  \psfrag{0.15}{\begin{scriptsize}$0.15$\end{scriptsize}}
  \psfrag{15}{\begin{scriptsize}$15$\end{scriptsize}}
  \psfrag{50}{\begin{scriptsize}$50$\end{scriptsize}}
  \psfrag{h}{\scalebox{0.85}{Hopf condition}}
  \psfrag{x}{\scalebox{0.85}{Sufficient condition}}
  \psfrag{y}{\scalebox{0.85}{Non-oscillatory convergence}}
  
  \includegraphics[width=2.75in,height=3.75in,angle=270]{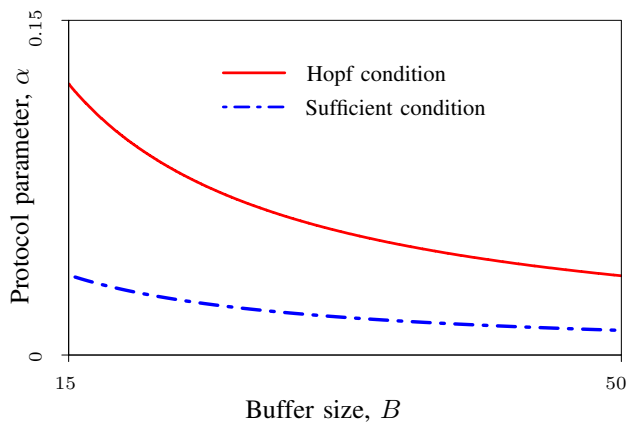}

  \caption {\emph{Stability Chart.} Shows the Hopf condition and sufficient condition for local stability with Compound TCP, with $\tau_1=0.01$ and $\tau_2=0.2$ seconds. The stability chart captures the trade off between the buffer size at the router $B$ and the Compound parameter $\alpha$ to ensure local stability.}
  \label{fig:stability_hrtt2}
 \end{center}
 \end{figure}

We now plot some stability charts to illustrate the impact of the system parameters on the local stability of \eqref{eq:modela_2}, see Figs. \ref{fig:stability_hrtt1} and \ref{fig:stability_hrtt2}. For this, we again consider two cases for the round trip times: $(i)$ $\tau_1=0.1$ seconds and $\tau_2=0.2$ seconds $(ii)$ $\tau_1=0.01$ seconds and $\tau_2=0.2$ seconds. We vary the buffer size $B$ in the interval $[15,50]$ and observe the variation in the protocol parameter $\alpha$ at the Hopf condition, and the boundary of the sufficient condition given by \eqref{eq:suff_hrtt2_final}. The remaining parameters are fixed as follows: $k=0.75,\,\beta = 0.5,\text{and}\,\,\widetilde{C} = 140$ packets/second. Note that the Hopf conditions are obtained numerically through DDE-BIFTOOL. In both cases, we observe a trade off between the parameters ensure stability. In particular, even if one round trip time is large, increasing buffer sizes would destabilise the system.
\subsection{Packet-level simulations}
Now, we present some packet-level simulations, which corroborate the insights obtained from our stability analysis. In particular, we show that emergence of limit cycles in the queue size dynamics, as the buffer size at the bottleneck router is increased.

\begin{figure}[t!]
\begin{center}
  \psfrag{cccc}{\hspace{-1mm}Queue size (pkts)}
  \psfrag{bbbb}{\hspace{2mm}Utilisation (\%)}
  \psfrag{275}{\begin{scriptsize}$275$\end{scriptsize}}
  \psfrag{300}{\begin{scriptsize}$300$\end{scriptsize}}
  \psfrag{60}{\begin{scriptsize}$60$\end{scriptsize}}
  \psfrag{100}{\begin{scriptsize}$100$\end{scriptsize}}
  \psfrag{15}{\begin{scriptsize}$15$\end{scriptsize}}
  \psfrag{eeee}{\begin{scriptsize}\hspace{8mm}
 Buffer size = $15$ packets\end{scriptsize}}
       \psfrag{dddd}{\begin{scriptsize}\hspace{8mm}
 Buffer size = $100$ packets\end{scriptsize}}
  \psfrag{aaaa}{\hspace{6mm}Time (seconds)}
  
  \includegraphics[width=2.75in,height=3.75in,angle=270]{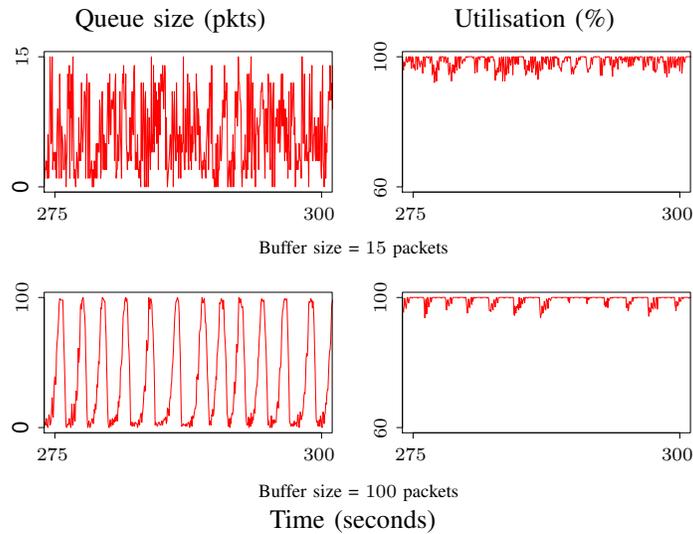}

  \caption {\emph{Queue size dynamics with heterogeneous round trip times.} Two sets of $30$ TCP flows, each with an access speed of $2$ Mbps, and feeding into a bottleneck router with a link capacity of $100$ Mbps. The round trip times of the two sets of flows are fixed at $100$ ms and $200$ ms. It can be easily seen that that the queue size dynamics exhibits limit cycles, as the buffer size at the bottleneck router is increased.}
  \label{fig:hetrtt_1}
 \end{center}
 \end{figure}

\begin{figure}[h!]
\begin{center}
   \psfrag{cccc}{\hspace{-1mm}Queue size (pkts)}
  \psfrag{bbbb}{\hspace{2mm}Utilisation (\%)}
  \psfrag{275}{\begin{scriptsize}$275$\end{scriptsize}}
  \psfrag{300}{\begin{scriptsize}$300$\end{scriptsize}}
  \psfrag{60}{\begin{scriptsize}$60$\end{scriptsize}}
  \psfrag{100}{\begin{scriptsize}$100$\end{scriptsize}}
  \psfrag{15}{\begin{scriptsize}$15$\end{scriptsize}}
  \psfrag{eeee}{\begin{scriptsize}\hspace{8mm}
 Buffer size = $15$ packets\end{scriptsize}}
       \psfrag{dddd}{\begin{scriptsize}\hspace{8mm}
 Buffer size = $100$ packets\end{scriptsize}}
  \psfrag{aaaa}{\hspace{6mm}Time (seconds)}
  \includegraphics[width=2.75in,height=3.75in,angle=270]{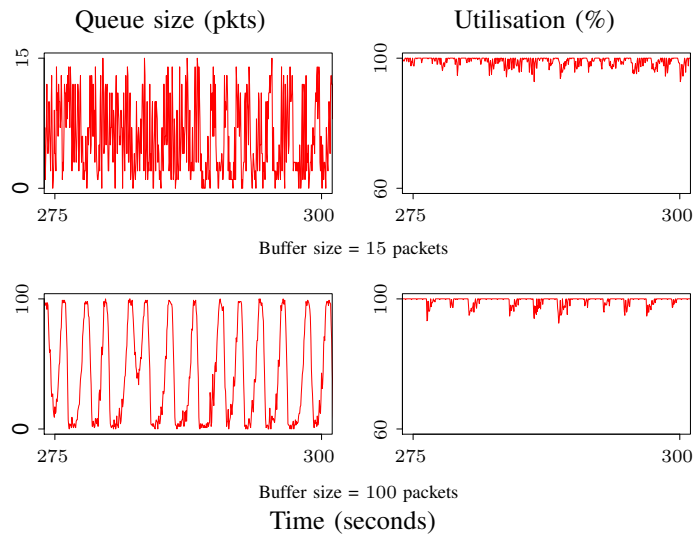}

  \caption {\emph{Queue size dynamics with heterogeneous round trip times.} Two sets of $30$ TCP flows, each with an access speed of $2$ Mbps, and feeding into a bottleneck router with a link capacity of $100$ Mbps. The round trip times of the two sets of flows are fixed at $10$ ms and $200$ ms. As the buffer size at the bottleneck router is increased, the queue size exhibits limit cycles.}
  \label{fig:hetrtt_2}
 \end{center}
 \end{figure}

We consider two sets of $30$ long-lived Compound TCP flows, each with an access speed of $2$ Mbps. The flows feed into a router with a link capacity of $100$ Mbps. We first consider the case when the average round trip times of both sets of flows are comparable to each other, and are fixed as $100$ ms and $200$ ms respectively. It can be seen from Fig. \ref{fig:hetrtt_1} that as the buffer size is increased from $15$ to $100$ packets, the queue size exhibits limit cycles. We then consider the case when one round trip time is much smaller than the other. In this case, we fix the average round trip times of the two sets as $10$ ms and $200$ ms respectively, see Fig. \ref{fig:hetrtt_2}. We can observe that a similar insight holds in this case also. Thus, even if one set of flows has a large round trip time, the underlying dynamical system loses stability if the buffer size at the bottleneck router is increased. This corroborates our analytical insights.


\section{ Single bottleneck with heavy-tailed files}
\label{model_aheavy}
Till now, we have shown the emergence of limit cycles in the queue dynamics in the presence of many TCP sources, each of which has an infinite file to transfer. At this juncture, a natural question that might arise is, are these limit cycles a consequence of the underlying assumptions on the network traffic? To answer this, we now analyse system \eqref{eq:modela2} under a different workload model, which caters for more realistic traffic scenarios.

Empirical studies on real time Ethernet LAN traffic have established the presence of high variability at the TCP connection level~\cite{Willinger}. It has been identified that heavy-tailed connections generated by individual TCP senders are a primary cause of this variability. To account for this, we consider the following workload model, which has been widely studied in the literature~\cite{Ayesta}. In this scenario, flows from each user arrive at the transport layer as a Poisson process, \emph{i.e.,} the interarrival time between any two flows is exponentially distributed. To cater for the heavy-tailed nature of TCP connections, we assume that file sizes follow a Pareto distribution. For an overview on Pareto distribution, see~\cite{Arnold}. The heavy-tailed nature of the file sizes ensures that each user generates a large file with a non-negligible probability. Hence, as the number of TCP senders increases, the number of long flows which are feedback controlled also increases. Apart from the presence of long flows, a salient feature of this workload model is that, a significant fraction of the total flows arriving at the transport layer is contributed by very short flows, or ``mice." However, consistent with real network traffic, a major portion of the traffic is still contributed by the long flows. This motivates us to model the window evolution of the long flows in the congestion avoidance phase as \eqref{eq:modela2}, to capture the dynamical properties of this system.

 Apart from high variability, it has been empirically shown that this source model also exhibits long-range dependence ~\cite{Willinger}. As argued in~\cite{Poisson_limit}, even with long-range dependent sources, as the number of multiplexed sources feeding into a small buffer grows large, the aggregate packet arrival process tends towards Poisson. For analytical purposes, this gives us the confidence to use the same functional form of the packet loss probability as with long-lived flows, in this scenario. Recall that  with long-lived flows, the packet loss probability at the core router is 
\begin{align}
\label{eq:loss_heavy}
p\left(w(t)\right)=\left(\frac{w(t)}{C'\tau}\right)^B.
\end{align}
 Let $V$ denote the random variable for the TCP connection sizes, and $\mathbb{E}(V)$ the expected file size in packets. Recall that the rate at which packets are transmitted is $x(t)=w(t)/\tau.$ We define $\mu(t)=x(t)/\mathbb{E}(V).$
\subsection{Local stability and Hopf bifurcation analysis with heavy-tailed files}
We now outline some stability conditions for Compound TCP, using the functional forms for the increase and decrease functions given by \eqref{eq:Compound}. We then obtain bounds on various model parameters, which would ensure local stability. Using\eqref{eq:loss_heavy}, and $w^{\ast}=\mathbb{E}(V)\mu^{\ast}\tau,$ the necessary and sufficient condition for local stability can be obtained as 
\begin{align}
\label{eq:NS_heavy}
&\alpha\left(\mathbb{E}(V)\mu^{\ast}\tau\right)^{k-1}\sqrt{B^2-\left(k-2\right)^2\left(1-\left(\frac{\mathbb{E}(V)\mu^{\ast}}{C'}\right)^B\right)^2}\notag\\
&<\cos^{-1}\left(\frac{\left(k-2\right)\left(1-\left(\frac{\mathbb{E}(V)\mu^{\ast}}{C'}\right)^B\right)}{B}\right).  
\end{align} 
When condition \eqref{eq:NS_heavy} is met with an equality, we obtain the Hopf condition. Further, in this scenario, a simple sufficient condition with Compound TCP flows is 
\begin{align}
\label{eq:S_heavy}
\alpha B\left(\mathbb{E}(V)\mu^{\ast}\tau\right)^{k-1}<\pi/2.
\end{align}
From conditions \eqref{eq:NS_heavy} and \eqref{eq:S_heavy}, we can easily observe that apart from protocol parameters and network parameters, local stability of \eqref{eq:modela2} now depends on the expected file size brought by the sessions. If we assume that $V$ follows a Pareto distribution with shape parameter $1<\chi<2,$ then 
\begin{align}
\label{eq:Ex_pareto}
\mathbb{E}(V)=\frac{\chi m}{\chi-1},
\end{align}
where $m$ is defined as the scale parameter of the Pareto distribution. Using \eqref{eq:Ex_pareto}, we can then re-write the necessary and sufficient condition as
\begin{align}
\label{eq:NS_heavy_shape}
&\alpha\left(\chi m\mu^{\ast}\tau\right)^{k-1}\sqrt{B^2-\left(k-2\right)^2\left(1-\left(\frac{\chi m\mu^{\ast}}{\left(\chi-1\right)C'}\right)^B\right)^2}\notag\\
&<\left(\chi-1\right)^{k-1}\cos^{-1}\left(\frac{\left(k-2\right)\left(1-\left(\frac{\chi m\mu^{\ast}}{\left(\chi-1\right)C'}\right)^B\right)}{B}\right).  
\end{align}
Similarly, condition \eqref{eq:S_heavy} can be re-written as
\begin{align}
\label{eq:S_heavy_shape}
\alpha B\left(\frac{\chi m\mu^{\ast}\tau}{\chi-1}\right)^{k-1}<\pi/2.
\end{align}
Conditions \eqref{eq:NS_heavy_shape} and \eqref{eq:S_heavy_shape} clearly highlight the interdependence of the shape parameter with network and protocol parameters to ensure stability of the system.
\subsection{Non-oscillatory convergence with heavy-tailed files}
We have already shown that both protocol parameters and network parameters such as queue thresholds or buffer sizes have to be chosen rather carefully if stability is to be ensured. Additionally, the file size distribution also impacts stability. However, even if local stability is ensured, the convergence of the system can be oscillatory or non-oscillatory. If the convergence is oscillatory, there would be some temporary degree of synchronisation among TCP flows before they can desynchronise again. There would be loss in link utilisation and bursty packet losses whenever synchronisation happens. If the system shows oscillatory convergence, and the oscillations in the queue size dynamics persist for a long time, it would affect the network performance. Hence, it becomes imperative to design parameters such that non-oscillatory convergence of the system can be ensured. To that end, we seek bounds on various protocol and network parameters to ensure non-oscillatory convergence of \eqref{eq:modela2}. The following theorem outlines the necessary and sufficient condition for non-oscillatory convergence of the linearised system \eqref{eq:lineara}. This would then yield the necessary and sufficient condition for non-oscillatory convergence of the original system \eqref{eq:modela2} in a neighbourhood of its equilibrium. 

We now show that the solution of system \eqref{eq:lineara} shows non-oscillatory convergence if and only if the parameters $a$, $b$ and $\tau$ satisfy the condition $\ln\left(b\tau\right)+a\tau+1< 0$. 

The boundary condition for the solution of \eqref{eq:lineara} to be non-oscillatory is the point at which the curve $f(s)=s+a+be^{-s\tau}$ touches the real axis. If this point is $\sigma$, then
\begin{eqnarray}
f(\sigma) =& \sigma+a+be^{-\sigma\tau} &= 0, \label{conv1} \hspace{1ex} \mbox{and}\\
f^{'}(\sigma) =& 1-b\tau e^{-\sigma\tau} &= 0 \label{conv2}.
\end{eqnarray}
From \eqref{conv2}, we get
\begin{align}
be^{-\sigma\tau}=\frac{1}{\tau} \hspace{1ex} \mbox{and} \hspace{1ex} \sigma= \frac{\ln\left(b\tau\right)}{\tau}.
\end{align}
Substituting values of $\sigma$ and $be^{-\sigma\tau}$ in \eqref{conv1} gives
\begin{align}
\ln\left(b\tau\right)+a\tau +1=0.
\end{align}
We now claim that the necessary and sufficient condition for non-oscillatory convergence of the equilibrium point is
\begin{align}
\label{eq:oscillation_condition}
\ln\left(b\tau\right)+a\tau+1< 0.
\end{align}
Suppose the solution of \eqref{eq:lineara} exhibits non-oscillatory convergence to its equilibrium \emph{i.e.} all roots of \eqref{eq:chareq} are real and lie on the left half of the complex plane. Then, we prove that the region of non-oscillatory convergence is characterised by \eqref{eq:oscillation_condition}. We prove this claim by contradiction. We assume that the condition for non-oscillatory convergence is
\begin{align}
\ln\left(b\tau\right)+a\tau +1>0.\label{eq:condition_contradiction}
\end{align}
Let $\sigma=-\alpha,$ where $\alpha>0$ is a root of \eqref{eq:chareq}. Then, substituting $\sigma=-\alpha$ in \eqref{eq:chareq}, we obtain
\begin{align}
\alpha = a+be^{\alpha\tau}.\label{eq:eqn_beg}
\end{align}
Multiplying both sides of \eqref{eq:eqn_beg} by $\tau e^{a\tau}$ yields
\begin{align*}
\alpha \tau e^{a\tau} =a \tau e^{a\tau}+b\tau e^{a\tau}e^{\alpha \tau}> a\tau e^{a\tau}+\frac{e^{\alpha\tau}}{e}.
\end{align*}
where, the inequality follows from \eqref{eq:condition_contradiction}. This leads to
\begin{align}
\ln\left(\alpha \tau -a \tau\right)>-1+\left(\alpha\tau-a\tau\right).\label{eq:contradiction}
\end{align}
Now, it can be easily observed form \eqref{eq:eqn_beg} that $\alpha>a.$ Hence, we obtain a contradiction \eqref{eq:contradiction}, since $\ln x\leq x-1, \hspace{2mm}\forall x>0.$ Thus, if the solution of system \eqref{eq:lineara} exhibits non-oscillatory convergence, then the parameters $a$ and $b$ satisfy the following condition:
\begin{align*}
\ln\left(b\tau\right)+a\tau +1<0.
\end{align*}
Now, we prove the converse statement \emph{i.e.} if the system parameters satisfy \eqref{eq:oscillation_condition} and all roots of \eqref{eq:chareq} have negative real parts, then all roots are real. To prove this by contradiction, we assume that all roots of \eqref{eq:chareq} have non-zero imaginary part and are of the form $s=-\sigma-j\omega,$ where $\sigma>0.$ Substituting $s$ in \eqref{eq:chareq} and separating real and imaginary parts, we obtain
\begin{align}
\sigma &=a+be^{\sigma \tau}\cos \omega \tau, \hspace{2ex} \text{and}\label{eq:eqn_1}\\
\omega &=be^{\sigma \tau}\sin \omega \tau. \label{eq:eqn_2}
\end{align}
From \eqref{eq:eqn_1} and \eqref{eq:eqn_2}, we get
\begin{align}
\frac{\tan \omega \tau}{\omega \tau}=\frac{1}{\left(a-\sigma\right)\tau}.\label{eq:eqn_3}
\end{align}
Now, the condition given by \eqref{eq:oscillation_condition} implies that $\left(\sigma-a\right)\tau\geq 1.$ This in turn implies that $\omega=0$ is the unique solution to the equation \eqref{eq:eqn_3}. Hence, the necessary and sufficient condition for non-oscillatory convergence of the solution of  \eqref{eq:lineara} is
\begin{align*}
\ln\left(b\tau\right)+a\tau +1<0.
\end{align*} 
With Compound TCP, the necessary and sufficient condition for non-oscillatory convergence of \eqref{eq:modela2} around its equilibrium is 
\begin{align}
\label{eq:osccompound}
&\alpha B \left(\chi m \mu^{\ast}\tau\right)^{k-1}\notag\\<
&\left(\chi-1\right)^{k-1}\exp \left(\alpha\left(k-2\right)\left(\frac{\chi m\mu^\ast\tau}{\chi-1}\right)^{k-1}\left(1-\left(\frac{\chi m\mu^{\ast}}{\left(\chi-1\right)C'}\right)^B\right)-1\right).
\end{align}
The above condition captures the dependence of protocol parameters $\alpha$ and $k$, network parameters like buffer size $B$, and shape parameter $\chi$ in ensuring the non-oscillatory convergence of the solution of \eqref{eq:modela2} to its equilibrium point, for Compound TCP.
\begin{figure}[t!]
\begin{center}
\vspace{-5mm}
  \psfrag{b}{\hspace{-6.5mm}Buffer size, $B$}
  \psfrag{T}{\hspace{2mm}Protocol parameter, $\alpha$}
  \psfrag{0}{\begin{scriptsize}$0$\end{scriptsize}}
  \psfrag{1}{\begin{scriptsize}$1$\end{scriptsize}}
  \psfrag{5}{\begin{scriptsize}$5$\end{scriptsize}}
  \psfrag{50}{\begin{scriptsize}$50$\end{scriptsize}}
  \psfrag{h}{Hopf condition}
  \psfrag{x}{Sufficient condition}
  \psfrag{y}{Non-oscillatory convergence}
  
  \includegraphics[width=2.75in,height=3.75in,angle=270]{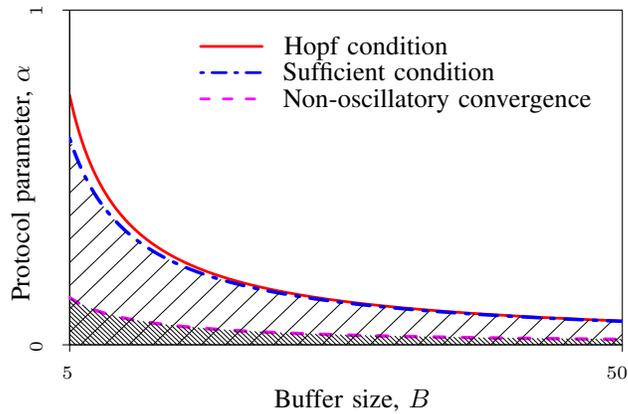}

  \caption {\emph{Stability chart for single bottleneck topology with heavy-tailed files.} The Hopf condition, the sufficient condition for local stability and condition for non-oscillatory convergence for compound TCP are outlined. The stability chart highlights the impact of the buffer threshold $B$ and the protocol parameter $\alpha$ to ensure local stability as well as non-oscillatory convergence.}
  \label{fig:stability_heavy1}
 \end{center}
 \end{figure}
 \begin{figure}[t!]
\begin{center}
\vspace{-5mm}
  \psfrag{T}{\hspace{-2cm}Expected file size, $\mathbb{E}(V)$}
  \psfrag{b}{\hspace{-2mm}Protocol parameter, $\alpha$}
  \psfrag{0}{\begin{scriptsize}$0$\end{scriptsize}}
  \psfrag{0.7}{\begin{scriptsize}$0.7$\end{scriptsize}}
  \psfrag{60}{\begin{scriptsize}$60$\end{scriptsize}}
  \psfrag{100}{\begin{scriptsize}$100$\end{scriptsize}}
  \psfrag{h}{Hopf condition}
  \psfrag{x}{Sufficient condition}
  \psfrag{y}{Non-oscillatory convergence}
  
  \includegraphics[width=2.75in,height=3.75in,angle=270]{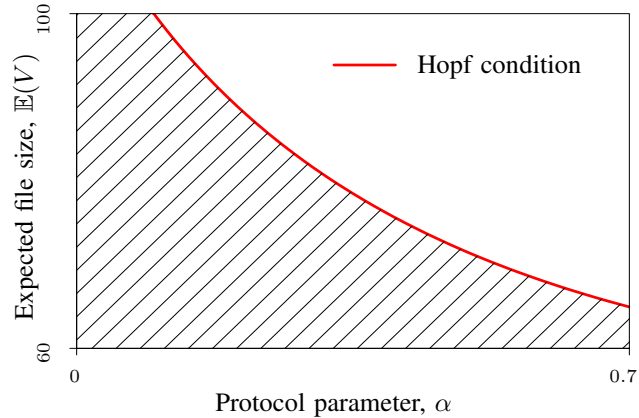}

  \caption {\emph{Stability chart for single bottleneck topology with heavy-tailed files.} The stability chart highlights the impact of the expected file size $\mathbb{E}(V)$ and the protocol parameter $\alpha$ on local stability. Observe the trade-off among the parameters to ensure stability.}
  \label{fig:stability_heavy2}
 \end{center}
 \end{figure}
  \begin{figure}[t!]
\begin{center}
\vspace{-5mm}
  \psfrag{b}{\hspace{6mm}Round trip time, $\tau$}
  \psfrag{T}{\hspace{4mm}Shape parameter, $\chi$}
  \psfrag{0.01}{\begin{scriptsize}$0.01$\end{scriptsize}}
  \psfrag{1.5}{\begin{scriptsize}$1.5$\end{scriptsize}}
  \psfrag{1.4}{\begin{scriptsize}$1.4$\end{scriptsize}}
  \psfrag{1.5}{\begin{scriptsize}$1.5$\end{scriptsize}}
  \psfrag{h}{Hopf condition}
  \psfrag{x}{Sufficient condition}
  \psfrag{y}{Non-oscillatory convergence}
  
  \includegraphics[width=2.75in,height=3.75in,angle=270]{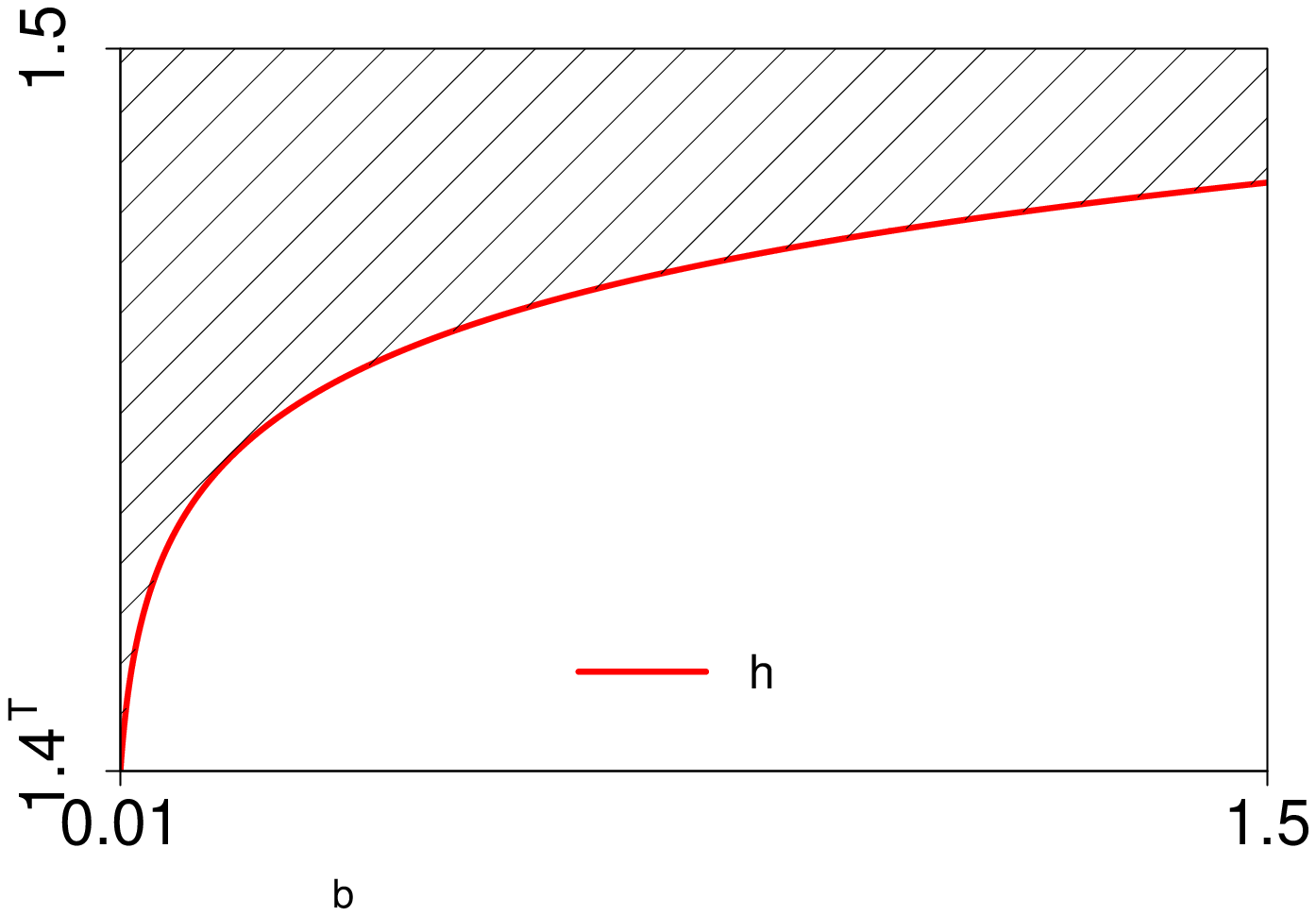}

  \caption {\emph{Stability chart for single bottleneck topology with heavy-tailed files.} The stability chart highlights the impact of the Pareto sized files characterised by the shape parameter $\chi$ and the round trip time $\tau$ on local stability. Note that a higher $\chi$ can accommodate a larger feedback delay.}
  \label{fig:stability_heavy3}
 \end{center}
 \end{figure}
 \subsubsection*{Stability charts} Given these conditions, we now aim to gain a better understanding of how various model parameters impact local stability. To that end, we plot some stability charts, as shown in Figs. \ref{fig:stability_heavy1}, \ref{fig:stability_heavy2} and \ref{fig:stability_heavy3}. Further, we assume $1-p^{\ast}\approx 1,$ to simplify the equilibrium structure of \eqref{eq:modela2}. 
 
Fig. \ref{fig:stability_heavy1} shows the stability chart with respect to two parameters, the buffer size at the core router $B,$ and the increase parameter $\alpha.$ We first fix the values of the remaining system parameters as $C'=140\,\text{packets/second}, \,\tau=0.2\,\text{seconds},\,\mathbb{E}(V)=100\,\text{packets}, \,\text{and} \,\,\chi=1.5.$  The protocol parameters $k$ and $\beta$ are fixed at their default values $0.75$ and $0.5$ respectively. We now vary the buffer size $B$ in $[5,50],$ and observe the variation in the protocol parameter $\alpha$ at the stability boundary, which is obtained when \eqref{eq:NS_heavy} is met with an equality. The stability chart also illustrates the sufficient condition for local stability given by \eqref{eq:S_heavy}, and the necessary and sufficient condition for non-oscillatory convergence given by \eqref{eq:osccompound}. It can be easily observed that if $\alpha$ is fixed, then increasing the buffer size $B$ would destabilise the system. Hence, from  a design point of view, buffer sizes at core routers should be dimensioned carefully to ensure local stability, as well as non-oscillatory convergence. 

We next present a stability chart which characterises the Hopf condition for Compound TCP with respect to two parameters, the expected file size $\mathbb{E}(V)$ and the protocol parameter $\alpha,$ see Fig. \ref{fig:stability_heavy2}. For this, we vary $\alpha$ in the interval $[0,0.7],$ and find the corresponding value of $B$ which satisfies the Hopf condition. Then, using the equilibrium condition for \eqref{eq:modela2}, and $w^{*}=\mathbb{E}(V)\mu^{*}\tau,$ we find $\mathbb{E}(V)$ at the Hopf boundary. We fix the values of the remaining system parameters as $\beta=0.5,\,k=0.75,\,C'=140\,\text{packets/second}, \,\tau=0.2\,\text{seconds},\text{and} \,\,\chi=1.5.$ Additionally, we choose $\mu^{*}=1$. We can see that keeping $\alpha$ fixed, increasing $\mathbb{E}(V)$ would prompt the system to lose local stability via a Hopf bifurcation and transit into the unstable region. 

At this juncture, a natural question which arises is, how does the file size distribution impact local stability? As argued before, TCP connection sizes are well modelled by the Pareto distribution, which is characterised by its shape parameter or the tail index. Hence, to better understand the dependence of local stability on the file size distribution, we  plot a stability chart with respect to the shape parameter $\chi$ and the feedback delay $\tau,$ as shown in Fig.~\ref{fig:stability_heavy3}. For this, we vary the $\tau$ in $[0.01,1.5],$ and find the corresponding value of $\alpha$ which satisfies the Hopf condition. Then, using the equilibrium condition for \eqref{eq:modela2}, $w^{*}=\mathbb{E}(V)\mu^{*}\tau$ and \eqref{eq:Ex_pareto}, we find $\chi$ at the Hopf boundary. We fix the remaining parameters as $C'=140\,\text{packets/second},\, B=100\,\text{packets},\,\beta=0.5 \,\text{and}\,\,k=0.75.$ Further, we choose $\mu^{*}=1,$ and $m=40$ packets. We can observe that for a fixed round trip time, increasing the shape parameter $\chi$ would have a stabilising effect on the system. Further, for larger values of $\chi,$ the system can accommodate larger feedback delays. However, if the round trip times become too large, the system could still lose local stability via a Hopf bifurcation. We now present some packet-level simulations to corroborate our analytical insights.
\subsection{Packet-level simulations} 
\begin{figure}[t!]
\begin{center}
\vspace{-5mm}
  \psfrag{cccc}{\hspace{1mm}Queue size (pkts)}
  \psfrag{bbbb}{\hspace{1mm}Utilization (\%)}
  \psfrag{275}{\begin{scriptsize}$275$\end{scriptsize}}
  \psfrag{300}{\begin{scriptsize}$300$\end{scriptsize}}
  \psfrag{60}{\begin{scriptsize}$60$\end{scriptsize}}
  \psfrag{100}{\begin{scriptsize}$100$\end{scriptsize}}
  \psfrag{15}{\begin{scriptsize}$15$\end{scriptsize}}
  \psfrag{eeee}{\hspace{8mm}\begin{scriptsize}Buffer size = $15$ packets\end{scriptsize}}
  \psfrag{dddd}{\hspace{8mm}\begin{scriptsize}Buffer size = $100$ packets\end{scriptsize}}
  \psfrag{aaaa}{\hspace{6mm}Time (seconds)}
  
  \includegraphics[width=2.75in,height=3.75in,angle=270]{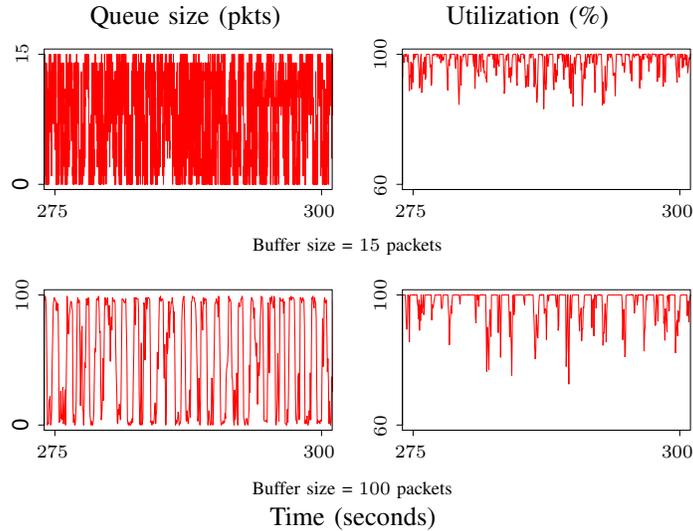}

  \caption {\emph{Compound TCP with heavy-tailed files.} $100$ TCP sources each with an access link speed of $2$ Mbps feeding into a core router of $100$ Mbps. The round trip time is fixed at $200$ ms. Observe the emergence of limit cycles in the queue size dynamics for a buffer size of $100$ packets.}
  \label{fig:buffer}
 \end{center}
 \end{figure}
 \begin{figure}[t!]
\begin{center}
\vspace{-5mm}
  \psfrag{cccc}{\hspace{2.5cm}Queue size (pkts)}
  \psfrag{bbbb}{\hspace{2mm}Utilisation (\%)}
  \psfrag{275}{\begin{scriptsize}$275$\end{scriptsize}}
  \psfrag{300}{\begin{scriptsize}$300$\end{scriptsize}}
  \psfrag{60}{\begin{scriptsize}$60$\end{scriptsize}}
  \psfrag{100}{\begin{scriptsize}$100$\end{scriptsize}}
  \psfrag{15}{\begin{scriptsize}$15$\end{scriptsize}}
  \psfrag{eeee}{\begin{scriptsize}\hspace{-20mm}
  (a) $\chi=1.5,$ $\tau = 10$ ms\end{scriptsize}}
       \psfrag{ffff}{\begin{scriptsize}\hspace{-10mm}
  (b) $\chi=1.5,$ $\tau = 200$ ms\end{scriptsize}}
   \psfrag{dddd}{\begin{scriptsize}\hspace{-20mm}
  (c) $\chi=1.9,$ $\tau = 200$ ms\end{scriptsize}}
  \psfrag{gggg}{\begin{scriptsize}\hspace{-10mm}
  (d) $\chi=1.9,$ $\tau = 500$ ms\end{scriptsize}}
  \psfrag{aaaa}{\hspace{8mm}Time (seconds)}
  
  \includegraphics[width=2.75in,height=3.75in,angle=270]{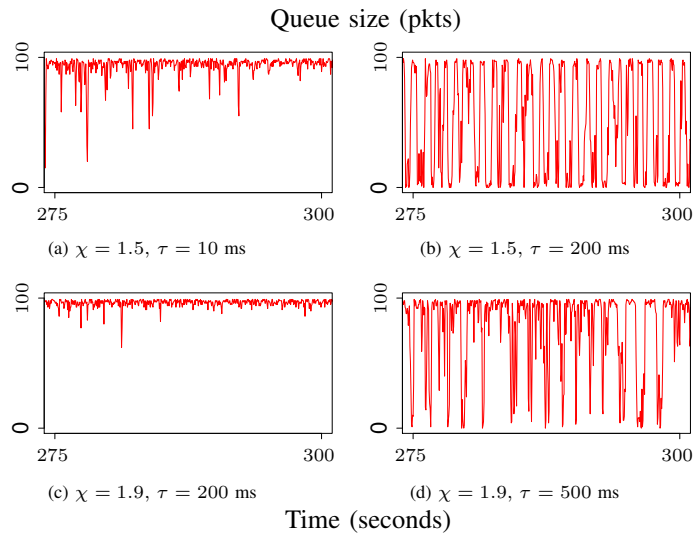}

  \caption {\emph{Compound TCP with heavy-tailed files.} $100$ TCP sources each with an access link speed of $2$ Mbps feeding into a core router of $100$ Mbps. The buffer size is fixed at $100$ packets. Observe the impact of the shape parameter $\chi$ and the round trip time $\tau$ on stability. As $\chi$ increases, the system stabilises. However, as the round trip time gets larger, the system again destabilises.}
  \label{fig:alpha_tau}
 \end{center}
 \end{figure}
\subsubsection*{Dynamical properties}
For our packet-level simulations, we consider the following setup. Each user generates finite volume files with sizes drawn from a Pareto distribution. The files arrive at the access link as a Poisson process with rate $\lambda.$ All users are connected to the bottleneck router via the access links. If there are $N$ users in the system, and the bottleneck link has a capacity $C,$ then the offered load at the bottleneck link is $\rho=\mathbb{E}(V)N\lambda/C.$ For our simulations, we vary the offered load to the system by varying $\lambda.$  

For our simulations, we fix the number of TCP senders at $100,$ and the bottleneck capacity at $100$ Mbps. Each access link has a speed of $2$ Mbps. Further, the packet size is $1500$ bytes.

\emph{Impact of the buffer size:} To capture the impact of buffer sizes on stability, we observe the queue size dynamics for two values of the buffer size, $15$ and $100$ packets, see Fig. \ref{fig:buffer}. The expected file size is chosen to be $100$ kB, and the shape parameter is fixed at $1.5$~\cite{Vishwanath}. Further, the average round trip time is fixed at $200$ ms. With this set of parameter values, the offered load to the bottleneck link is $0.9.$ It can be observed that, when the buffer size at the core router is fixed at $15$ packets, the queue size does not exhibit any periodic oscillations, and hence is stable. However, as the buffer size is increased to $100,$ the system destabilises, and we can observe the emergence of periodic oscillations in the queue size dynamics. This lends credence to the fact that these oscillations are not an artefact of the underlying assumptions for the traffic. Even if we take into account the high variability at the TCP connection level, the queue size dynamics could still exhibit periodic oscillations, if buffers are not dimensioned carefully.

\emph{Impact of the shape parameter:} We now present a set of simulations which captures the impact of the file size distribution, \emph{i.e.,} the shape parameter $\chi$ on stability. For this, we fix the buffer size at the core router at $100$ packets. Fig. \ref{fig:alpha_tau}(a) shows the queue size dynamics when the shape parameter is $1.5,$ the expected file size is $100$ kB, and the round trip time is $10$ ms. For such a short round trip time, the feedback would be fast, and the queue does not have time to drain completely. Hence, in this case, the queue size is stable.  As the round trip time is increased to $200$ ms, we can observe the emergence of oscillations in the queue size dynamics, see Fig.~\ref{fig:alpha_tau}(b). This is because, the long flows present in the network would experience synchronisation effects due to increase in the feedback delay. This would result in TCP senders backing off simulataneously and draining the queue repeatedly before it can become full again. 

We now increase the shape parameter to $1.9,$ keeping the round trip time and the offered load fixed. We observe that increasing the shape parameter to $1.9$ stabilises the system, see Fig.~\ref{fig:alpha_tau}(c). A plausible explanation for this change in the qualitative behaviour of the system is as follows. As the shape parameter is increased from $1.5$ to $1.9,$ the tail of the Pareto distribution becomes lighter. This implies that the traffic now constitutes more ``mice" in the latter case. Hence, even if the long flows are synchronised and decreasing their sending rates, ``mice" can arrive and depart the system without experiencing any drop. This leads to the effective utilisation of the available bandwidth and prevents the bottleneck queue from draining completely. This in turn ensures  that there are no periodic oscillations in the queue size, and it is stable. However, with the shape parameter $\chi=1.9,$ if the delay gets too large, for example $500$ ms, the system de-stabilises. This is because, in the congestion avoidance phase, the long flows would experience large delays before they can increase their sending rates after each packet drop. Hence, the feedback effects on the long flows would be more pronounced for larger round trip times. Additionally, a large feedback delay would affect the sending rates of the short flows. This would lead to the emergence of oscillations in the queue size, as shown in \ref{fig:alpha_tau}(d). Note that the interdependence of the shape parameter and the round trip time to ensure stability of the system was indeed predicted by the necessary and sufficient condition, given by \eqref{eq:NS_heavy_shape}. 

\emph{Impact of the expected file size:} Figs.~\ref{fig:alpha_tau}(b) and \ref{fig:alpha_tau}(c) also encapsulate the impact of the expected file size $\mathbb{E}(V)$ on stability. As we increase $\chi$ from $1.5$ to $1.9,$ keeping the round trip time fixed at $200$ ms, $\mathbb{E}(V)$ decreases from $100$ kB to $70$ kB, and the system stabilises. This is consistent with the insight obtained from our stability analysis, see Fig.~\ref{fig:stability_heavy2}. 
\subsubsection*{Statistical properties}
\begin{figure}[t!]
\begin{minipage}[b]{0.45\linewidth}
\hspace{15mm}
\psfrag{T}{\hspace{6mm}CCDF}
\psfrag{0}{\begin{scriptsize}$0$\end{scriptsize}}
\psfrag{1}{\begin{scriptsize}$1$\end{scriptsize}}
\psfrag{15}{\begin{scriptsize}$15$\end{scriptsize}}
\psfrag{h}{\begin{tiny}Empirical\end{tiny}}
\psfrag{x}{\begin{tiny}$M/M/1/B$\end{tiny}}
\psfrag{y}{\begin{tiny}$M/D/1/B$\end{tiny}}
\psfrag{c}{}
\psfrag{b}{\hspace{-15mm}\begin{scriptsize}Buffer size = $15$ packets\end{scriptsize}}
\includegraphics[width=1.7in,height=2.58in,angle=270]{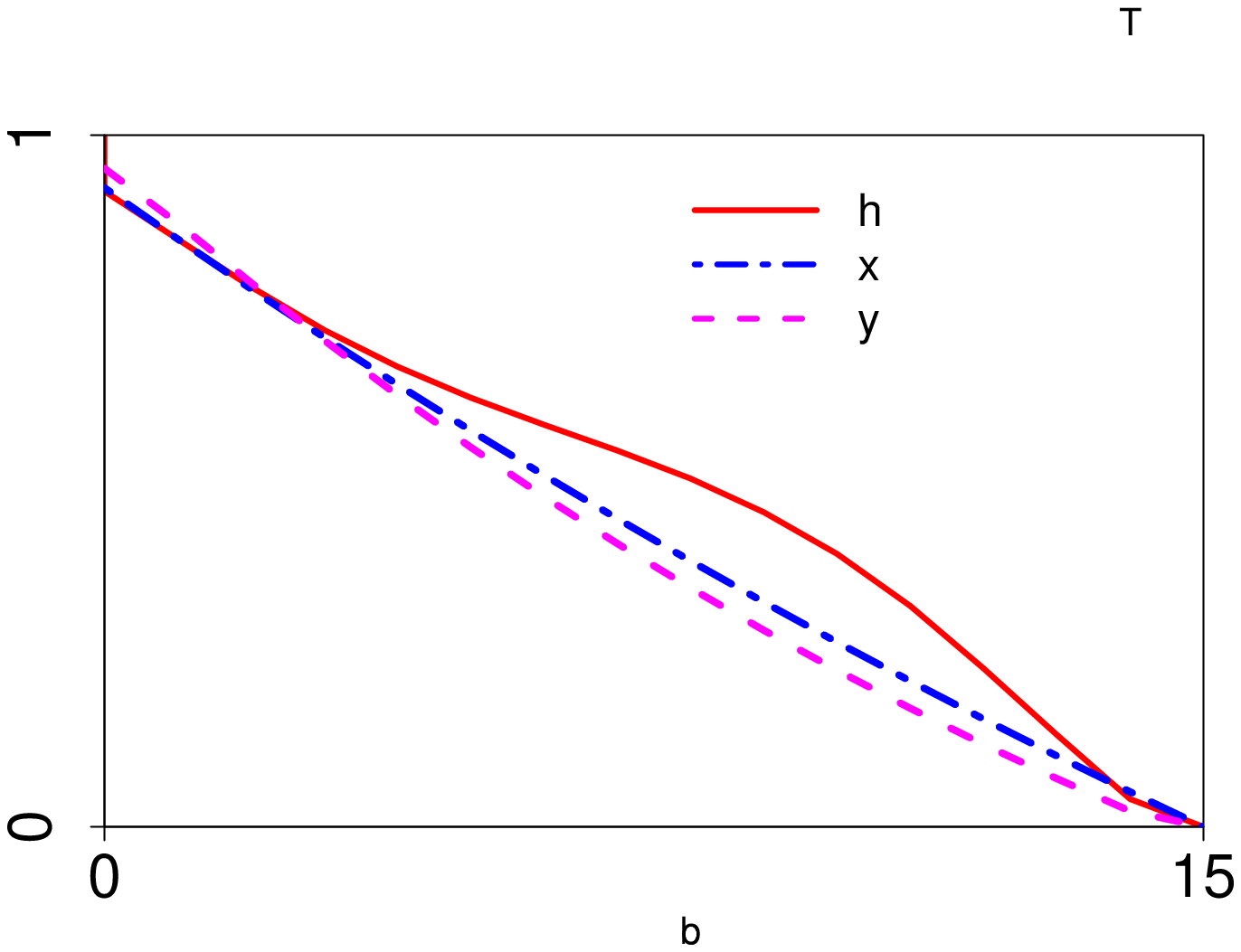}
\end{minipage}
\begin{minipage}[b]{0.45\linewidth}
\centering
\psfrag{0}{\begin{scriptsize}$0$\end{scriptsize}}
\psfrag{1}{\begin{scriptsize}$1$\end{scriptsize}}
\psfrag{10}{\begin{scriptsize}$10$\end{scriptsize}}
\psfrag{h}{\begin{tiny}Empirical\end{tiny}}
\psfrag{x}{\begin{tiny}$M/M/1/B$\end{tiny}}
\psfrag{y}{\begin{tiny}$M/D/1/B$\end{tiny}}

\psfrag{b}{\hspace{-3cm}\begin{scriptsize}Buffer size = $10$ packets\end{scriptsize}}
\psfrag{c}{\hspace{-3.3cm}\normalsize Queue threshold}
\includegraphics[width=1.7in,height=2.58in,angle=270]{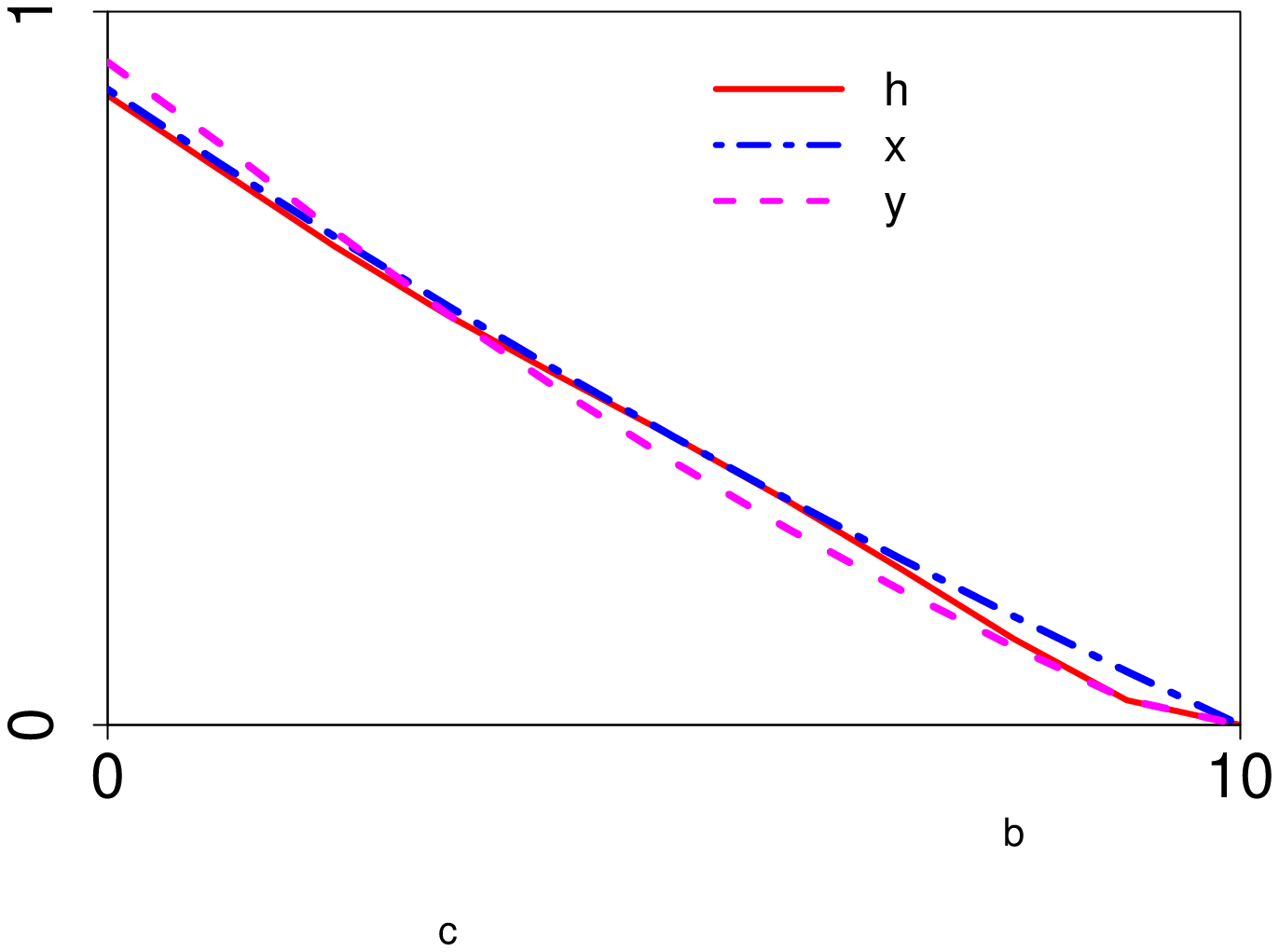}
\end{minipage}
\caption {\emph{Compound TCP with heavy-tailed files.} $100$ TCP sources each with an access link speed of $2$ Mbps feeding into a core router of $100$ Mbps. The round trip time is fixed at $200$ ms. We plot the empirical queue distribution of the bottleneck queue for two buffer sizes $B = 15$ and $B=10$ packets, for which the underlying dynamical system is stable, and compare it with that of $M/M/1/B$ and $M/D/1/B$ queues. We can observe that  $M/M/1/B$ and $M/D/1/B$ approximations for the bottleneck queue seem reasonably justified.}
  \label{fig:stat_heavy}
\end{figure}

We now perform an empiricial study of the statistical properties of the bottleneck queue under this traffic scenario, as shown in Fig.~\ref{fig:stat_heavy}. For this, we again consider $100$ TCP senders feeding into the bottleneck router via access links with a speed of $2$ Mbps. The expected file size is fixed at $100$ kB, and the shape parameter is chosen to be $1.5.$ 

We illustrate  the queue distributions for two values of buffer sizes at the core router in the small buffer regime, $15$ and $10$ packets, for which the underlying dynamical system is stable. We can observe that for a buffer size of $15$ packets, the queue distribution of the core router can be reasonably approximated by that of $M/M/1/B$ or an $M/D/1/B$ queue. An interesting observation is that this approximation holds remarkably well at a smaller buffer size of $10$ packets. This strongly suggests that even with high variability at the source level, an $M/M/1/B$ approximation for the bottleneck queue is still valid in the asymptotic regime wherein, a large number of senders are present in the network, the bandwidth-delay product is high, and the buffer at the core router is dimensioned small enough to mitigate synchronisation effects. Further, from a theoretical perspective, this approximation seems benign, since packet-level simulations match our theoretical predictions well.   

In the next section, we consider a multiple bottleneck topology which depicts a more realistic network scenario as opposed to the simple single bottleneck topology.

\section{Multiple bottlenecks}
\label{model_b}

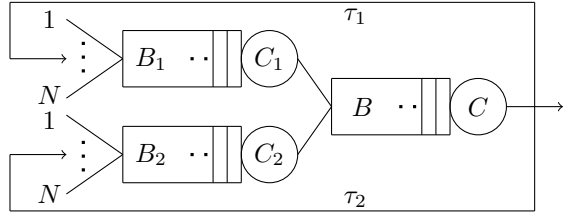
\begin{figure}
\vspace{2mm}
\begin{center}
{
\begin{tikzpicture}[scale=0.75]
\draw (-1,0) -- (-2,0.7);
\draw (-2,-0.7) -- (-1,0);
\draw (-1,-0.5) -- (1.1,-0.5);
\draw (1.1,-0.5) -- (1.1,0.5);
\draw (1.1,0.5) -- (-1,0.5);
\draw (-1,0.5) -- (-1,-0.5);
\draw (0.85,-0.5) -- (0.85,0.5);
\draw (0.60,-0.5) -- (0.60,0.5);

\draw (1.6,0) circle(0.5 cm);
\draw [loosely dotted,very thick] (0.50,0) -- (0.10,0);
\draw [loosely dotted,very thick] (-1.7,0.3) -- (-1.7,-0.3);

\node at (-0.5,0) {$B_{1}$};
\node at (1.6,0) {$C_{1}$};
\node at (-2.3,0.7) {$1$};
\node at (-2.3,-0.65) {$N$};

\node at (-2.3,-1.1) {$1$};
\node at (-2.3,-2.4) {$N$};

\draw (-1,-1.7) -- (-2,-1);
\draw (-2,-2.4) -- (-1,-1.7);
\draw (-1,-2.2) -- (1.1,-2.2);
\draw (1.1,-2.2) -- (1.1,-1.2);
\draw (1.1,-1.2) -- (-1,-1.2);
\draw (-1,-1.2) -- (-1,-2.2);
\draw (0.85,-2.2) -- (0.85,-1.2);
\draw (0.60,-2.2) -- (0.60,-1.2);

\draw (1.6,-1.7) circle(0.5 cm);
\draw [loosely dotted,very thick] (0.50,-1.7) -- (0.10,-1.7);
\draw [loosely dotted,very thick] (-1.7,-2) -- (-1.7,-1.4);

\node at (-0.5,-1.7) {$B_{2}$};
\node at (1.6,-1.7) {$C_{2}$};

\draw (2.1,0) -- (2.7,-0.85);
\draw (2.1,-1.7) -- (2.7,-0.85);
\draw (2.7,-1.35) -- (4.8,-1.35);
\draw (4.8,-1.35) -- (4.8,-0.35);
\draw (4.8,-0.35) -- (2.7,-0.35);
\draw (2.7,-0.35) -- (2.7,-1.35);
\draw (4.55,-0.35) -- (4.55,-1.35);
\draw (4.30,-0.35) -- (4.30,-1.35);

\draw (5.3,-0.85) circle(0.5 cm);
\draw [loosely dotted,very thick] (4.2,-0.85) -- (3.8,-0.85);

\node at (3.25,-0.85) {$B$};
\node at (5.3,-0.85) {$C$};

\draw[->] (5.8,-0.85) -- (6.8,-0.85);
\draw (6.3,-0.85) -- (6.3,1);
\draw (6.3,1) --(-3,1);
\draw (-3,1) --(-3,0);
\draw[->] (-3,0) --(-2,0);
\node at (3.125,0.75) {$\tau_{1}$};

\draw (6.3,-0.85) -- (6.3,-2.7);
\draw (6.3,-2.7) --(-3,-2.7);
\draw (-3,-2.7) --(-3,-1.7);
\draw[->] (-3,-1.7) --(-2,-1.7);
\node at (3.125,-2.5) {$\tau_{2}$};

\end{tikzpicture}
}
\caption{Multiple bottleneck topology with two distinct sets of TCP flows, regulated by two edge routers, having round trip times $\tau_{1}$  and $\tau_{2}$ and feeding into a core router. The edge routers have link capacities $C_1$ and $C_2,$ and buffer sizes $B_1$ and $B_2$ packets. The link capacity of the core router is $C$ and the buffer size at the core router is $B$ packets.}
\label{fig:multi-bottleneck}
\end{center}
\end{figure}

The model consists of two distinct sets of \emph{many} TCP flows having different round trip times $\tau_{1}$ and $\tau_{2}$ and regulated by two edge routers, as shown in Fig.~\ref{fig:multi-bottleneck}. For this model, our focus will be on long-lived flows. The average window sizes of the two sets of flows are $w_{1}(t)$ and $w_{2}(t)$ respectively. The outgoing flows from both edge routers feed into a common core router. The buffer sizes of the edge routers are $B_{1}$ and $B_{2}$ respectively, and buffer size of the core router is $B$. The link capacities of the edge routers are $C_{1}$ and $C_{2}$ respectively. Let $C'_1$ and $C'_2$ denote the service rates per flow for the edge routers. We consider the case where both edge routers and the core router have small buffer sizes and employ a Drop-Tail queue policy. The link capacity of the core router is $C$. The service rate per flow for the core router is denoted by $C'$. Suppose $p_{1}(t)$ and $p_{2}(t)$ are the packet loss probabilities at the two edge routers for the packets sent at time instant $t$, for the two distinct sets of flows respectively. The packet loss probability at the core router is denoted as $q(t,\tau_{1},\tau_{2})$. For a generalised TCP flavour, the non-linear, time-delayed, fluid model of the system is given by the following equations:
\begin{align}
\dot{w}_j(t) =& \frac{w_{j}(t-\tau_{j})}{\tau_{j}}\bigg(i\left(w_{j}(t)\right)\Big(1-p_{j}(t-\tau_{j})-q(t,\tau_{1},\tau_{2})\Big)\notag\\
&- d\left(w_{j}(t)\right)\Big(p_{j}(t-\tau_{j})+q(t,\tau_{1},\tau_{2})\Big)\bigg), j=1,2.
\label{eq:modelb}
\end{align}
The loss probabilities at the three routers are approximated as
$$
p_{1}(t)=\left(\frac{w_{1}(t)}{C'_{1}\tau_{1}}\right)^{B_{1}} \hspace{1ex},\hspace{1ex} p_{2}(t)=\left(\frac{w_{2}(t)}{C'_{2}\tau_{2}}\right)^{B_{2}},\ {\rm and}
$$$$q(t,\tau_{1},\tau_{2})= \left(\frac{w_{1}(t-\tau_{1})/\tau_{1}+w_{2}(t-\tau_{2})/\tau_{2}}{\widetilde{C}}\right)^{B}.$$
Here, $\widetilde{C}=2C'.$ In this section, we prove that even in the multiple bottleneck topology, the system loses stability through a \emph{Hopf bifurcation} \cite{Guckenheimer} as system parameters vary. This loss of local stability leads to the emergence of limit cycles in the queue size of the core router.

\subsection{Necessary and sufficient condition for stability}
For system \eqref{eq:modelb}, we will perform a local stability analysis to derive a necessary and sufficient condition for stability. Suppose the equilibrium of the system is $(w_1^\ast,w_2^\ast)$. Let $u_{1}(t) = w_{1}(t)-w_{1}^{*}$ and $u_{2}(t) = w_{2}(t)-w_{2}^{*}$ be small perturbations about $w_{1}^{*}$ and $w_{2}^{*}$ respectively. Linearising system \eqref{eq:modelb} about its equilibrium $(w_{1}^{*},w_{2}^{*})$, we get
\begin{align}
\label{eq:linearb}
&\dot{u}_1(t) = -\mathcal{M}_{1}u_{1}(t)-\mathcal{N}_{1}u_{1}(t-\tau_{1})-\mathcal{P}_{1}u_{2}(t-\tau_{2}),\notag\\
&\dot{u}_2(t) = -\mathcal{M}_{2}u_{2}(t)-\mathcal{N}_{2}u_{2}(t-\tau_{2})-\mathcal{P}_{2}u_{1}(t-\tau_{1}),
\end{align}
where,
for Compound TCP, the increase and decrease functions \eqref{eq:Compound} yield the following coefficients 
\begin{align}
&\mathcal{M}_{j} =-\frac{\alpha}{\tau_{j}}\left(k-2\right) \left(w_{j}^{*}\right)^{k-1}\Bigg(1-\bigg(\frac{w_j^*}{C'_j\tau_j}\bigg)^{B_j}-\frac{1}{(2C')^B}\bigg(\frac{w_1^*}{\tau_1}+\frac{w_2^*}{\tau_2}\bigg)^{B}\Bigg),\notag\\
&\mathcal{N}_{j} =\left(\alpha\left(w_{j}^{*}\right)^{k-1}+\beta w_{j}^{*}\right)\Bigg(\frac{B_{j}}{\tau_j}\left(\frac{w_{j}^{*}}{C'_j \tau_j}\right)^{B_j}+\frac{B \left(w_{j}^{*}\right)^{2}}{(2C')^{B}\tau_{j}^{2}}\left(\frac{w_{1}^{*}}{\tau_{1}}+\frac{w_{2}^{*}}{\tau_{2}}\right)^{B-1}\Bigg),\notag\\
&\mathcal{P}_{j}=\left(\alpha\left(w_{j}^{*}\right)^{k-1}+\beta w_{j}^{*}\right)\frac{B w_{j}^{*}}{\tau_{1}\tau_{2}(2C')^{B}}\left(\frac{w_{1}^{*}}{\tau_{1}}+\frac{w_{2}^{*}}{\tau_{2}}\right)^{B-1}, \hspace{0.5cm}j=1,2. \label{eq:coefficients}
\end{align}
At equilibrium, the following equations are satisfied
\begin{align*}
\alpha \left(w_{j}^{*}\right)&^{k-1}\left(1-\left(\frac{w_{j}^{*}}{C'_{j}\tau_{j}}\right)^{B_{j}}-\frac{1}{(2C')^{B}}\left(\frac{w_{1}^{*}}{\tau_{1}}+\frac{w_{2}^{*}}{\tau_{2}}\right)^{B}\right)\\
&=\beta w_{j}^{*}\left(\left(\frac{w_{j}^{*}}{C'_{j}\tau_{j}}\right)^{B_{j}}+\frac{1}{(2C')^{B}}\left(\frac{w_{1}^{*}}{\tau_{1}}+\frac{w_{2}^{*}}{\tau_{2}}\right)^{B}\right),\hspace{1ex} j=1,2.
\end{align*}
For analytical tractability, we consider two different scenarios with simple assumptions. 
\subsection*{Case I}
In this scenario, we assume that the network parameters for all routers are the same \emph{i.e.} $B_{1}=B_{2}=B, C'_{1}=C'_{2}=C'.$ We further assume that the round trip times of both sets of TCP flows are identical \emph{i.e.} $\tau_{1}=\tau_{2}=\tau$. Then, $w_{1}^{*}=w_{2}^{*}=w^{*}$ will be an equilibrium of the system, and satisfies the following equation:
\begin{align*}
\alpha\left(w^{*}\right)^{k-2}=2\left(\alpha\left(w^{*}\right)^{k-2}+\beta\right)\left(\frac{w^{*}}{C'\tau }\right)^{B}.
\end{align*}

\noindent Let $\mathcal{M}=\frac{\left(\alpha\left(w^{*}\right)^{k-2}+\beta\right)B\left(w^{*}\right)^{B+1}}{\tau^{B+1}C'^{B}}$, then the coefficients $\mathcal{M}_{1}$, $\mathcal{M}_{2}$, $\mathcal{N}_{1}$, $\mathcal{N}_{2}$, $\mathcal{P}_{1}$, $\mathcal{P}_{2}$ reduce to
\begin{align}
\label{eq:reduced}
\mathcal{M}_{1}&=\mathcal{M}_{2}=\frac{2\mathcal{M}\beta w^{*}}{\left(\alpha\left(w_{j}^{*}\right)^{k-1}+\beta w_{j}^{*}\right)B}\left(2-k\right)=a,\notag\\
\mathcal{N}_{1}&=\mathcal{N}_{2}=\frac{3}{2}\mathcal{M}=b,\notag\\
\mathcal{P}_{1}&=\mathcal{P}_{2}=\frac{1}{2}\mathcal{M}=c. 
\end{align}
Note that $a$, $b$, $c$ $>$ 0. With these assumptions the linearised system \eqref{eq:linearb} becomes
\begin{align}
\label{eq:linearb_wkappa}
&\dot{u}_{1}(t) =  -a u_{1}(t)-b u_{1}(t-\tau)-c u_{2}(t-\tau),\notag\\
&\dot{u}_{2}(t) = -a u_{2}(t)-b u_{2}(t-\tau)-c u_{1}(t-\tau).
\end{align}

We now show that system \eqref{eq:linearb_wkappa} is stable if and only if the parameters $a$, $b$, $c$ and $\tau$ satisfy the condition $\tau<\frac{1}{\omega_1}\cos^{-1}\left(\frac{-a}{b+c}\right)$ with crossover frequency $\omega_1 = \sqrt{\left(b+c\right)^2-a^2}$.

Looking for exponential solutions, we get the characteristic equation for the linearised system \eqref{eq:linearb_wkappa} as
\begin{align}
\label{eq:characb}
\left(s +a+b e^{-s \tau}\right)^2 -c^2 e^{-2s \tau}=0,
\end{align}
which can be written as 
\begin{align*}
g_{1}\left(s\right)g_{2}\left(s\right)=0,
\end{align*}
where,
\begin{align}
g_{1}\left(s\right)&= s +a+\left(b+c\right)e^{-s \tau}, \hspace{1ex} \text{and} \notag\\
g_{2}\left(s\right)&= s +a+\left(b-c\right)e^{-s \tau}.
\end{align}
For stability, all roots of \eqref{eq:characb} should have negative real parts. For negligible values of delay $\tau$, system \eqref{eq:linearb_wkappa} is stable, \emph{i.e.} all roots of the characteristic equation lie of the left half of the complex plane. As the delay is increased, the system becomes unstable if one pair of complex conjugate roots of either $g_1(s)$ or $g_2(s)$ or both crosses over the imaginary axis. We aim to determine the values of delay $\tau$ at which one pair of complex conjugate roots of $g_1(s)$ and $g_2(s)$ cross over the imaginary axis. Let $\tau_{1,c}$ and $\tau_{2,c}$ denote the values of $\tau$ at which $g_1(s)$ and $g_2(s)$ have exactly one pair of purely imaginary roots. Then, the critical value of $\tau$, denoted by $\tau_c$, at which \eqref{eq:characb} has one pair of purely imaginary roots is $\tau_c=\min(\tau_{1,c},\tau_{2,c})$~\cite{Campbell}. Substituting $s=j\omega_1$ in $g_1(s)$ and separating real and imaginary parts we get
\begin{align}
\left(b+c\right)\sin \omega_1\tau &= \omega_1, \hspace{1ex} \text{and} \label{eq:equation_1}\\
\left(b+c\right)\cos \omega_1 \tau &= -a.\label{eq:equation_2}
\end{align}
Solving \eqref{eq:equation_1} and \eqref{eq:equation_2} for $\omega_1$ we get 
\begin{align}
\omega_1 = \sqrt{\left(b+c\right)^2-a^2},
\end{align}
and under the condition $b+c>a$, $\omega_1^2$ is strictly positive. This implies that there exists a cross over frequency $\omega_1$ at which one pair of complex conjugate roots of $g_1(s)$ crosses over to the right half of the complex plane. Solving \eqref{eq:equation_1} and \eqref{eq:equation_2} for $\tau$, we get the critical value of delay at which the system transits from stability to instability as 
\begin{align}
\label{eq:delay_critical}
\tau_{1,c}=\frac{1}{\omega_1}\cos^{-1}\left(\frac{-a}{b+c}\right).
\end{align}
Similarly, substituting $s=j\omega_2$ in $g_2(s)$ we get the cross over frequency as 
\begin{align}
\label{eq:omega_2}
\omega_2=\sqrt{\left(b-c\right)^2-a^2}.
\end{align}
and under the condition $b-c>a$, $\omega_2^2$ is strictly positive. Hence, there exists a cross over frequency $\omega_2$ at which
one pair of complex conjugate roots of $g_2(s)$ crosses over to the right half of the complex plane. Solving for $\tau$, we obtain the critical value of the delay at which the system having the characteristic equation $g_2(s)$ has exactly one pair of purely imaginary roots as
 \begin{align}
\label{eq:delay_critical_2}
\tau_{2,c}=\frac{1}{\omega_2}\cos^{-1}\left(\frac{-a}{b-c}\right).
\end{align} 
Observe that $\omega_1>\omega_2.$ Since $\cos^{-1}(x)$ is monotonically decreasing for $x\in [-1,1]$, it can be shown that $\tau_1<\tau_2.$ This implies that $\tau_c=\tau_{1,c}$. Consequently, all roots of the characteristic equation \eqref{eq:characb} lie on the left half of the complex plane for all $\tau<\tau_{c}$. Hence, system \eqref{eq:linearb_wkappa} is asymptotically stable for $\tau<\tau_c$, and unstable for $\tau>\tau_c$. Further, we will analytically show that this loss of stability occurs via a Hopf bifurcation when one pair of complex conjugate roots of \eqref{eq:characb} crosses over the imaginary axis with non-zero velocity at $\tau=\tau_c$. Therefore, the necessary and sufficient condition for local stability of \eqref{eq:linearb_wkappa} is
\begin{align}
\label{eq:ns}
\tau<\frac{1}{\omega_1}\cos^{-1}\left(\frac{-a}{b+c}\right).
\end{align}
Substituting values of $\omega_1$, $a$, $b$ and $c$ in \eqref{eq:ns}, we get the necessary and sufficient condition for local stability of \eqref{eq:modelb}, with Compound TCP as
\begin{align}
\label{eq:condition}
\alpha \left(w^{*}\right)^{k-1}\sqrt{B^{2}
-(k-2)^2\left(1-2p(w^{*})\right)^{2}}<\cos^{-1}
\left(\frac{(k-2)\left(1-2p(w^*)\right)}{B}\right),
\end{align}
\noindent where $p(w^{*})=\left(\frac{w^{*}}{C'\tau}\right)^B$. This condition captures the relationship between the equilibrium window size, protocol parameters $k$ and $\alpha$, and buffer size $B$ of the core router to ensure stability of the system. 

We now derive a sufficient condition for local stability of system \eqref{eq:modelb}. To that end, we show that system \eqref{eq:linearb_wkappa} is stable if the parameters $a,$ $b,$ $c$ and the feedback delay $\tau$ satisfy $\tau<\frac{\pi}{2(b+c)}.$

A sufficient condition for stability for a system with the characteristic equation $g_1(s)=0$ is $\left(b+c\right)\tau<\frac{\pi}{2}$. Similarly, a sufficient condition for stability for a system with the characteristic equation $g_2(s)=0$ is 
$\left(b-c\right)\tau<\frac{\pi}{2}$. Hence, a sufficient condition for system \eqref{eq:linearb_wkappa} to be asymptotically stable is
 \begin{align}
 \label{eq:s}
 \left(b+c\right)\tau<\frac{\pi}{2}.
 \end{align}
Substituting $b$ and $c$ in \eqref{eq:s}, a sufficient condition for stability of system \eqref{eq:modelb} with Compound TCP flows is 
\begin{align}
\label{eq:s_wstar}
\alpha B \left(w^{\ast}\right)^{k-1}<\frac{\pi}{2}.
\end{align}

We now show that a simple sufficient condition for local stability of system \eqref{eq:modelb} is 
\begin{align}
\label{eq:s_wwstar}
\alpha B<\frac{\pi}{2}.
\end{align}
We note that at equilibrium, trivially $w^{\ast}\geq 1.$ Since $k=0.75,$ it is easy to see that $h(w^{\ast})=\left(w^{\ast}\right)^{k-1}$ is a decreasing function of $w^{\ast}.$ This implies that $\alpha B \left(w^{\ast}\right)^{k-1}\leq\alpha B,$ $\forall\,w^{\ast}.$ Hence, if we ensure $\alpha B<\pi/2,$ then local stability of \eqref{eq:modelb} would be ensured. This yields a rather simple sufficient condition \eqref{eq:s_wwstar} for local stability of \eqref{eq:modelb}.

\emph{Discussion:} Observe that condition \eqref{eq:s_wwstar} is independent of the equilibrium window size, and hence provides a \emph{decentralised} design guideline for a network designer to dimension router buffers. Interestingly, this condition ensures that a network designer need not have the exact knowledge of the network parameters, such as the feedback delay and the capacity of the network, to dimension router buffers. 
\subsection*{Case II}
In this scenario, we assume that the network parameters for all routers are distinct, and the average round trip time of the first set of TCP flows is much larger than the other. Further, we consider that the average round trip time of the second set of TCP flows is negligible. This implies that $\tau_1>>\tau_2$ and $\tau_2\approx 0$. As a consequence of this assumption, the dynamics of the second set of TCP flows will appear to be almost instantaneous. This leads to the following non-linear, time-delayed fluid model of the system:
\begin{align}
\dot{w}_1(t) =& \frac{w_{1}(t-\tau_{1})}{\tau_{1}}\bigg(i\left(w_{1}(t)\right)\Big(1-p_{1}(t-\tau_{1})-q(t,\tau_{1},\tau_{2})\Big)\notag\\
&- d\left((w_{1}(t)\right)\Big(p_{1}(t-\tau_{1})+q(t,\tau_{1},\tau_{2})\Big)\bigg),\notag \\
\dot{w}_2(t) =& \frac{w_{2}(t)}{\tau_{2}}\bigg(i\left(w_{2}(t)\right)\Big(1-p_{2}(t)-q(t,\tau_{1},\tau_{2})\Big)\notag\\
&- d\left((w_{2}(t)\right)\Big(p_{2}(t)+q(t,\tau_{1},\tau_{2})\Big)\bigg).
 \label{eq:modelb_1}
\end{align}
The loss probabilities at the three routers are approximated as
$$
p_{1}(t)=\left(\frac{w_{1}(t)}{\widetilde{C}_{1}\tau_{1}}\right)^{B_{1}} \hspace{1ex},\hspace{1ex} p_{2}(t)=\left(\frac{w_{2}(t)}{\widetilde{C}_{2}\tau_{2}}\right)^{B_{2}},\ {\rm and}
$$$$q(t,\tau_{1},\tau_{2})= \left(\frac{w_{1}(t-\tau_{1})/\tau_{1}+w_{2}(t)/\tau_{2}}{C'}\right)^{B}.$$ 
Here, $C'=2\widetilde{C}.$ We now outline local stability conditions for system \eqref{eq:modelb_1}. This will enable us to characterise the stability of the system in the presence of heterogeneity in network parameters. Suppose $(w_{1}^{*},w_{2}^{*})$ be a non-trivial equilibrium of system \eqref{eq:modelb_1}. Let $u_{1}(t)=w_{1}(t)-w_{1}^{*}$ and $u_{2}(t)=w_{2}(t)-w_{2}^{*}$ represent small perturbations about $w_1^{*}$ and $w_{2}^{*}$ respectively. Linearising system \eqref{eq:modelb_1} about its equilibrium, we get the following:
\begin{align}
\label{eq:linearb1}
&\dot{u}_1(t) = -\mathcal{M}_{1}u_{1}(t)-\mathcal{N}_{1}u_{1}(t-\tau_{1})-\mathcal{P}_{1}u_{2}(t),\notag\\
&\dot{u}_2(t) = -\big(\mathcal{M}_{2}+\mathcal{N}_{2}\big)u_{2}(t)-\mathcal{P}_{2}u_{1}(t-\tau_{1}),
\end{align}
where, the coefficients are given by \eqref{eq:coefficients}. Looking for exponential solutions, we get the characteristic equation for the linearised system \eqref{eq:linearb1} as
\begin{align}
\label{eq:charac_assum2}
s^2 + as +bs e^{-s \tau_1} + c e^{-s \tau_1} + d=0,
\end{align}
where, 
\begin{align}
\label{eq: abcd}
&a = \mathcal{M}_{1}+\mathcal{M}_{2}+\mathcal{N}_{2} \hspace{1ex}, \hspace{7.5ex} b=\mathcal{N}_{1},\notag\\
&c=\mathcal{N}_{1}\left(\mathcal{M}_{2}+\mathcal{N}_{2}\right)-
\mathcal{P}_{1}\mathcal{P}_{2}, \hspace{2ex}d=\mathcal{M}_{1}\left(\mathcal{M}_{2}+\mathcal{N}_{2}\right).
\end{align}
\begin{result}
The stability boundary of system \eqref{eq:linearb1}, having the characteristic equation \eqref{eq:charac_assum2} is characterised by \text{\cite{TCNS}}
$\tau_1=\frac{1}{\omega}\cos^{-1}\Big(\frac{\omega^2(d-ab)-cd}{b^2\omega^2+d}\Big)$ with the cross over frequency as
\begin{align*}
\omega = \sqrt{\frac{(2c-a^2+b^2)}{2}+ \frac{\sqrt{(2c-a^2+b^2)^2-4(c^2-d^2)}}{2}}.
\end{align*} 
\end{result}
We will show later that if this boundary condition just gets violated, the underlying dynamical system loses local stability via a Hopf type bifurcation. Hence, system \eqref{eq:modelb_1} is locally stable if and only if $\tau_1<\frac{1}{\omega}\cos^{-1}\Big(\frac{\omega^2(d-ab)-cd}{b^2\omega^2+d}\Big).$ Substituting $a,$ $b,$ $c$ and $d$ in the above would yield a condition which captures the interdependence among different network parameters and Compound TCP parameters to ensure stability of the system.    
\subsection{Hopf Condition}
We have seen that protocol parameters, buffer thresholds and feedback delay all play an important role to ensure local stability. If the local stability conditions get violated, the system could transit from a locally stable to an unstable regime Varying any of the system parameters beyond the critical value can also drive the system to instability. Thus, instead of treating delay or any of the system parameters as the bifurcation parameter, we introduce an exogenous non-dimensional parameter $\kappa$ which can act as the bifurcation parameter. If $\kappa$ is varied keeping the values of the system parameters constant at their critical values, the system loses stability at $\kappa_c=1$. To show that this loss of stability occurs via a Hopf bifurcation, we proceed to verify the transversality condition of the Hopf spectrum \cite[Chapter 11, Theorem 1.1]{Hale} for both scenarios. To verify the transversality condition, we need to show that $\mathrm{Re}(\mathrm{d}s/\mathrm{d}\kappa)\neq 0$ at $\kappa=\kappa_c$.

\subsection*{Case I} 
\indent In this scenario, the linearised system, with the non-dimensional parameter $\kappa$, becomes
\begin{align}
\label{eq:linearb_kappa}
&\dot{u}_1(t) = \kappa \Big(-a u_{1}(t)-b u_{1}(t-\tau)-c u_{2}(t-\tau)\Big),\notag\\
&\dot{u}_2(t) = \kappa \Big(-a u_{2}(t)-b u_{2}(t-\tau)-c u_{1}(t-\tau)\Big).
\end{align}
Looking for exponential solutions of \eqref{eq:linearb_kappa} we get 
\begin{align}
\label{eq:characb_kappa}
\left(s +\kappa a +\kappa \left(b+c\right)e^{-s \tau}\right)\left(s +\kappa a +\kappa \left(b-c\right)e^{-s \tau}\right)=0.
\end{align}
Differentiating \eqref{eq:characb_kappa} with respect to $\kappa$, we get
\begin{align}
\label{eq:dl_dk}
\frac{\mathrm{d}s}{\mathrm{d}\kappa}=\frac{-\kappa a^2-s a-s b e^{-s \tau}-2\kappa abe^{-s \tau}-\kappa \left(b^2-c^2\right)e^{-2 s \tau}}{s + \kappa a + \kappa b e^{-s \tau}-s \kappa b \tau e^{-s \tau}-\kappa ^2ab\tau e^{-s \tau}-\kappa ^2 \tau \left(b^2-c^2\right)e^{-2 s \tau}}.
\end{align}
\normalsize
From \eqref{eq:characb_kappa} we get,
 \begin{align}
\label{eq:exp}
e^{-s \tau}= -\frac{s+\kappa a}{\kappa \left(b+c\right)}.
\end{align}
Next, substituting \eqref{eq:exp} in \eqref{eq:dl_dk}, we get
\begin{align}
\label{eq:prove_real}
\frac{\mathrm{d}s}{\mathrm{d}\kappa}=\frac{ s}{\kappa\left(1+s \tau+\kappa a \tau\right)}. 
\end{align}
At $\tau=\tau_0$, $\kappa=\kappa_c$. Substituting $s=j\omega_1$ in \eqref{eq:prove_real} we get 
\begin{align*}
\mathrm{Re}\left(\frac{\mathrm{d}s}{\mathrm{d}\kappa}\right)_{s=j \omega_1}= \frac{\omega_1^2 \tau_0}{\kappa_c\left(\left(1+\kappa_c a \tau_0\right)^2+\left(\omega_1 \tau_0\right)^2\right)}>0.
\end{align*}
In particular, we have proved that, $\mathrm{Re}(\mathrm{d}s/\mathrm{d}\kappa)>0$, which implies that the roots cross over the imaginary axis with positive velocity at $\kappa=\kappa_c$.
\subsection*{Case II}
For the second scenario, we observe that the linearised system, with the non-dimensional exogenous parameter $\kappa$ is given as 
\begin{align}
\label{eq:linearb1_kappa}
&\dot{u}_1(t) = \kappa \Big(\-\mathcal{M}_{1}u_{1}(t)-\mathcal{N}_{1}u_{1}(t-\tau_{1})-\mathcal{P}_{1}u_{2}(t-\tau_{2})\Big),\notag\\
&\dot{u}_2(t) = \kappa \Big(-\big(\mathcal{M}_{2}+\mathcal{N}_{2}\big)u_{2}(t)-\mathcal{P}_{2}u_{1}(t-\tau_{1})\Big).
\end{align}
To show that system \eqref{eq:linearb1_kappa} loses stability via a Hopf bifurcation as the non-dimensional parameter $\kappa$ is increased, we need to verify the transversality of the Hopf spectrum. Note that, for any complex number $z$, $\mathrm{Re}(z)\neq 0$ if and only if $\mathrm{Re}(z^{-1})\neq 0$. Hence, for ease of analysis, we proceed to verify that $\mathrm{Re}(\mathrm{d}s/\mathrm{d}\kappa)\neq 0$ at $\kappa=\kappa_c$. Looking for exponential solutions of \eqref{eq:linearb1_kappa} leads us to the following characteristic equation:
\begin{align}
\label{eq:case2_kappa}
s^2 + \kappa as +\kappa bs e^{-s \tau_1} + \kappa^2 c e^{-s \tau_1} + \kappa^2 d=0.
\end{align}
Differentiating \eqref{eq:case2_kappa} with respect to $\kappa$, we get
\begin{align}
\label{eq:dl_dk_2}
\frac{\mathrm{d}s}{\mathrm{d}\kappa} = \frac{-a s - bs e^{-s \tau_1}-2\kappa c e^{-s \tau_1}-2kd}{2s+ \kappa a +\kappa be^{-s\tau_1}-\kappa b s \tau_1 e^{-s \tau_1}-\kappa^2 c \tau_1 e^{-s\tau_1}}
\end{align}
From the characteristic equation \eqref{eq:case2_kappa}, we get
\begin{align}
e^{-s \tau_1}=-\frac{s^2+ \kappa as +\kappa^2 d}{\kappa b s +\kappa^2 c}.
\end{align}
Now, substituting the value of $e^{-s\tau_1}$ in \eqref{eq:dl_dk_2} and performing some algebraic manipulations, we obtain
\begin{align*}
\Bigg(\frac{\mathrm{d}s}{\mathrm{d}\kappa}\Bigg)^{-1}=\Bigg(\frac{\mathrm{d}s}{\mathrm{d}\kappa}\Bigg)_{1}^{-1}+\Bigg(\frac{\mathrm{d}s}{\mathrm{d}\kappa}\Bigg)_{2}^{-1}+\Bigg(\frac{\mathrm{d}s}{\mathrm{d}\kappa}\Bigg)_{3}^{-1},
\end{align*}

where,
\begin{align}
\label{eq:dl_dk_3}
&\Bigg(\frac{\mathrm{d}s}{\mathrm{d}\kappa}\Bigg)_{1}^{-1} = \frac{\kappa}{s}, \hspace{4ex} \Bigg(\frac{\mathrm{d}s}{\mathrm{d}\kappa}\Bigg)_{2}^{-1} = \kappa \tau_1, \notag\\
&\Bigg(\frac{\mathrm{d}s}{\mathrm{d}\kappa}\Bigg)_{3}^{-1} = \frac{\kappa^2 \big(s^2 ab\tau_1 - s^2 c\tau_1 +2\kappa s bd\tau_1+\kappa^2 cd \tau_1\big)}{s \Big(s^2 b+\kappa^2 ac+ 2\kappa s c -\kappa^2 bd\Big)}.
\end{align}
Recall that, at the crossover point, the system has one pair of complex conjugate roots on the imaginary axis. Hence, substituting $s=j\omega$ in \eqref{eq:dl_dk_3}, we obtain $\left(\frac{\mathrm{d}s}{\mathrm{d}\kappa}\right)_{1,s=j\omega}^{-1} = \frac{\kappa}{j\omega}$, which is purely imaginary. Similarly, we see that $\left(\frac{\mathrm{d}s}{\mathrm{d}\kappa}\right)_{2,s=j\omega}^{-1} = \kappa \tau_1$ which is strictly positive. Thus, to verify that $\mathrm{Re}\left(\frac{\mathrm{d}s}{\mathrm{d}\kappa}\right)_{s=j\omega}^{-1}>0,$ verifying $\mathrm{Re}\left(\frac{\mathrm{d}s}{\mathrm{d}\kappa}\right)_{3,s=j\omega}^{-1}>0$ suffices. Now,
\begin{align}
\label{eq:dl_dk_3real}
\mathrm{Re}\Bigg(\frac{\mathrm{d}s}{\mathrm{d}\kappa}\Bigg)_{3,s=j\omega}^{-1}= \frac{2\omega^2 \kappa^3\tau_1\big(abc-c^2-b^2d\big)\big(\omega^2+\kappa^2d\big)}{4\omega^4\kappa^2 c^2 +\big(\kappa^2 \omega ac-\omega^3 b -\kappa^2 \omega  bd\big)^2}.
\end{align} 
Recall that $d$ is positive. Hence, the expression $\omega^2+\kappa^2 d$ is positive. Thus, it suffices to verify that $(abc-c^2-b^2d)>0$. Substituting the values of  $a,d,c$ and $d$ from \eqref{eq: abcd}, we get
\begin{align*}
abc-c^2-b^2d =\, \mathcal{P}_1\mathcal{P}_2\left(\mathcal{N}_1\mathcal{N}_2-\mathcal{P}_1\mathcal{P}_2\right)+ \mathcal{N}_1\mathcal{P}_1\mathcal{P}_2\left(2\mathcal{M}_2-\mathcal{M}_1\right).
\end{align*} 
Note that $\mathcal{M}_j$, $\mathcal{N}_j$ and $\mathcal{P}_j$ are strictly positive for $j=1,2$. Now, it can be easily concluded that $\mathcal{N}_1\mathcal{N}_2>\mathcal{P}_1\mathcal{P}_2$. Hence, the first term in the above expression is positive. Recall that we consider a regime wherein the router buffers are small. Consequently, the average window sizes $w_1^{\ast}$ and $w_2^{\ast}$ would also be small. Additionally, we consider that the bandwidth-delay product is high. Hence, we assume that the per flow capacities of the edge routers $C'_1$ and $C'_2$ are large enough such that $1-p^{\ast}_j-q^{\ast}\approx 1\,\,\forall\,j=1,\,2.$  With this approximation, in the regime wherein $\tau_1\gg\tau_2$ and $\tau_2 \approx 0$, we can conclude that $\mathcal{M}_2>\mathcal{M}_1.$ This ensures that $abc-c^2-b^2d>0$. Hence, we can conclude that $\mathrm{Re}\left(\frac{\mathrm{d}s}{\mathrm{d}\kappa}\right)_{3,s=j\omega}^{-1}>0$, which in turn ensures that
\begin{align*}
\mathrm{Re}\Bigg(\frac{\mathrm{d}s}{\mathrm{d}\kappa}\Bigg)_{\kappa=\kappa_c}^{-1}>0.
\end{align*}    
Thus, we observe that, the system undergoes a \emph{Hopf Bifurcation} at $\kappa=\kappa_c$ for both scenarios. This implies that the system loses stability, as the system parameters vary, leading to the emergence of limit cycles. These limit cycles could in turn induce synchronisation among the Compound TCP flows which leads to periodic packet losses and loss in link utilisation. In turn, we expect the downstream traffic to be bursty.

\begin{figure}
\begin{center}
  \psfrag{cccc}{\begin{small}\hspace{-7mm}Edge router $1$\end{small}}
  \psfrag{eeee}{\begin{small}\hspace{-5.5mm}Core router\end{small}}
  \psfrag{aaaa}{\begin{small}\hspace{-13mm}Buffer size = 15 pkts \end{small}}
   \psfrag{yyyy}{\begin{small}\hspace{-6mm}Edge route $2$ \end{small}}
  \psfrag{275}{\begin{scriptsize}$275$\end{scriptsize}}
  \psfrag{300}{\begin{scriptsize}$300$\end{scriptsize}}
  \psfrag{100}{\begin{scriptsize}$100$\end{scriptsize}}
  \psfrag{15}{\begin{scriptsize}$15$\end{scriptsize}}
  \psfrag{ffff}{\hspace{-10mm}Time (seconds)}
  \psfrag{xxxx}{\begin{small}\hspace{-13.25mm}Buffer size = 100 pkts\end{small}}
  \psfrag{gggg}{\hspace{-0.87cm}Queue size (pkts)}
  \psfrag{0}{\begin{scriptsize}$0$\end{scriptsize}}
  \psfrag{aaa}{\begin{scriptsize}NS2\end{scriptsize}}
   \psfrag{xyz}{\begin{scriptsize}Numerical\end{scriptsize}}
  
  \includegraphics[width=2.85in,height=3.85in,angle=270]{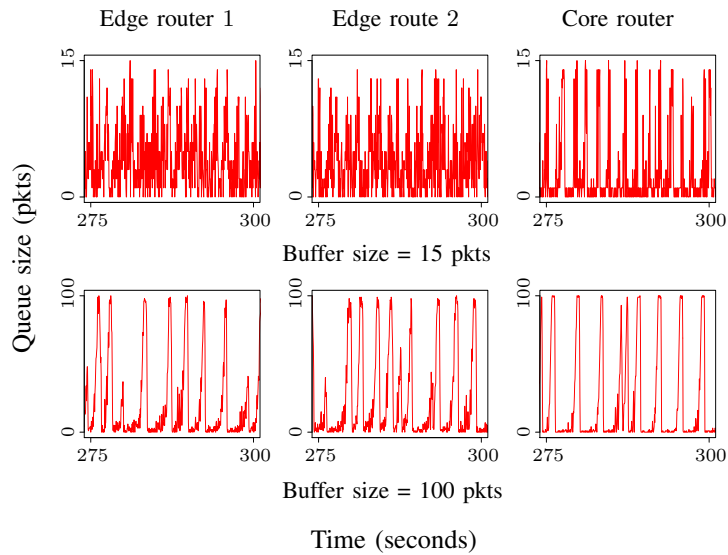}

  \caption {\emph{Queue size dynamics with long-lived flows:} Two sets of $60$ Compound TCP flows each with an access speed of $2$ Mbps, regulated by two edge routers each with a link capacity of $100$ Mbps, and feeding into a core router with a link capacity of $197$ Mbps. The flows in each set have an average round trip time of $200$ ms. We can easily see that as buffer thresholds at routers are increased, the queue size exhibits limit cycles.}
  \label{fig:multi2200}
 \end{center}
\end{figure}

\begin{figure}[t!]
\begin{center}
  \psfrag{cccc}{\begin{small}\hspace{-7mm}Edge router $1$\end{small}}
  \psfrag{eeee}{\begin{small}\hspace{-5.5mm}Core router\end{small}}
  \psfrag{aaaa}{\begin{small}\hspace{-13mm}Buffer size = 15 pkts \end{small}}
   \psfrag{yyyy}{\begin{small}\hspace{-6mm}Edge route $2$ \end{small}}
  \psfrag{275}{\begin{scriptsize}$275$\end{scriptsize}}
  \psfrag{300}{\begin{scriptsize}$300$\end{scriptsize}}
  \psfrag{100}{\begin{scriptsize}$100$\end{scriptsize}}
  \psfrag{15}{\begin{scriptsize}$15$\end{scriptsize}}
  \psfrag{ffff}{\hspace{-10mm}Time (seconds)}
  \psfrag{xxxx}{\begin{small}\hspace{-13.25mm}Buffer size = 100 pkts\end{small}}
  \psfrag{gggg}{\hspace{-0.87cm}Queue size (pkts)}
  \psfrag{0}{\begin{scriptsize}$0$\end{scriptsize}}
  \psfrag{aaa}{\begin{scriptsize}NS2\end{scriptsize}}
   \psfrag{xyz}{\begin{scriptsize}Numerical\end{scriptsize}}
  
  \includegraphics[width=2.85in,height=3.85in,angle=270]{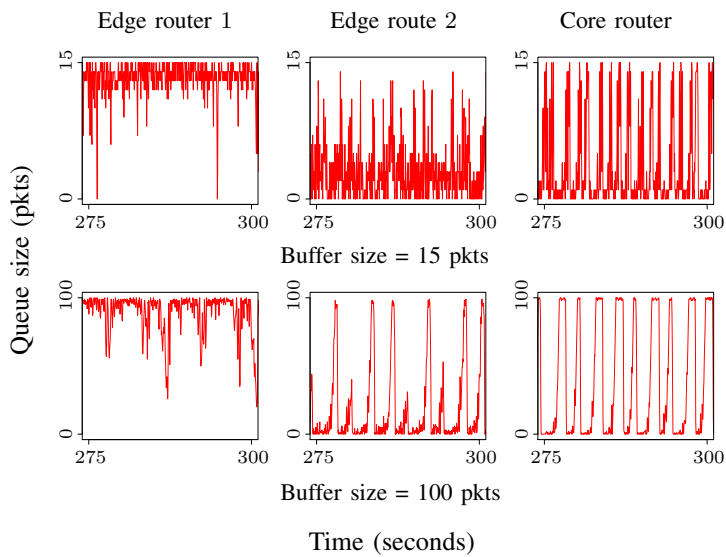}

  \caption {\emph{Queue size dynamics with long-lived flows:} Two sets of $60$ Compound TCP flows each with an access speed of $2$ Mbps, regulated by two edge routers each with a link capacity of $100$ Mbps, and feeding into a core router with a link capacity of $197$ Mbps. The flows in the two sets have average round trip times $10$ ms and $200$ ms respectively. It  can be easily noted that as buffer sizes are increased from $15$ to $100$ packets, the queue size dynamics exhibits limit cycles. }
  \label{fig:multi210200}
 \end{center}
\end{figure}

\begin{figure}[t!]
\begin{center}
  \psfrag{cccc}{\begin{small}\hspace{-7mm}Edge router $1$\end{small}}
  \psfrag{eeee}{\begin{small}\hspace{-5.5mm}Core router\end{small}}
  \psfrag{aaaa}{\begin{small}\hspace{-13mm}Buffer size = 15 pkts \end{small}}
   \psfrag{yyyy}{\begin{small}\hspace{-6mm}Edge route $2$ \end{small}}
  \psfrag{275}{\begin{scriptsize}$275$\end{scriptsize}}
  \psfrag{300}{\begin{scriptsize}$300$\end{scriptsize}}
  \psfrag{100}{\begin{scriptsize}$100$\end{scriptsize}}
  \psfrag{15}{\begin{scriptsize}$15$\end{scriptsize}}
  \psfrag{ffff}{\hspace{-10mm}Time (seconds)}
  \psfrag{xxxx}{\begin{small}\hspace{-13.25mm}Buffer size = 100 pkts\end{small}}
  \psfrag{gggg}{\hspace{-0.87cm}Queue size (pkts)}
  \psfrag{0}{\begin{scriptsize}$0$\end{scriptsize}}
  \psfrag{aaa}{\begin{scriptsize}NS2\end{scriptsize}}
   \psfrag{xyz}{\begin{scriptsize}Numerical\end{scriptsize}}
  
  \includegraphics[width=2.85in,height=3.85in,angle=270]{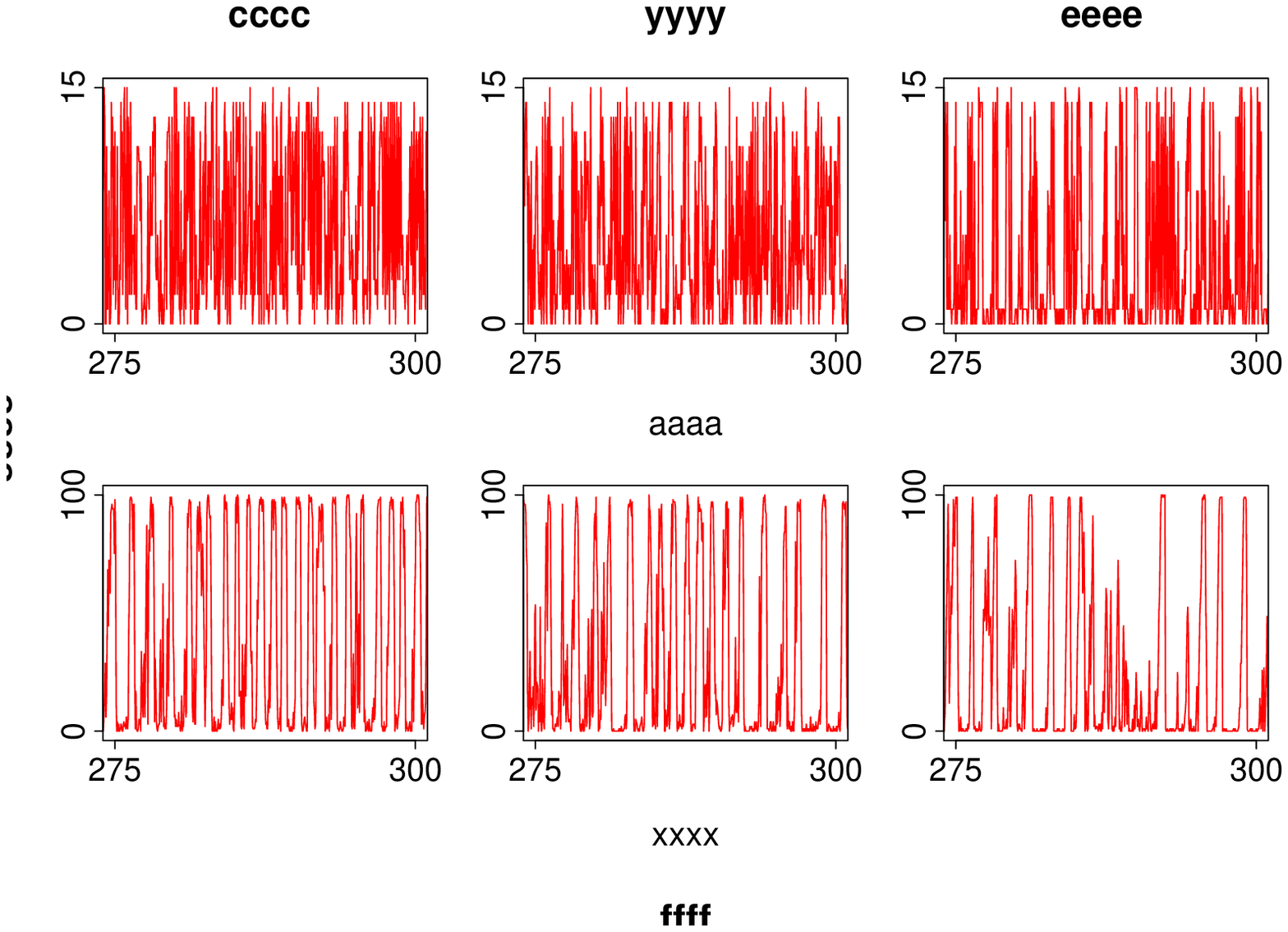}

  \caption {\emph{Queue size dynamics with heavy-tailed files:} Two sets of $100$ Compound TCP sources, regulated by two edge routers each with a link capacity of $100$ Mbps, and feeding into a core router with a link capacity of $197$ Mbps. The flows in both sets have an average round trip time of $200$ ms. The file sizes are drawn from a Pareto distribution with shape parameter 1.5. The average file size is fixed at $100$ KB. Observe that the queue size exhibits limit cycles, as buffer sizes are increased from $15$ to $100$ packets.}
  \label{fig:multi2h200}
 \end{center}
\end{figure}


\subsection{Simulations}
To validate our analytical insights, we simulate two scenarios in the multiple bottleneck topology: only long-lived flows, and with heavy-tailed files. 
\subsubsection{Dynamical Properties}
The system consists of two distinct sets of 60 Compound TCP flows each with an access speed of $2$ Mbps, regulated by two edge routers and feeding into a common core router. Each edge router has a link capacity of 100 Mbps and the core router has a link capacity of $197$ Mbps. 

To illustrate the impact of increasing buffer sizes on the queue size dynamics, we consider two cases: all routers have a buffer size of $(i)$ $15$ packets, and $(ii)$ $100$ packets. For the round trip times, we consider the following two cases: $(i)$ both sets of flows have same average round trip times, $200$ ms, and $(ii)$ the average round trip time of one set is much smaller compared to the other. In this case, we choose the average round trip times as $10$ ms and $200$ ms respectively. 

Figs. \ref{fig:multi2200}, and \ref{fig:multi210200} show the queue size dynamics, with long-lived flows. It is evident that as the buffer thresholds are increased from $15$ to $100$ packets, limit cycles emerge in the queue size. In particular, even if one round trip time is large, the underlying dynamical systems lose stability if buffer sizes increase.  This corroborates our analysis. Fig. \ref{fig:multi2h200} shows the impact of increasing buffer sizes on the queue size dynamics, with heavy-tailed TCP connections. We can see that even with high variability at the connection level, increasing buffer thresholds would induce limit cycles in the queue size dynamics, an insight consistent with that obtained for a single bottleneck topology.
\subsubsection{Statistical Properties}
 \begin{figure}[t!]
\begin{subfigure}{0.33\linewidth}
\hspace{-4mm}
\psfrag{c}{\hspace{3.2cm}Coefficient of variation versus $\log_2$ (Time in $\mu s$)}
\psfrag{20}{\begin{scriptsize}$20$\end{scriptsize}}
\psfrag{12}{\begin{scriptsize}$12$\end{scriptsize}}
\psfrag{0.24}{\begin{scriptsize}$0.24$\end{scriptsize}}
\psfrag{0.01}{\begin{scriptsize}$0.01$\end{scriptsize}}
\psfrag{h}{\begin{tiny}Buffer size = 2084 pkts\end{tiny}}
\psfrag{x}{\begin{tiny}Buffer size = 100 pkts\end{tiny}}
\psfrag{y}{\begin{tiny}Buffer size = 50 pkts\end{tiny}}
\psfrag{z}{\begin{tiny}Buffer size = 15 pkts\end{tiny}}
\psfrag{b}{\begin{scriptsize}\hspace{0.5cm}Edge router $1$\end{scriptsize}}
\includegraphics[width=1.7in,height=2.5in,angle=270]{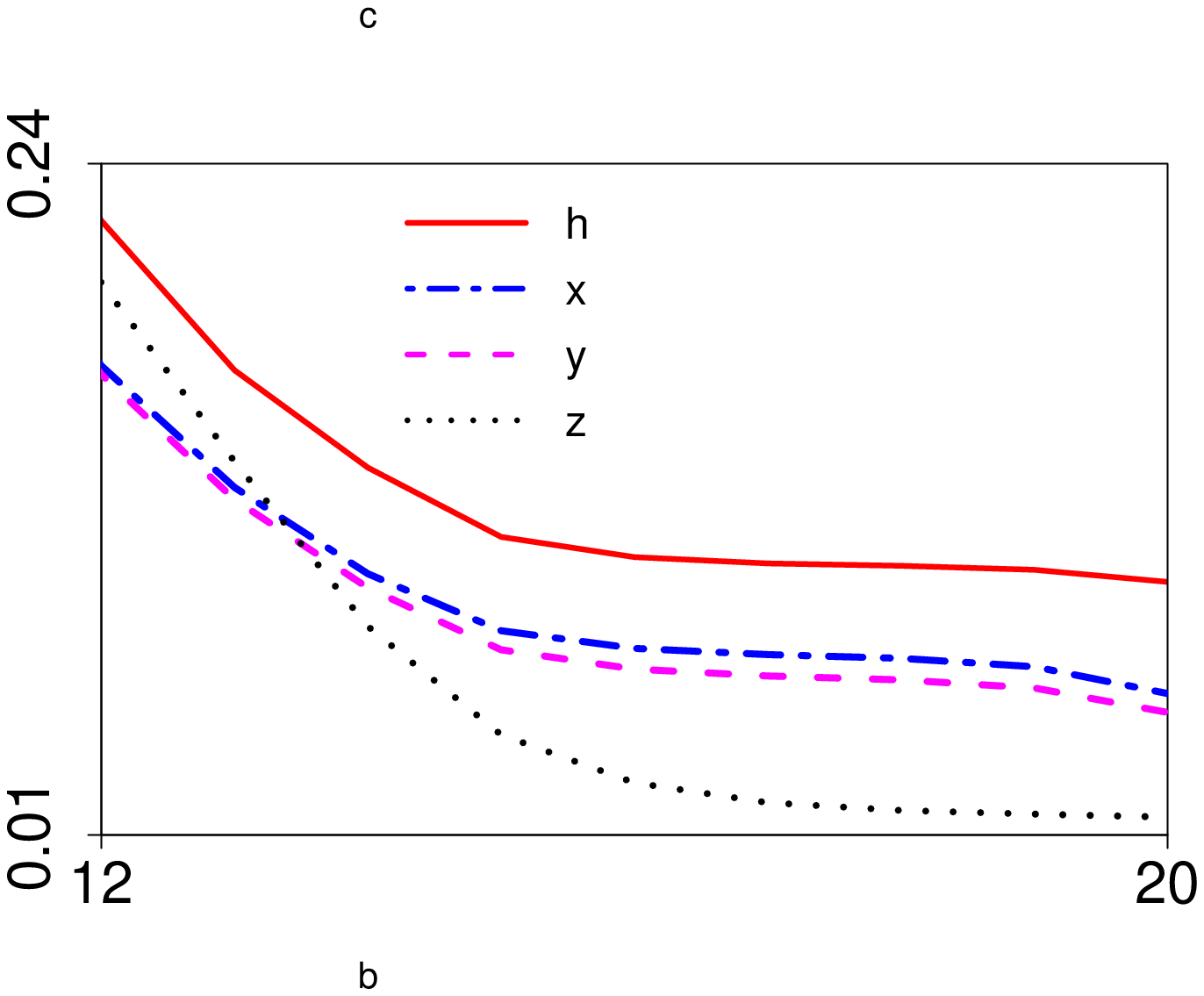}
\end{subfigure}
\begin{subfigure}{0.33\linewidth}
\hspace{-3mm}
\psfrag{T}{}
\psfrag{20}{\begin{scriptsize}$20$\end{scriptsize}}
\psfrag{12}{\begin{scriptsize}$12$\end{scriptsize}}
\psfrag{0.24}{\begin{scriptsize}$0.24$\end{scriptsize}}
\psfrag{0.01}{\begin{scriptsize}$0.01$\end{scriptsize}}
\psfrag{h}{\begin{tiny}Buffer size = 2084 pkts\end{tiny}}
\psfrag{x}{\begin{tiny}Buffer size = 100 pkts\end{tiny}}
\psfrag{y}{\begin{tiny}Buffer size = 50 pkts\end{tiny}}
\psfrag{z}{\begin{tiny}Buffer size = 15 pkts\end{tiny}}
\psfrag{b}{\begin{scriptsize}\hspace{0.5cm}Edge router $2$\end{scriptsize}}
\psfrag{c}{}
\includegraphics[width=1.7in,height=2.5in,angle=270]{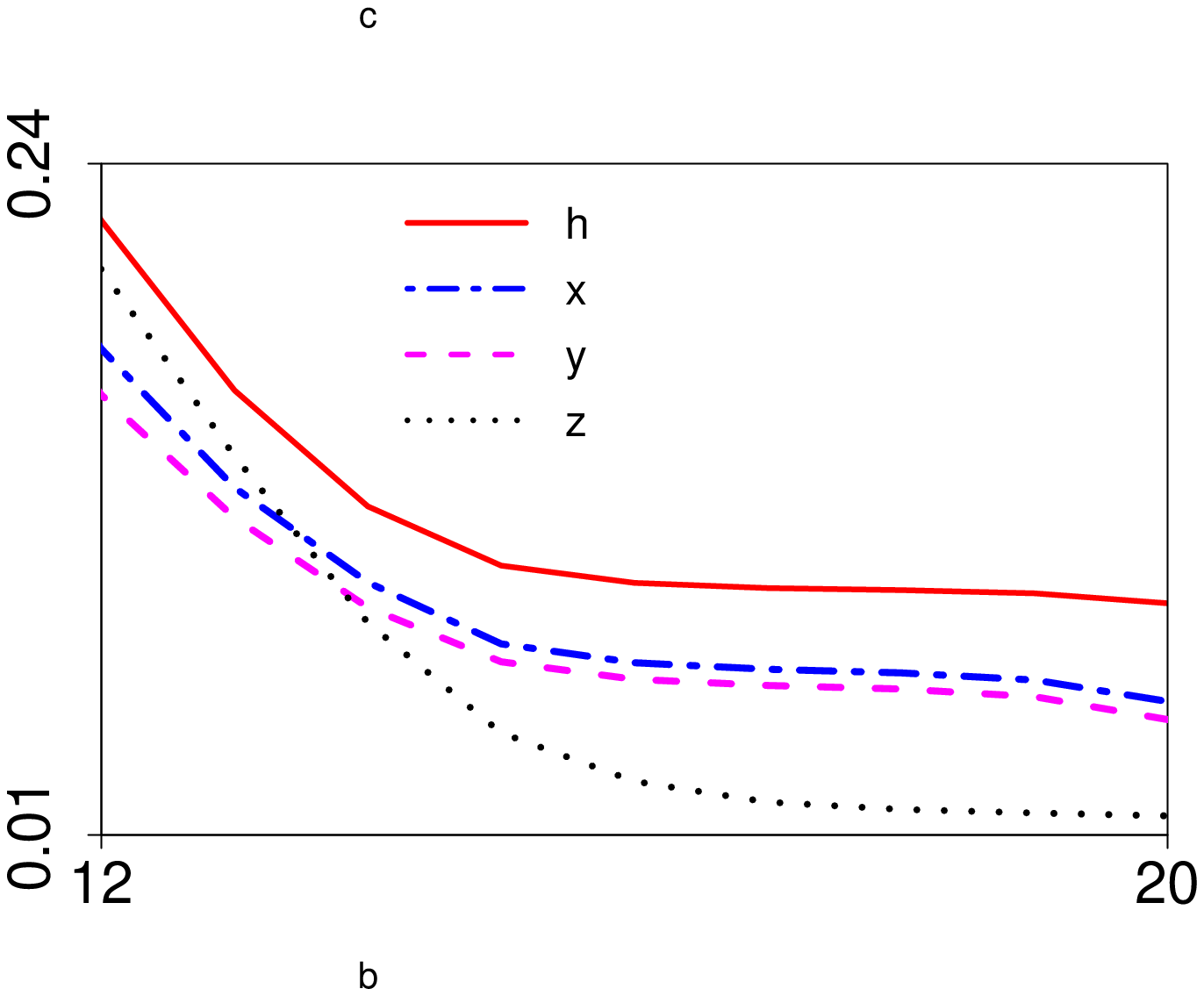}
\end{subfigure}
\begin{subfigure}{0.33\linewidth}
\hspace{-2mm}
\psfrag{0.01}{\begin{scriptsize}$0.01$\end{scriptsize}}
\psfrag{12}{\begin{scriptsize}$12$\end{scriptsize}}
\psfrag{20}{\begin{scriptsize}$20$\end{scriptsize}}
\psfrag{0.08}{\begin{scriptsize}$0.08$\end{scriptsize}}
\psfrag{h}{\begin{tiny}Buffer size = 4040 pkts\end{tiny}}
\psfrag{x}{\begin{tiny}Buffer size = 100 pkts\end{tiny}}
\psfrag{y}{\begin{tiny}Buffer size = 50 pkts\end{tiny}}
\psfrag{z}{\begin{tiny}Buffer size = 15 pkts\end{tiny}}
\psfrag{b}{\begin{scriptsize}\hspace{0.5cm}Core router\end{scriptsize}}
\psfrag{T}{}
\psfrag{c}{}
\includegraphics[width=1.7in,height=2.5in,angle=270]{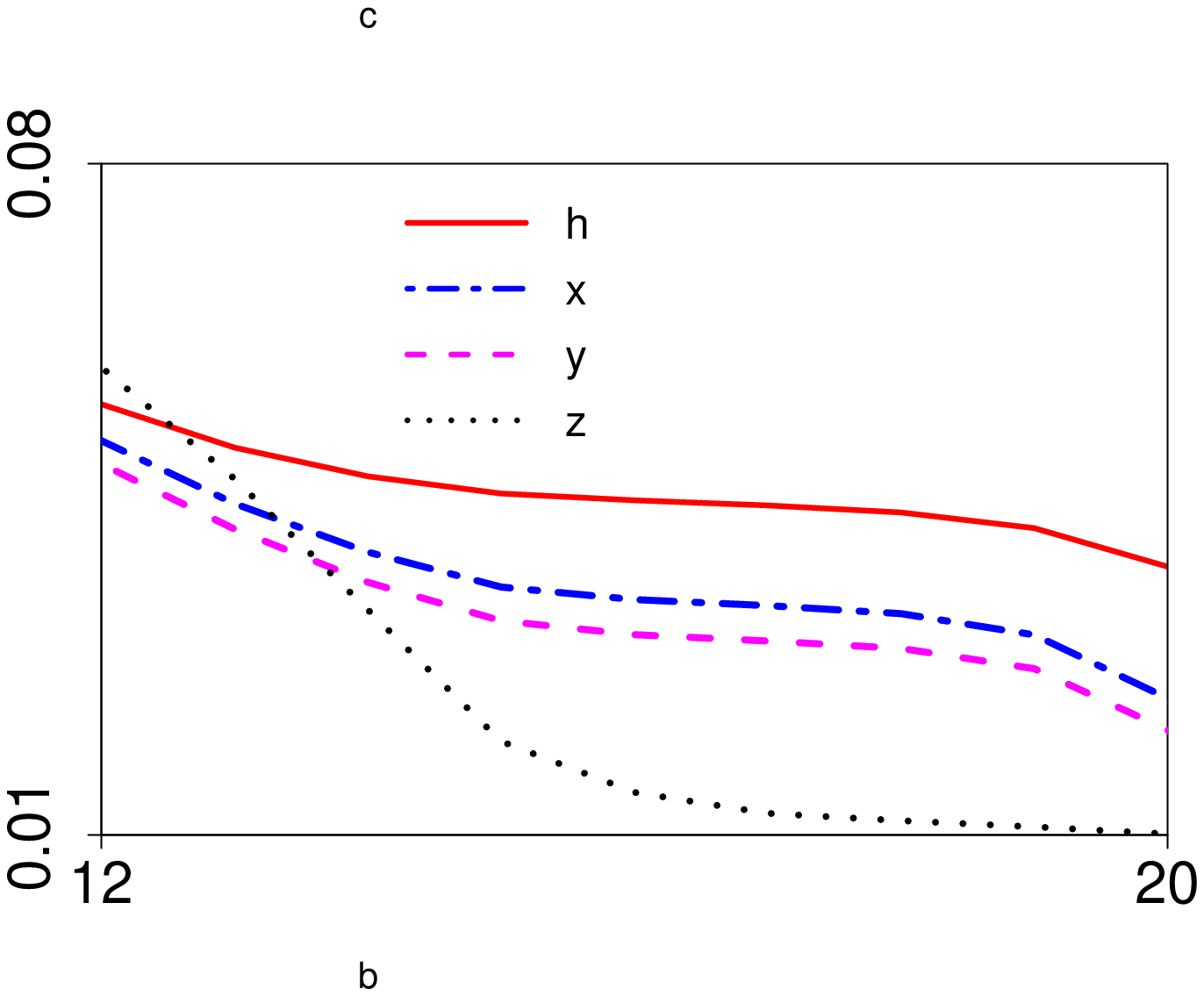}
\end{subfigure}
\caption {\emph{Statistics of the arrival process.} Two sets of $480$ long-lived Compound TCP flows each with an access link speed of $0.25$ Mbps, regulated by two edge routers and feeding into a core router of $194$ Mbps. The average round trip of each set is chosen as $300$ ms. We consider three representative regimes: $(i)$ stable (each router has a buffer size of $15$ packets, $(ii)$ presence of synchronisation (each router has a buffer size of $50$ and $100$ packets respectively) and $(iii)$ each router follows the bandwidth-delay product rule, used in practice. Observe that with smaller buffers ($15$ packets), the aggregate arrival process at each queue exhibits reduced burstiness.}
  \label{fig:statarrival_multi}
\end{figure}
 \begin{figure}[t!]
\begin{subfigure}{0.33\linewidth}
\hspace{-4mm}
\psfrag{T}{\hspace{1.25cm}CCDF versus queue threshold}
\psfrag{0}{\begin{scriptsize}$0$\end{scriptsize}}
\psfrag{1}{\begin{scriptsize}$1$\end{scriptsize}}
\psfrag{15}{\begin{scriptsize}$15$\end{scriptsize}}
\psfrag{h}{\begin{tiny}Empirical\end{tiny}}
\psfrag{x}{\begin{tiny}$M/M/1/B$\end{tiny}}
\psfrag{y}{\begin{tiny}$M/D/1/B$\end{tiny}}
\psfrag{e}{\begin{scriptsize}\hspace{0.25cm}Edge router $1$\end{scriptsize}}
\includegraphics[width=1.7in,height=2.5in,angle=270]{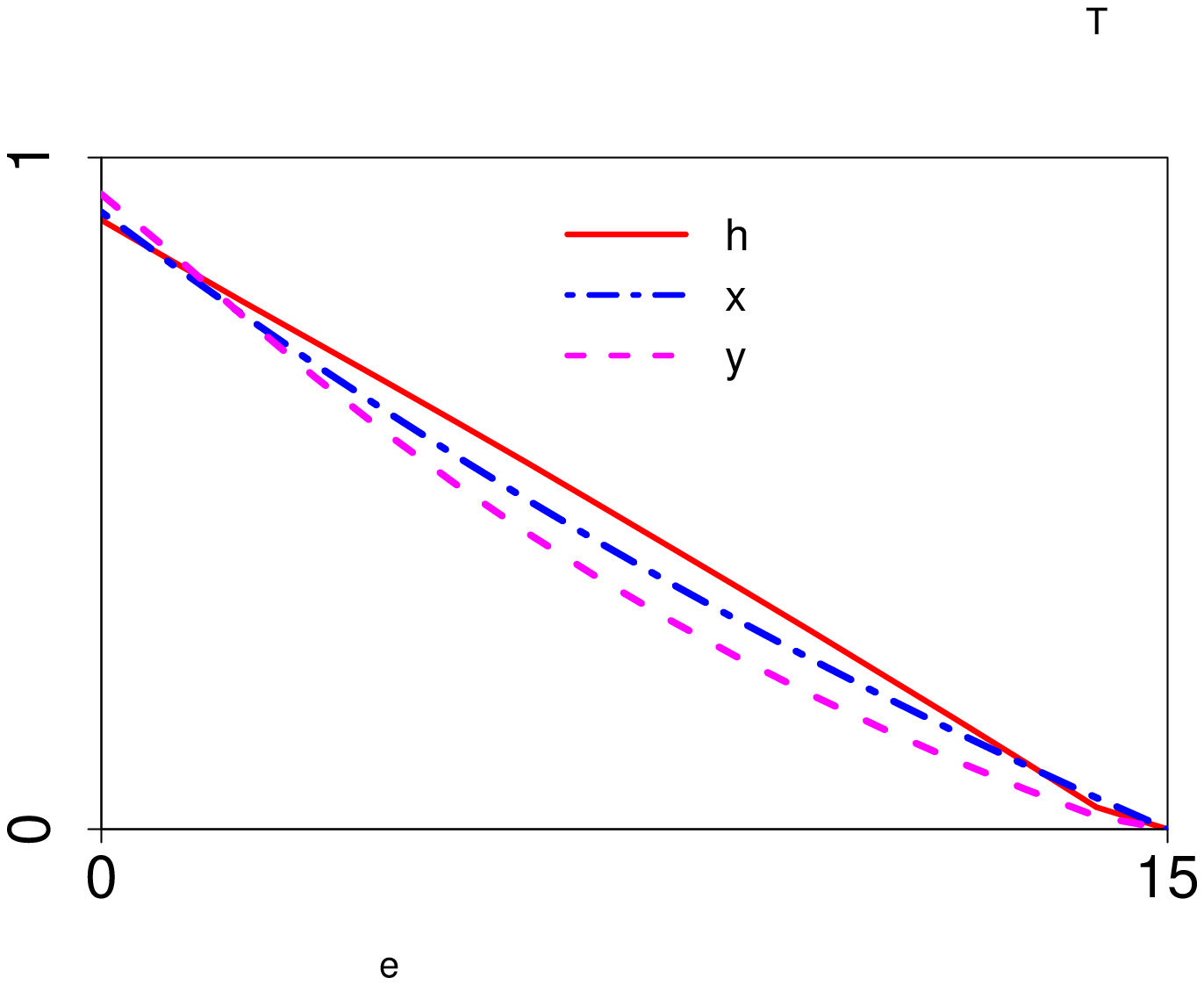}
\end{subfigure}
\begin{subfigure}{0.33\linewidth}
\hspace{-3mm}
\psfrag{0}{\begin{scriptsize}$0$\end{scriptsize}}
\psfrag{1}{\begin{scriptsize}$1$\end{scriptsize}}
\psfrag{15}{\begin{scriptsize}$15$\end{scriptsize}}
\psfrag{h}{\begin{tiny}Empirical\end{tiny}}
\psfrag{x}{\begin{tiny}$M/M/1/B$\end{tiny}}
\psfrag{y}{\begin{tiny}$M/D/1/B$\end{tiny}}
\psfrag{b}{\hspace{-3cm}\begin{scriptsize}Buffer size = $10$ packets\end{scriptsize}}
\psfrag{c}{\hspace{-2.5cm}\begin{scriptsize}Time scale (seconds)\end{scriptsize}}
\psfrag{e}{\begin{scriptsize}\hspace{0.25cm}Edge router $2$\end{scriptsize}}
\psfrag{T}{}
\includegraphics[width=1.7in,height=2.5in,angle=270]{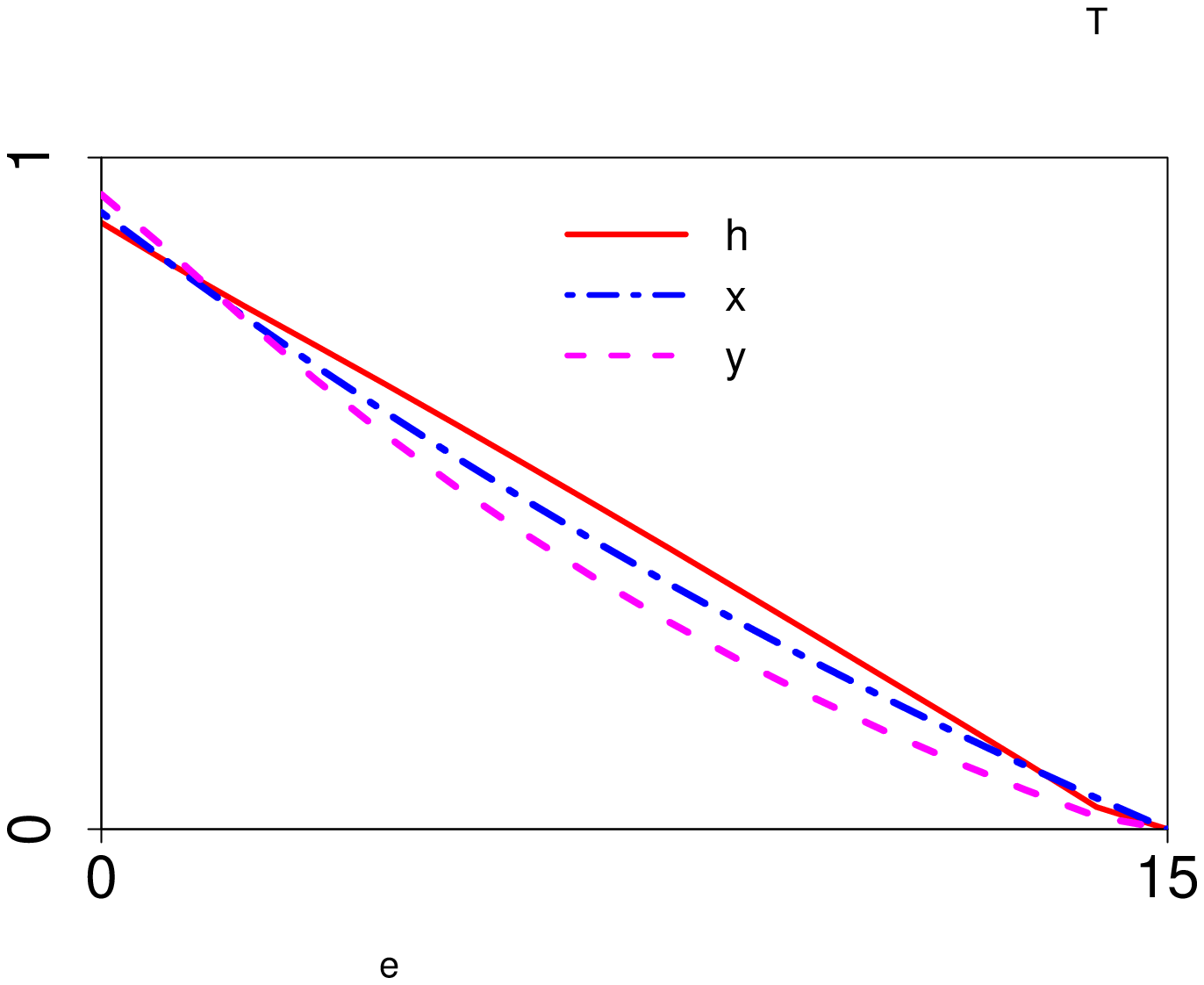}
\end{subfigure}
\begin{subfigure}{0.33\linewidth}
\hspace{-2mm}
\psfrag{0}{\begin{scriptsize}$0$\end{scriptsize}}
\psfrag{1}{\begin{scriptsize}$1$\end{scriptsize}}
\psfrag{15}{\begin{scriptsize}$15$\end{scriptsize}}
\psfrag{h}{\begin{tiny}Empirical\end{tiny}}
\psfrag{x}{\begin{tiny}$M/M/1/B$\end{tiny}}
\psfrag{y}{\begin{tiny}$M/D/1/B$\end{tiny}}
\psfrag{b}{\begin{scriptsize}\hspace{-0.75cm}Core router\end{scriptsize}}
\psfrag{c}{\hspace{-2.5cm}\begin{scriptsize}Queue threshold\end{scriptsize}}
\psfrag{e}{\begin{scriptsize}\hspace{0.5cm}Core router \end{scriptsize}}
\psfrag{T}{}
\includegraphics[width=1.7in,height=2.5in,angle=270]{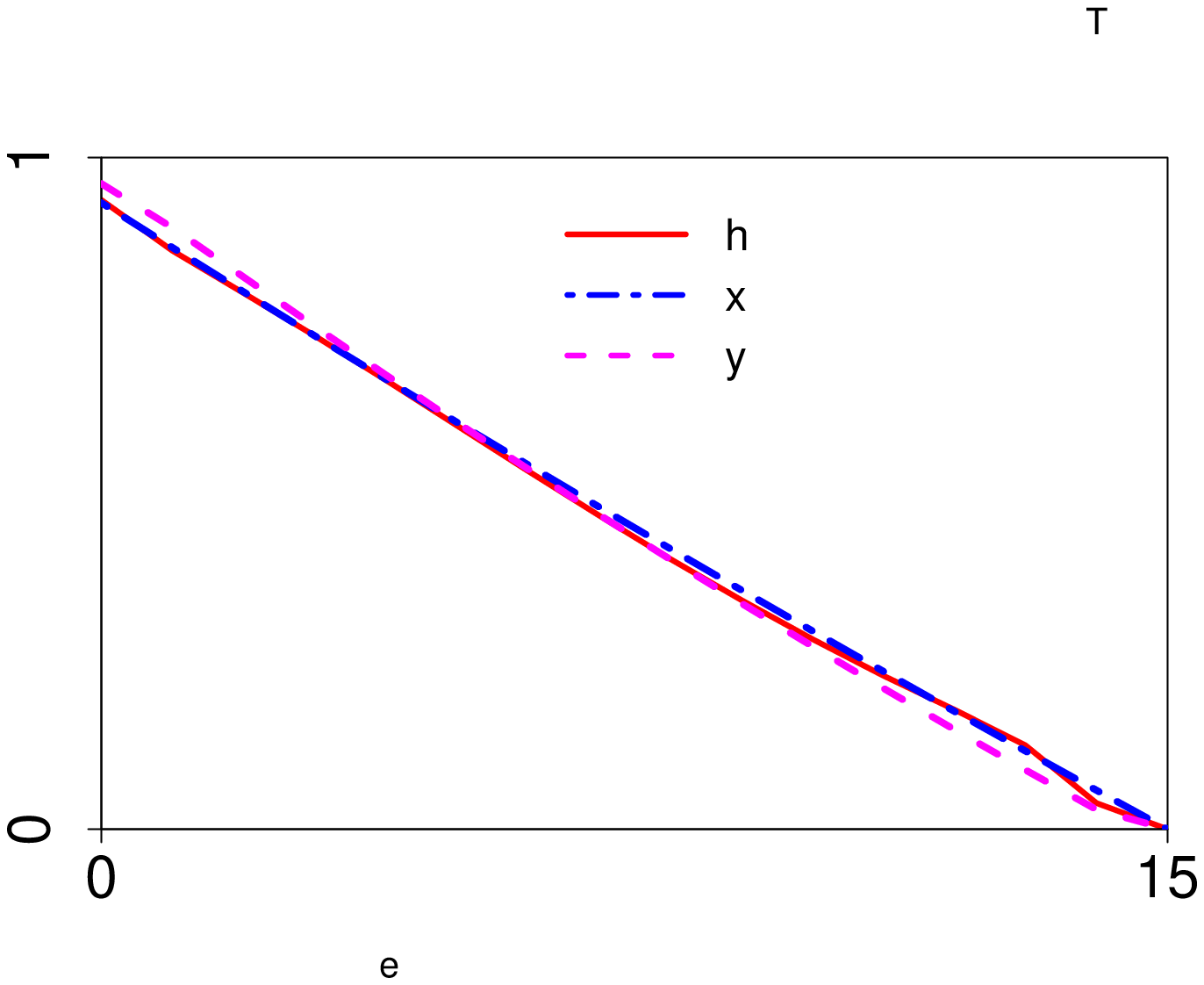}
\end{subfigure}
\caption {\emph{Statistics of the queue size.} Two sets of $480$ long-lived Compound TCP flows each with an access link speed of $0.25$ Mbps, regulated by two edge routers and feeding into a core router of $194$ Mbps. The buffer size at each router is fixed at $15$ packets, and the average round trip of each set is chosen as $300$ ms. We compare the queue distribution at each queue with that of $M/M/1/B$ and $M/D/1/B$ queues. Observe that either an $M/M/1/B$ or $M/D/1/B$ approximation seems reasonable with a large number of long-lived flows and high bandwidth-delay product.}
  \label{fig:stat_multi_long}
\end{figure}
To establish the validity of our theoretical approximation that the packet loss probability at each bottleneck queue can be approximated by the corresponding blocking probability of an $M/M/1/B$ queue, we now empirically study the statistical properties of the arrival process, and the queue length distribution at each queue, for both topologies.

Note that in packet-level simulations, we can easily observe the loss of stability and hence a qualitative change in the dynamical properties of the system, with a
reasonable number of TCP flows ($60$). However, for our statistical analyses, we need a larger number of long-lived flows for the statistical properties to hold. Hence, we consider two sets of $480$ flows, each over an access link with a speed of $0.25$ Mbps. Each set of flows is regulated by an edge router with a link capacity of $100$ Mbps. The outputs of the edge routers feed into a core router with a link capacity of $194$ Mbps. We choose the average round trip time of each set of flows as $300$ ms.
 
\emph{Statistics of the arrival process:} We first conduct an empirical study on the statistical properties of the traffic arrival at each bottleneck queue in a similar spirit as done for the single bottleneck topology. Specifically, we measure the burstiness of the arrival process at each queue in terms of their coefficient of variation at different time scales. 

For the empirical study, we consider three representative scenarios. In the first scenario, the buffer sizes at all routers are fixed at $15$ packets, which ensures that the underlying dynamical system is stable. In the second scenario, all buffers are dimensioned at $50$ and $100$ packets respectively. In both these cases, the system dynamics exhibits limit cycles, and synchronisation among TCP windows. In the third scenario, the buffer sizes at all routers are chosen according to the bandwidth-delay product rule, which leads to $2084$ packets at the edge routers, and $4040$ packets at the core router. Since we are interested in measuring the burstiness of the arrival process at short time scales, we aggregate the arrival traffic over time scales ranging from $2^{12}\,\mu s=4$ ms to $2^{20}\,\mu s=1$ second.

Fig. \ref{fig:statarrival_multi} depicts the coefficient of variation curves at each router, at various
time scales, for this topology. We can easily observe that similar qualitative insights obtained in the single bottleneck topology carry forward to the multiple bottleneck topology also. In particular, when all buffers are dimensioned at $15$ packets, the coefficient of variation curves exhibits a relatively faster decay as the aggregation increases, as opposed to larger buffer thresholds. Further, we can observe that for larger buffer thresholds, the coefficient of variation curves for the traffic arrival at each queue flattens significantly at larger time scales. This indicates that larger buffers maintain higher variability or burstiness in the traffic arrival, in the presence of synchronisation. On the contrary, in the absence of synchronisation, we can observe reduced variability or burstiness in the traffic arrival at each queue. This suggests that the aggregate traffic arrival behaves qualitatively similar to short range dependent processes, when buffers are sized small enough to mitigate synchronisation effects. Hence, the approximation that the aggregate traffic arrival at each bottleneck queue is Poisson in the presence of a large number of TCP flows seems reasonable, in the regime considered.


\emph{Statistics of the queue size:} We now perform a comparative study on the queue size distribution of each bottleneck queue in each topology, with that of an $M/M/1/B$ and an $M/D/1/B$ queue. For our study, we consider the regime when all buffers are dimensioned at $15$ packets. This is because, we have already established that only in this regime, the arrival process at each queue can be reasonably approximated by a Poisson process.

As shown in Fig.~\ref{fig:stat_multi_long},for a buffer size of $15$ packets, the queue distribution at each queue can be well approximated by that of an $M/M/1/B$ or an $M/D/1/B$ queue, in this topology. This strongly suggests that even with TCP controlled flows in a multiple bottleneck topology, each bottleneck queue can be approximated as either an $M/M/1/B$ or an $M/D/1/B$ queue, thus validating our modelling assumptions, in the asymptotic regime considered in this work. 


\section{Impact of buffer sizing on system performance}
\label{performance_multi}

Through a combination of stability analysis and extensive packet-level simulations, we have highlighted smaller buffer thresholds play an important role to ensure stability. Hence, it is imperative to study the impact of such small buffers on the system performance. To that end, we consider the multiple bottleneck topology, and choose two performance metrics, throughput and flow completion time. We conduct packet-level simulations for the same. Note that for a single bottleneck topology also, we observe similar qualitative behaviour.
\subsection{Throughput}
\begin{figure}[t!]
\begin{center}
   \psfrag{5}{\begin{scriptsize}$5$\end{scriptsize}}
  \psfrag{100}{\begin{scriptsize}$100$\end{scriptsize}}
  \psfrag{200}{\begin{scriptsize}$200$\end{scriptsize}}
   \psfrag{300}{\begin{scriptsize}$300$\end{scriptsize}}
    
      \psfrag{fffff}{Throughput (Mbps)}
  \psfrag{eeeee}{\hspace{0mm}Buffer size (pkts)}
\psfrag{aaaaaaaaaaaaaaaa}{\begin{scriptsize}Capacity = $500$ Mbps\end{scriptsize}} 
\psfrag{cccccccccccccccc}{\begin{scriptsize}Capacity = $100$ Mbps\end{scriptsize}}
\psfrag{xyzxyz}{\begin{scriptsize}Capacity = $300$ Mbps\end{scriptsize}}

  \includegraphics[width=2.75in,height=3.75in,angle=270]{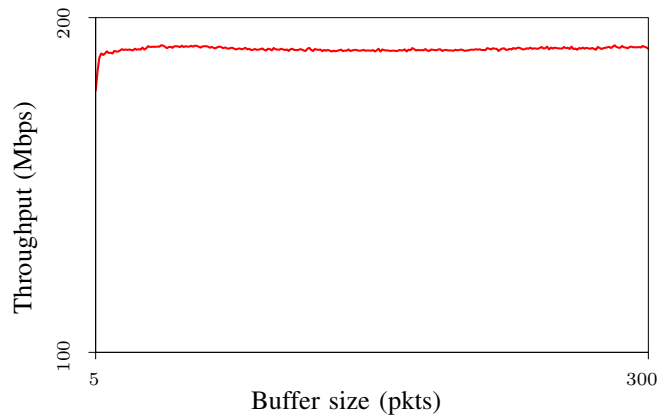}
  \caption{ \emph{Impact of buffer sizing on throughput}. Two sets of $60$ long-lived Compound TCP flows, regulated by two edge routers each with a link capacity of $100$ Mbps. The outputs of the edge routers feed into a core router with a link capacity of $197$ Mbps. The round trip time of each set is $200$ ms. We vary the core router buffer size. Observe that smaller buffers do not degrade throughput significantly. }
  \label{fig:multi_throughput}
  \end{center}
\end{figure}

For this, we consider two sets of $60$ long-lived Compound TCP flows, each over a $2$ Mbps access link. Both sets of flows are regulated by two edge router with a link capacity of $100$ Mbps. The outputs of the edge routers feed into a core router with a link capacity of $197$ Mbps. To study the impact of buffer sizing on throughput, we fix the buffer sizes at the edge routers at $15$ packets, and vary the buffer size at the core router from $5$ to $300$ packets. We fix the average round trip time of each set of flows as $200$ ms. Fig. \ref{fig:multi_throughput} shows the variation in throughput as the buffer size at the core router varies. It is evident that even with smaller buffers, we achieve fairly good throughput. 
\subsection{Average Flow Completion Time (AFCT)}
While throughput is undoubtedly an important performance metric from a network operator point of view, flow completion time is more important from a user perspective \cite{Dukkipati}. In particular, users would want their flows to complete in the shortest time possible. Hence, if buffer sizes are made much smaller that what they are today (bandwidth-delay product rule), they should not degrade flow completion times significantly.

\begin{figure}[t!]
\begin{center}
   \psfrag{0}{\begin{scriptsize}$0$\end{scriptsize}}
  \psfrag{900}{\begin{scriptsize}$900$\end{scriptsize}}
  \psfrag{25000}{\begin{scriptsize}$25000$\end{scriptsize}}
   \psfrag{510}{\begin{scriptsize}$510$\end{scriptsize}}
    
      \psfrag{T}{\hspace{8mm}AFCT (seconds)}
  \psfrag{b}{\hspace{23mm}Flow size (pkts)}
\psfrag{h1}{\begin{small}$B_1,$ $B_2$, $B$ = $15$ pkts\end{small}} 
\psfrag{x1}{\begin{small}$B_1,$ $B_2$ = $2084$, $B$ = $4100$ pkts\end{small}}
\psfrag{xyzxyz}{\begin{scriptsize}Capacity = $300$ Mbps\end{scriptsize}}

  \includegraphics[width=2.75in,height=3.75in,angle=270]{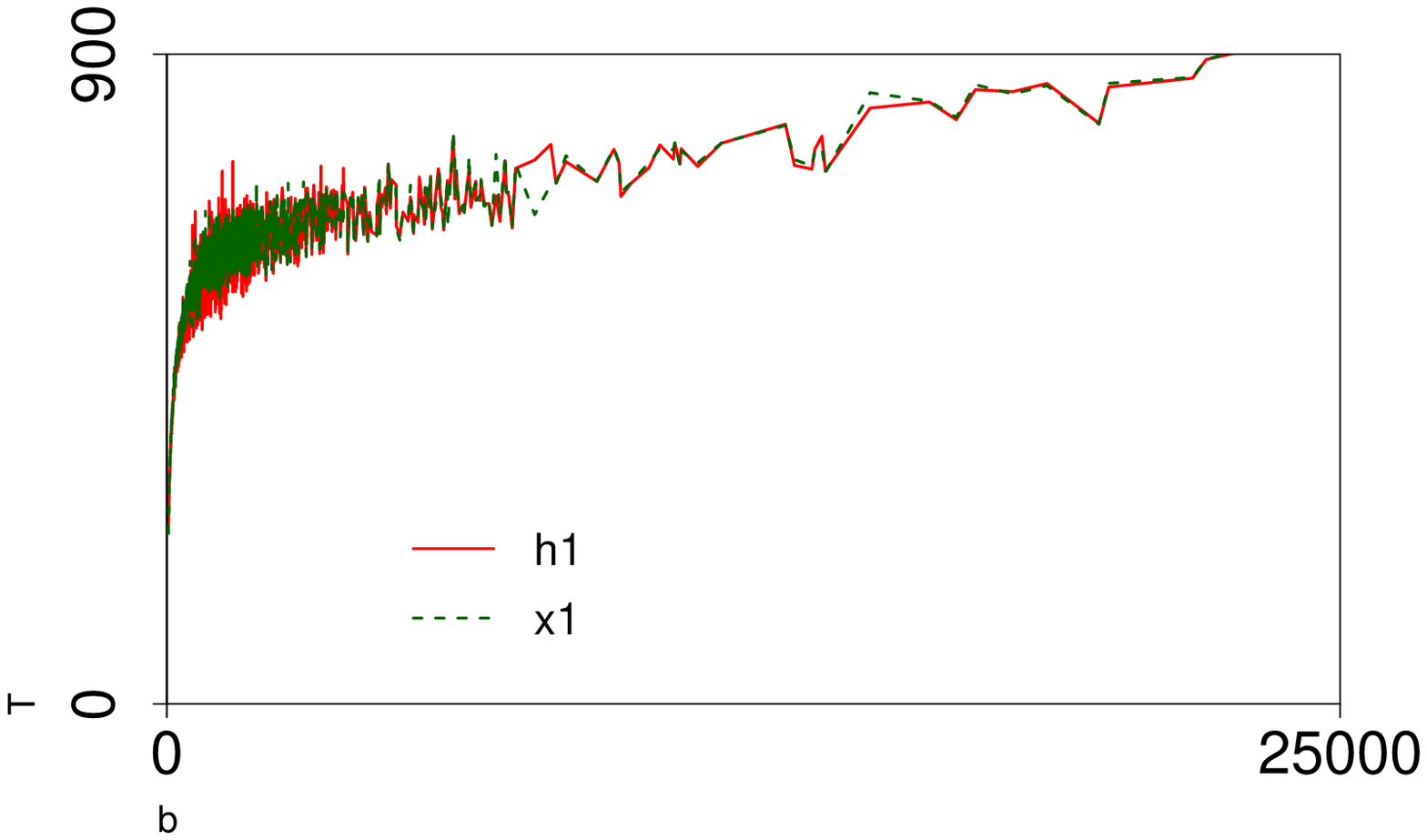}
  \caption{ \emph{Impact of buffer sizing on AFCT}. Two sets of $100$ Compound TCP sources, regulated by two edge routers each with a link capacity of $100$ Mbps. The outputs of the edge routers feed into a core router with a link capacity of $197$ Mbps. The round trip time of each set is $200$ ms. The file sizes are drawn from a Pareto distribution with mean $100$ KB and shape $1.5$. We consider two cases: $(i)$ buffer sizes at all routers are fixed at $15$ packets, $(ii)$ buffer sizes at all routers follow the bandwidth-delay product rule, used in practice. Observe that smaller buffers yield comparable AFCT as bandwidth-delay product worth of buffering. }
  \label{fig:multi_afct}
  \end{center}
\end{figure}

For this, we consider two sets of $100$ Compound TCP sources, regulated by two edge routers, each with a link capacity of $100$ Mbps. The outputs of the edge routers feed into a core router with a link capacity of $197$ Mbps. Each TCP source is connected to an edge router via an access link with a speed of $2$ Mbps. Further, each TCP source performs successive transfer of files according to a Poisson process, and file sizes are drawn from a Pareto distribution. The expected file size is $100$ KB and the shape parameter is $1.5.$ We consider the average duration between each transfer to be $0.1$ seconds. To study the impact of buffer sizing on flow completion times, we consider two cases: $(i)$ the buffer size at each router is fixed at $15$ packets, and $(ii)$ the buffer size at each router follows the bandwidth-delay product rule. With this, the buffer size is $2084$ packets at each router, and $4100$ packets at the core router. Fig. \ref{fig:multi_afct} shows the AFCT for these two cases. We can observe that smaller buffers yield comparable AFCTs as that with bandwidth-delay worth of buffering. Hence, it is indeed possible to significantly reduce buffer sizes without affecting flow completion times.

In summary, smaller buffers ensure stability without degrading the system performance. 

\section{Conclusions}
\label{conclusions}
In this paper, we conducted a performance evaluation of Compound TCP in two different topologies, with Drop-Tail queues, in a small buffer regime. For the traffic, we considered three scenarios. In the first scenario, we assume that only long-lived flows are present in the system. The second scenario constitutes a combination of long and short flows. The third scenario aims to capture the high variability present in real Internet traffic and considers heavy-tailed connections at the source level. Using a combination of analysis and packet-level simulations, we explored numerous dynamical and some statistical properties to obtain a few key insights.       

From a \emph{dynamical} perspective, we emphasised the interplay between buffer sizes and stability. In particular, we showed that smaller buffers tend to favour stability. On the other hand, larger buffer thresholds may help link utilisation, but they would increase queuing delay and are also prone to inducing limit cycles, via a Hopf bifurcation, in the queue size dynamics. However, such limit cycles can in turn lead to a drop in link utilisation, induce synchronisation among TCP flows, and make the downstream traffic bursty. We also noted that when a network has high variability in terms of the connections generated at the source level, such limit cycles continue to exist in the queue size despite the heterogenity in the incoming traffic. Some design considerations for protocol and network parameters, to ensure stability and low-latency queues, are also outlined. We repeatedly witnessed the existence of limit cycles in the queue size. To that end, it would be of both theoretical and practical interest, to establish the \emph{asymptotic orbital stability} of the bifurcating limit cycles and to the determine the \emph{type} of the Hopf bifurcation. One way to approach this analytically would be via the theory of normal forms and the centre manifold analysis \cite{Guckenheimer}, and such an analysis is conducted in~\cite{TCNS}.      

In terms of the \emph{statistical} properties, we examined the arrival process and the empirical queue distribution at the bottleneck queues, both in single and multiple bottleneck topologies. We observed that in the regime considered, each bottleneck queue can be well approximated by an $M/M/1/B$ or an $M/D/1/B$ queue, in the absence of synchronisation. Further, this approximation holds reasonably well even in the presence of high variability at the TCP connection level. This makes our system models amenable to analysis, and thus gives confidence in using the underlying models to better understand network performance and quality of service. 

In summary, our work recommends that buffer sizes at routers should be significantly reduced to ensure stability as well as low latency. We showed that design of such small buffers is indeed possible without compromising the system performance, namely, throughput and flow completion times.

The insights obtained in this thesis could have important consequences for the modelling and the performance evaluation of communication networks. From a theoretical
perspective, this opens many challenging questions centered around the development of accurate fluid models for different versions of TCP and queue management policies, and their interaction with different buffer sizing strategies. From a practical perspective,
of immediate interest would be to understand buffer sizing requirements with CUBIC TCP, which is the default protocol in the Linux OS.

\section*{References}
\label{references}


\end{document}